\documentclass[10pt]{article}
\usepackage{amssymb}\usepackage{amsmath}\usepackage{amsthm}
\usepackage{bbm}
\usepackage[arrow, matrix, curve]{xy}
\def\sqr#1#2{{\vcenter{\vbox{\hrule height.#2pt\hbox{\vrule width.#2pt
height#1pt \kern#1pt \vrule width.#2pt}\hrule height.#2pt}}}}
\def\d{\partial}

\def\=d{\,{\buildrel\rm def\over =}\,}

\def\1{{\leavevmode{\rm 1\ifmmode\mkern -4.8mu\else\kern -.3em\fi l}}}
\newcommand{\del}{\partial}

\newcommand{\Rcl}{R_\mathrm{cl}}
\newcommand{\ets}{e_\otimes^S}

\newcommand{\gr}{\Gamma_{\hspace{-2pt} \mathrm{ret}}}
\newcommand{\defi}{\stackrel{\scriptscriptstyle{\mathrm{def}}}{=}}

\newcommand{\dA}{\delta_A}
\newcommand{\ffgx}{\frac{\delta S_0}{\delta \varphi(x)}}

\newcommand{\beq}{\begin{equation}}
\newcommand{\eeq}{\end{equation}}
\newcommand{\bg}{\begin{gather}}
\newcommand{\eg}{\end{gather}}
\newcommand{\CC}{\mathbb C}
\newcommand{\RR}{\mathbb R}

\newcommand{\NN}{\mathbb N}
\newcommand{\MM}{\mathbb M}
\newcommand{\TT}{\mathbb T}
\newcommand{\supp}{{\rm supp\>}}

\newtheorem{prop}{Proposition}
\newtheorem{thm}[prop]{Theorem}
\newtheorem{cor}[prop]{Corollary}
\newtheorem{lemma}[prop]{Lemma}
\newtheorem{definition}{Definition}
\theoremstyle{remark}

\begin{document}
\thispagestyle{empty}
\bibliographystyle{diplbibl}
\title{Removal of violations of the Master Ward Identity in perturbative QFT}
\author{Ferdinand Brennecke\\[0mm]
{\small Institut f\"ur Quantenelektronik, ETH Z\"urich}\\
{\small CH-8093 Z\"urich, Switzerland}\\
{\tt \small brennecke@phys.ethz.ch}\\[2mm]
Michael D\"utsch\\[0mm] 
{\small Institut f\"ur Theoretische Physik, Universit\"at Z\"urich}\\
{\small CH-8057 Z\"urich, Switzerland}\\
{\small and Max Planck Institute for Mathematics in the Sciences}\\
{\small D-04103 Leipzig, Germany}\\
{\tt \small duetsch@physik.unizh.ch}\\[2mm]} 

\date{}
\maketitle
\begin{abstract}
We study the appearance of anomalies of the Master Ward Identity, which is a universal renormalization condition 
in perturbative QFT. The main insight of the present paper 
is that any violation of the Master Ward Identity can be expressed 
as a {\it local} interacting field; this is a version of the well-known Quantum Action Principle of 
Lowenstein and Lam. Proceeding in a proper field formalism by induction on 
the order in $\hbar$, this knowledge about 
the structure of possible anomalies as well as techniques of algebraic renormalization are used to remove possible
anomalies by finite renormalizations. As an example the method is applied to prove the Ward identities of the $O(N)$ scalar field model.
 

\end{abstract}

\tableofcontents
\section{Introduction}\setcounter{equation}{0}

In the quantization of a classical field theory symmetries and corresponding conservation laws are in general not maintained: due to the
distributional character of quantum fields the arguments valid for classical field theory are not applicable.
Therefore, in perturbative quantum field theory (pQFT) symmetries and conservation laws play the role of renormalization conditions 
(the 'Ward identities'). 


In \cite{Duetsch:2002yp} a universal formulation of Ward identities was studied
and termed 
Master Ward Identity (MWI). This identity --  which originally 
was proposed in \cite{Duetsch:2001sw} -- can be derived in the framework of 
classical field theory simply from the fact that classical fields can be multiplied pointwise.  
However, in pQFT the MWI serves as a highly non-trivial renormalization condition, which cannot 
be fulfilled in general due to the well known anomalies appearing in QFT.

In traditional renormalization theory (e.g.~BPHZ 
renormalization or dimensional renormalization) 
the question whether certain Ward identities can be fulfilled, is usually treated by means of 
algebraic renormalization\footnote{There is a huge literature about
algebraic renormalization. For shortness we cite only some of the founding articles 
and a few textbooks and reviews, which should suffice to understand this paper.}. 
This method relies on the Quantum Action Principle (QAP), which 
was derived by Lowenstein and Lam in the early seventies 
\cite{Lowenstein:1971jk, Lam:1972mb} and 
proved in several renormalization schemes \cite{Breitenlohner:1977hr}. 
The QAP characterizes the possible violations of Ward identities and,
hence, allows one to derive algebraic conditions 
whose solvability guarantees the existence of a renormalization maintaining 
the Ward identities. These conditions often lead to cohomological problems 
involving the infinitesimal symmetry operators which appear in the considered 
Ward identities. The application of this procedure to Yang-Mills theories
lead to a detailed study of BRST cohomology 
\cite{Becchi:1974md, Becchi:1975nq, Zinn-Justin:1999,
Henneaux:1992ig, Piguet:1995er, Barnich:2000zw}.

Traditionally, perturbation theory is done in the functional formulation of 
QFT which starts from the path integral. 
However, we take the point of view of algebraic pQFT 
\cite{Brunetti:1999jn, Duetsch:1998hf, Duetsch:2000de, Duetsch:2000nh, 
Duetsch:2004dd}, which is based on causal perturbation theory 
(Bogoliubov \cite{Bogoliubov:1959}, Epstein and Glaser \cite{Epstein:1973gw}) 
and concentrates on the algebraic structure of interacting fields. Starting 
with some well defined free QFT one separates UV-problems from IR-problems
by considering solely interactions with compact support. Whereas the UV-problem 
concerns the construction of time ordered or retarded products 
(in terms of which interacting fields are formulated), the IR-problem 
appears only in the construction of states on the algebra of local observables. 
The restriction on compactly supported interactions leaves it possible to 
construct the whole net of local observables \cite{Brunetti:1999jn}. Therefore, 
this approach seems to be well suited for a rigorous perturbative construction 
of quantum Yang-Mills theories, for example, where an adiabatic limit seems to 
be out of reach. However, in the non-Abelian case the construction of the net 
of local observables is still an open problem within the framework of 
algebraic pQFT. As it was worked out in \cite{Duetsch:2001sw}, 
the decisive input to reach this goal is the MWI resp.~certain 
cases of it. Motivated by this, the 
aim of our present work is to transfer techniques from algebraic 
renormalization theory into the framework of 
algebraic pQFT in order to gain more insight into the violations 
of the MWI and to find concrete conditions for the 
solvability of the MWI in relevant cases.

The paper is organized as follows: Sect.~\ref{classFT} deals with classical 
field theory for localized interactions.
We generalize the treatment given in Sect.~2 of \cite{Duetsch:2002yp} 
to the off-shell formalism, i.e.~the values 
of the retarded products are off-shell fields. In Sect.~\ref{pQFT} we 
summarize the quantization of perturbative 
classical field theory worked out in \cite{Duetsch:2004dd}
and give some completions. Algebraic renormalization proceeds
in terms of the 'vertex functional' (or 'proper function') 
$\Gamma$ (which is usually derived in the functional 
formulation of QFT along a Legendre transformation). Hence, to make accessible techniques 
of algebraic renormalization, we develop a proper field formalism, which describes the combinatorics of 
1-particle-irreducible (1PI) diagrams in a purely algebraic setting (Sect.~\ref{propV}). This is done
by reformulating perturbative QFT as a classical field theory with a non-local interaction $\Gamma$ which is a formal 
power series in $\hbar$. 

After these preparatory sections we turn in Sect.~\ref{MWI} to the maintenance of the MWI in the process of renormalization.
Starting from a derivation of the MWI in the off-shell formalism \cite{Duetsch:2004dd}, we 
prove an identity ('anomalous MWI') which gives a characterization of the possible violations of the MWI. 
More precisely, we find that the most general violation can be expressed in terms of a local interacting field. Translation of the anomalous MWI into the proper field formalism (introduced in Sect.~\ref{propV})
yields an identity which contains solely the 'quantum part' (loop part) 
of the original version and which is shown to be formally equivalent to the QAP. 
Crucial properties of the violating local terms appearing in the anomalous MWI are proved,
in particular an upper bound for 
the mass dimension. Equipped with this new insight, we transfer basic ideas of algebraic 
renormalization into the setting of causal perturbation theory. Usually, in the latter the maintenance 
of Ward identities in renormalization is proved by induction 
on the power of the coupling constant. 
In contrast, we proceed by induction on the power of  $\hbar$ similarly 
to algebraic renormalization.  
We find explicit conditions, whose solvability guarantees the existence of a 
renormalization prescription satisfying corresponding cases of the MWI.  
In addition, we prove that the MWI can always be fulfilled 
to first order in the coupling constant (that is to 
second order of the corresponding time ordered products). We apply these 
results to models fulfilling 
(classically) a localized off-shell version of Noether's Theorem
and find simplifications of the mentioned conditions. Finally, as a 
simple application of the method, we prove the Ward identities of the $O(N)$ 
scalar field model by using cohomological arguments. 

\section{Classical field theory for localized interactions}\setcounter{equation}{0}
\label{classFT}
In this section we generalize the formalism developed in Sect.~2 of 
\cite{Duetsch:2002yp} to off-shell fields. This will provide us with 
the necessary framework to derive the off-shell version of the MWI in 
Sect.~\ref{classMWI}. In order to keep the formulas as simple as possible, 
we study the theory of a real scalar field $\varphi$ on $d$ dimensional 
Minkowski space $\MM$, $d>2$. We interpret $\varphi$ and partial derivatives 
thereof as evaluation functionals on the configuration space 
$\mathcal{C}\equiv\mathcal{C}^{\infty}(\MM,\RR)\,$:
$(\d^a\varphi) (x)(h)=\d^ah(x),\>a\in\NN_0^d$. Let $\mathcal{F}({\cal C})$ 
be the space of all functionals  
\beq
F(\varphi)\,:\,\mathcal{C}\longrightarrow\CC\,,\,\quad
F(\varphi)(h)=F(h)\ ,\label{functionals}
\eeq
which are localized polynomials\footnote{A generalization to 
non-polynomial (localized) interactions is possible, see e.g.~\cite{BDF}.} in $\varphi$:
\begin{equation}
  F(\varphi)=\sum_{n=0}^N\int\! dx_1\ldots dx_n\,\varphi(x_1)
  \cdots\varphi(x_n)f_n(x_1,\ldots,x_n)=:\sum_n\langle f_n\,,\,\varphi^{\otimes n}\rangle\ ,
  \label{F(phi)} 
\end{equation}
where $N<\infty$ and the $f_n$'s are $\CC$-valued 
distributions with compact support, which are symmetric under 
permutations of the arguments and whose wave front 
sets satisfy the condition\footnote{$\overline{V}_{\pm}$ denotes the closure 
of the forward and backward light-cones, respectively, and $\overline{V}_{\pm}^{\,n}$ their $n$-fold direct products.} 
\begin{equation}
  \mathrm{WF}(f_n)\cap \bigl(\MM^n\times
(\overline{V}_+^{\,n}\cup \overline{V}_-^{\,n})\bigr)
   =\emptyset \label{WF}
\end{equation}
and $f_0\in\CC$. 
$\mathcal{F}$ is a commutative algebra with
the {\it classical product} $(F_1\cdot F_2)(h):=F_1(h)\cdot F_2(h)$. By the support of a functional 
$F\in\mathcal{F}$ we mean the support of $\tfrac{\delta F}{\delta\varphi}$.

The space of {\it local functionals} $\mathcal{F}_{\rm loc}({\cal C})\subset \mathcal{F}({\cal C})$ is defined as
\begin{equation}
\mathcal{F}_{\rm loc}({\cal C})\=d
  \Big\{\int\! dx\,\sum_{i=1}^N A_i(x)h_i(x) \equiv \sum_{i=1}^{N} 
  A_{i}(h_{i})\,|\, A_i\in {\cal P}\, ,\,h_i\in {\cal D}(\MM)\Big\}\ ,\label{F_loc}
\end{equation}
where ${\cal P}$ is the space of all polynomials of the field $\varphi$ and its 
partial derivatives. This representation of local functionals as smeared fields 
can be made unique by introducing the subspace of balanced fields ${\cal P}_{\rm bal}\subset{\cal P}$ 
\cite{Duetsch:2004dd,Dutsch:2005bm}:
\begin{gather}
{\cal P}_{\rm bal}\=d\Big\{P(\d_1,...,\d_n)\varphi(x_1)\cdots\varphi(x_n)
|_{x_1=\ldots=x_n=x}\,\vert\,P(\d_1,...,\d_n)\in P_n^{\rm rel}\ ,\,n\in\NN_0\Big\}
\end{gather}
where $P_n^{\rm rel}$ is the space of all polynomials in the 
'relative derivatives' $(\d_k-\d_l)\,,\,1\leq k<l\leq n\,$.
With that it holds:
given $F\in\mathcal{F}_{\rm loc}$ there exists a {\it unique}
$h\in{\cal D}(\MM,\mathcal{P}_{\text{bal}})$ with $h\vert_{\varphi=0}=0$ and
$F-F(0)=\int\! dx\, h(x)$.
Proofs are given in Proposition 3.1 of \cite{Duetsch:2004dd} and in Lemma 1 of \cite{Dutsch:2005bm}.

Since we are mainly interested in perturbation theory we consider action functionals of the form 
$S_{\mathrm{tot}}= S_0+\lambda\,S$ where $S_0 \defi \int \! dx \frac{1}{2}(\del_\mu \varphi 
\del^\mu \varphi -m^2 \varphi^2)$ denotes the free action, $\lambda$ a real parameter 
and $S\in \mathcal{F}(\mathcal{C})$ is some compactly supported interaction,
which may be non-local.\footnote{Note that the free action is not an element of $\mathcal{F}(\mathcal{C})$. 
Therefore we interpret $S_0$ as functional on the subspace of compactly supported functions in 
configuration space $\mathcal{C}$. However, this restriction is not necessary (and, hence, 
will not be done) for the free field equation: $\frac{\delta S_0}{\delta \varphi}$ is an element of $\mathcal{P}$.} 
We denote by $\Delta^\mathrm{ret}_{S_\mathrm{tot}}$ the retarded Green function corresponding to the action $S_\mathrm{tot}$, which is defined by
\begin{equation}\label{def:green}
\int\! dy\, \Delta^\mathrm{ret}_{S_\mathrm{tot}}(x, y)\frac{\delta^2 S_\mathrm{tot}}{\delta \varphi(y) 
\delta \varphi(z)}=\delta(x-z)=\int \!dy\, \frac{\delta^2 S_\mathrm{tot}}{\delta \varphi(x) 
\delta \varphi(y)}\Delta^\mathrm{ret}_{S_\mathrm{tot}}(y, z)
\end{equation}
and $\Delta^\mathrm{ret}_{S_\mathrm{tot}}(x,y)=0$ for $x$ sufficiently early. In the following we will assume 
that for all actions $S_\mathrm{tot}$ under consideration the retarded Green function exists and is unique
in the sense of formal power series in $\lambda$. Analytic expressions for $\Delta^\mathrm{ret}_{S_\mathrm{tot}}$ 
are in general unknown. However, perturbatively the retarded solution of (\ref{def:green})
can be given in terms of the (unique) retarded Green function 
$\Delta^\mathrm{ret}_{S_0}(x,y)=\Delta^\mathrm{ret}_m
(x-y)$ of the Klein Gordon operator:
\begin{lemma}
In the sense of formal power series in $\lambda$, the retarded Green function $\Delta^\mathrm{ret}_{S_0+\lambda S}$ is given by the following formula \cite{Duetsch:2002yp}:
\begin{gather}
\Delta^\mathrm{ret}_{S_0+\lambda S}(x, y)=\Delta^\mathrm{ret}_{S_0}(x, y)+\sum_{n=1}^\infty (-\lambda)^n \int\! d(u_1, \ldots, u_n)d(v_1, \ldots, v_n)\Delta^\mathrm{ret}_{S_0}(x, u_1)\notag \\
\cdot\frac{\delta^2 S}{\delta \varphi(u_1)\delta\varphi(v_1)}\cdots \Delta^\mathrm{ret}_{S_0}(v_{n-1}, u_n)\frac{\delta^2 S}{\delta \varphi(u_n)\delta\varphi(v_n)}\Delta_{S_0}^\mathrm{ret}(v_n, y).\label{retgreenformal}
\end{gather}
Its support is contained in the set
\begin{equation}
\{(x, y)|x\in \mathrm{supp}(\tfrac{\delta S}{\delta\varphi})+\overline{V}_+ \wedge y\in \mathrm{supp}(\tfrac{\delta S}{\delta\varphi})
+\overline{V}_-\}\cup \{(x, y)|x\in y+\overline{V}_+\}.
\end{equation}
In case of a local interaction $S\in \mathcal{F}_\mathrm{loc}(\mathcal{C})$ 
the support can be limited even stronger:
\begin{equation}\label{supplocal}
\mathrm{supp}\big(\tfrac{\delta\Delta^\mathrm{ret}_{S_0+S}}{\delta\varphi}\big)\subset \{(x,y)|x\in y+\overline{V}_+\}.
\end{equation}
\end{lemma}
Our assumption ensures that the pointwise product of distributions in (\ref{retgreenformal}) exists.
\begin{proof}
We multiply the left equation in \eqref{def:green} by $\Delta_{S_0}^\mathrm{ret}(z, u)$ and integrate 
over $z$ afterwards to obtain the relation
\begin{equation}
\Delta^\mathrm{ret}_{S_0+\lambda S}(x, u)=\Delta^\mathrm{ret}_{S_0}(x, u)-
\lambda\int \! dy dz \,\Delta^\mathrm{ret}_{S_0+\lambda S}(x, y)\frac{\delta^2 S}{\delta \varphi(y)\delta\varphi(z)}\Delta^\mathrm{ret}_{S_0}(z, u),
\end{equation}
which can be solved by recursion on the powers of $\lambda$ \eqref{retgreenformal}. For 
local interactions $S\in \mathcal{F}_\mathrm{loc}(\mathcal{C})$, the support property 
\eqref{supplocal} follows immediately from $\mathrm{supp}\big(\tfrac{\delta\Delta_{S_0}^\mathrm{ret}}{\delta\varphi}\big)
\subset \{(x, y)|x\in y+\overline{V}_+\}$ and $\tfrac{\delta^2 S}{\delta \varphi(u)\delta\varphi(v)}=0$ for
$u\ne v$. In the general case 
$S\in \mathcal{F}(\mathcal{C})$ we use $\mathrm{supp}\big(\frac{\delta^2 S}{\delta \varphi^2}\big)\subset 
\mathrm{supp}\big(\frac{\delta S}{\delta \varphi}\big)\times \mathrm{supp}\big(\frac{\delta S}
{\delta \varphi}\big)$ to conclude that $x$ has to lie in the future of $u_1\in \mathrm{supp}\big(\tfrac{\delta S}{\delta \varphi}\big)$ 
and $y$ in the past of $v_n\in \mathrm{supp}\big(\tfrac{\delta S}{\delta \varphi}\big)$ to get a non vanishing contribution of 
the integral in \eqref{retgreenformal}.
\end{proof}
The space of all smooth solutions of the Euler-Lagrange equation with respect to the action $S_\mathrm{tot}$ will be denoted by $\mathcal{C}_{S_\mathrm{tot}}\subset \mathcal{C}$. Interacting fields $F_S$, corresponding to some functional $F\in \mathcal{F}(\mathcal{C})$, could be defined by restricting $F$ to the space of solutions $\mathcal{C}_{S_0+S}\,$:
$F_S\defi F|_{\mathcal{C}_{S_0+S}}\,$. However, the idea of perturbative algebraic classical field theory is to introduce interacting fields as functionals on the space $\mathcal{C}_{S_0}$ of \emph{free} solutions. This corresponds to the usual interaction picture known from QFT where interacting fields are constructed as operators on the Fock space of the underlying free theory. Therefore one introduces retarded wave operators which map solutions of the free theory to solutions of the interacting theory \cite{Duetsch:2002yp}. As it turns out, it is convenient to construct such a map on the space $\mathcal{C}$ of all field configurations (\emph{off-shell formalism}) and not only on the space of solutions, as it was done in \cite{Duetsch:2002yp}:

\begin{definition}\label{def:waveop}
A retarded wave operator is a family of maps $(r_{S_0+S, S_0})_{S\in \mathcal{F}(\mathcal{C})}$ from $\mathcal{C}$ into itself with the properties
\begin{enumerate}
\item $r_{S_0+S, S_0}(f)(x)=f(x)$ for $x$ sufficiently early
\item $\frac{\delta(S_0+S)}{\delta \varphi}\circ r_{S_0+S, S_0}=\frac{\delta S_0}{\delta \varphi}$.
\end{enumerate}
\end{definition}
\begin{lemma}
The retarded wave operator $(r_{S_0+ S, S_0})_{S\in \mathcal{F}(\mathcal{C})}$ exists and is unique 
and invertible in the sense of formal power series in the interaction $S$.
\end{lemma}
\begin{proof}
To determine $r_{S_0+S, S_0}$ we multiply $S$ in Def.~\ref{def:waveop},\emph{(ii)} 
by a real parameter $\lambda$ and differentiate with respect to this parameter, ending up with
\begin{eqnarray*}
\int\! dy \frac{\delta(S_0+\lambda S)}{\delta \varphi(x) \delta \varphi(y)}\circ r_{S_0+\lambda S, S_0}\cdot \varphi(y)\circ \frac{d}{d\lambda}r_{S_0+\lambda S, S_0}+\frac{\delta S}{\delta \varphi(x)}\circ r_{S_0+\lambda S, S_0}=0.
\end{eqnarray*}
After multiplication of this equation with $\Delta_{S_0+\lambda S}^\mathrm{ret}(z, x)$ and integration over $x$, we obtain the following differential equation for $r_{S_0+\lambda S, S_0}$
\begin{equation}
\varphi(z)\circ \frac{d}{d \lambda}r_{S_0+\lambda S, S_0}=-\int\! dx \Big(\Delta^\mathrm{ret}_{S_0+\lambda S}(z, x)\cdot \frac{\delta S}{\delta \varphi(x)}\Big)\circ r_{S_0+\lambda S, S_0}.
\end{equation}
Finally, integration over $\lambda$ leads to the equation
\begin{equation}\label{waveoperator}
\varphi(z)\circ r_{S_0+S, S_0}=\varphi(z)-\int_0^1 \!d\lambda \int\! dx \Big(\Delta^\mathrm{ret}_{S_0+\lambda S}(z, x)\cdot\frac{\delta S}{\delta \varphi(x)}\Big)\circ r_{S_0+\lambda S, S_0},
\end{equation}
which can be solved iteratively in the sense of formal power series in the interaction $S$.
\end{proof}

We define the retarded wave operator $r_{S_0+S_1,S_0+S_2}$ connecting two interacting theories by 
\beq
r_{S_0+S_1,S_0+S_2}:=r_{S_0+S_1, S_0}\circ (r_{S_0+S_2, S_0})^{-1}\ .
\eeq
Obviously it fulfills  Def.~\ref{def:waveop}\emph{(i)} and 
$\frac{\delta(S_0+S_1)}{\delta \varphi}\circ 
r_{S_0+S_1, S_0+S_2}=\frac{\delta (S_0+S_2)}{\delta \varphi}$ and
\beq
r_{S_0+S_1,S_0+S_2}\circ r_{S_0+S_2,S_0+S_3}=r_{S_0+S_1,S_0+S_3}\ .
\eeq

Retarded fields (to the interaction $S$ and the free theory $S_0$
and corresponding to the functional $F\in \mathcal{F}(\mathcal{C})$)
are defined by
\begin{equation}
F^{\mathrm{ret}}_{S_0, S}\defi F\circ r_{S_0+ S, S_0}: \mathcal{C}\longrightarrow \CC.
\end{equation}
A crucial property of classical interacting fields -- which does \emph{not} hold anymore for interacting quantum fields -- is their factorization with respect to the classical product
\begin{equation}\label{interfactor}
(F\cdot G)^{\mathrm{ret}}_{S_0, S}=F^{\mathrm{ret}}_{S_0, S}\cdot G^{\mathrm{ret}}_{S_0, S}.
\end{equation}
This is why certain symmetry properties of classical field theory in general cannot be transferred directly into quantum field theory (see Sect.~\ref{classMWI}).

In classical field theory retarded products $R_\mathrm{cl}$ are defined as coefficients in the
expansion (with respect to the interaction) of interacting retarded fields \cite{Duetsch:2002yp}:
\begin{equation}\label{retprod}
R_\mathrm{cl}:\mathcal{F}(\mathcal{C})^{\otimes n}\otimes \mathcal{F}(\mathcal{C}) 
\rightarrow \mathcal{F}(\mathcal{C}),\quad R_\mathrm{cl}(S^{\otimes n}, F)\defi 
\frac{d^n}{d\lambda^n}\Big|_{\lambda=0}F\circ r_{S_0+\lambda S, S_0}.
\end{equation}
Interacting fields can then be written as
\begin{equation}
F^{\mathrm{ret}}_{S_0, S}\simeq\sum_{n=0}^\infty \frac{1}{n!}R_\mathrm{cl}(S^{\otimes n}, F)\equiv R_\mathrm{cl}(e_\otimes^S, F)
\end{equation}
where the r.h.s. of $\simeq$ is interpreted as a formal power series in $S$ (we do not care about convergence of the series). 
In the last expression $R_\mathrm{cl}$ is viewed as a linear map
\begin{equation}
R_\mathrm{cl}:\mathbb{T}\mathcal{F}(\mathcal{C})\otimes \mathcal{F}(\mathcal{C}) \rightarrow \mathcal{F}(\mathcal{C}),\label{R:map}
\end{equation}
where $\mathbb{T}\mathcal{V}\defi \CC\oplus\bigoplus_{n=1}^\infty \mathcal{V}^{\otimes n}$ denotes the tensor algebra corresponding to some vector space $\mathcal{V}$.

By introducing the differential operator
\begin{equation}
\mathcal{D}_{S_0, S}(\lambda)\defi -\int \!dx\,dy\,\Delta_{S_0+\lambda S}^\mathrm{ret}(x,y)
\frac{\delta S}{\delta \varphi(y)}\frac{\delta}{\delta \varphi(x)},
\end{equation}
we obtain from \eqref{waveoperator} the following explicit expression for the interacting field:
\begin{equation}
F^{\mathrm{ret}}_{S_0, S}\simeq F+\sum_{n=1}^\infty \int_0^1 \!d\lambda_1\int_0^{\lambda_1} \!d\lambda_2\ldots \int_0^{\lambda_{n-1}}\!d\lambda_n \,\mathcal{D}_{S_0, S}(\lambda_n)\ldots \mathcal{D}_{S_0, S}(\lambda_1) F.
\end{equation}
To first order in $S$ this formula reads (see also \cite{Duetsch:2002yp})
\begin{equation}\label{retpro1}
R_\mathrm{cl}(S, F)=-\int \!dx\,dy\,\Delta_{S_0}^\mathrm{ret}(x,y)\frac{\delta S}{\delta \varphi(y)}\frac{\delta F}{\delta \varphi(x)}.
\end{equation}

We can now endow classical fields with a Poisson structure: 
we introduce the (off-shell) Poisson bracket using Peierls definition 
\cite{Peierls:1952} (see also \cite{Marolf:1993af} and \cite{Duetsch:2002yp})
\begin{definition}
The Poisson bracket associated to the action $S\in {\cal F}({\cal C})$ is the map
\begin{gather}
\{\cdot, \cdot\}_S\,:\,\mathcal{F}(\mathcal{C})\otimes \mathcal{F}(\mathcal{C})\rightarrow \mathcal{F}(\mathcal{C})\\
\{F, G\}_S\defi R_S(F, G)-R_S(G, F).
\end{gather}
where
\beq
R_S(F, G)\defi \frac{d}{d\lambda}\Big|_{\lambda=0}G\circ r_{S+\lambda F, S}\ .\label{R_S}
\eeq
\end{definition}
The properties
\begin{itemize}
\item $R_S(F, G)=-\int \!dx\,dy\, \frac{\delta G}{\delta \varphi(x)}\,\Delta_{S}^\mathrm{ret}(x,y)\,
\frac{\delta F}{\delta \varphi(y)}$ (which is a generalization of (\ref{retpro1})) and
\item $\{\cdot, \cdot\}_S$ is indeed a Poisson bracket, i.e.~it satisfies the Leibniz rule and the Jacobi identity,
\end{itemize}
are proved in  \cite{Duetsch:2002yp} for the on-shell restrictions
$R^\mathrm{on-shell}_S(F,G)=R_S(F,G)\vert_{{\cal C}_S}$ and\\$\{F, G\}^\mathrm{on-shell}_S=
\{F, G\}_S\vert_{{\cal C}_S}$. These proofs can easily be generalized to $R_S$
and $\{\cdot, \cdot\}_S$. The Jacobi identity is derived from
\begin{gather}
\{R_S(H, F), G\}_S+\{F,R_S(H, G) \}_S=
R_S(H,\{F,G\}_S)+ \frac{d}{d\lambda}\Big|_{\lambda=0}\{F,G\}_{S+\lambda\, H}\ ,\label{cantrafo:inf}
\end{gather}
which is the infinitesimal version of the statement that the map ${\cal F}({\cal C})\ni F\mapsto F\circ
r_{S_1,S_2}\in {\cal F}({\cal C})$ is a canonical transformation:
\beq
\{F\circ r_{S_1,S_2},G\circ r_{S_1,S_2}\}_{S_2}=\{F,G\}_{S_1}\circ r_{S_1,S_2}\ .\label{cantrafo}
\eeq
In \cite{Duetsch:2002yp} only the  proof of the infinitesimal version (\ref{cantrafo:inf}) is given.
We are going to show that integration of (\ref{cantrafo:inf}), written in a suitable form, yields indeed 
(\ref{cantrafo}).\footnote{This proof is due to Klaus Fredenhagen.}
First note 
\beq
 \frac{d}{d\lambda}\Big|_{\lambda=0}\, F\circ r_{S,S+\lambda H}=
-\frac{d}{d\lambda}\Big|_{\lambda=0}\, F\circ r_{S+\lambda H,S}=-R_S(H,F)\ .\label{R_S1}
\eeq
Let $H:=S_2-S_1\,,\,\,S(\lambda):=S_1+\lambda H$ and $F_\lambda:=F\circ r_{S_1,S(\lambda)}$. With that we obtain
\begin{gather}
\frac{d}{d\lambda}\Big|_{\lambda=\lambda_0}\,\{F\circ r_{S_1,S(\lambda)},G\circ r_{S_1,S(\lambda)}\}_{S(\lambda)}
\circ  r_{S(\lambda),S_1}=\notag\\
\frac{d}{d\lambda}\Big|_{\lambda=\lambda_0}\Bigl(\{F_{\lambda_0}\circ r_{S(\lambda_0),S(\lambda)},
G_{\lambda_0}\circ r_{S(\lambda_0),S(\lambda)}\}_{S(\lambda)}\circ  
r_{S(\lambda),S(\lambda_0)}\Bigr)\circ  r_{S(\lambda_0),S_1}=0
\end{gather}
by using (\ref{R_S}), (\ref{R_S1}) and the infinitesimal version (\ref{cantrafo:inf}). Integrating this equation 
over $\lambda_0$ from $\lambda_0 =0$ to $\lambda_0 =1$ it results the assertion. 

Due to the perturbative expansion around the free theory only the Poisson bracket 
associated to the free action, $\{\cdot, \cdot\}_\mathrm{cl}\equiv \{\cdot, \cdot\}_{S_0}$, will be used 
in the following sections.


As one can easily check, the retarded products \eqref{retprod} have the same properties as the on-shell 
retarded products in \cite{Duetsch:2002yp} (which are related to (\ref{retprod}) by
$R^\mathrm{on-shell}_\mathrm{cl}(S^{\otimes n},F)=R_\mathrm{cl}(S^{\otimes n},F)\vert_{{\cal C}_{S_0}}$).
These are the properties which are used to define retarded products in perturbative QFT in an axiomatic way.
\section{Perturbative Quantum Field Theory}\setcounter{equation}{0}\label{pQFT}
We summarize the quantization of perturbative classical 
fields as it is worked out in \cite{Duetsch:2004dd}
on the basis of causal perturbation theory \cite{Bogoliubov:1959}, \cite{Epstein:1973gw} 
and work of Steinmann \cite{Steinmann:1971}.
Since the direct quantization of an interacting theory is in general not solved,
we quantize the free theory, around which the perturbative expansion is done 
(see Sect.~\ref{classFT}), by using deformation quantization: we replace 
$\mathcal{F}({\cal C})$ and $\mathcal{F}_\mathrm{loc}({\cal C})$ by 
$\mathcal{F}\=d \mathcal{F}({\cal C})[[\hbar]]$ and
$\mathcal{F}_{\mathrm{loc}}\=d \mathcal{F}_{\mathrm{loc}}({\cal C})[[\hbar]]$ 
resp.~(i.e.~all functionals are formal power 
series in $\hbar$) and deform the classical product into the 
$\star$-product, $\star\,:\, {\cal F}\times {\cal F}\rightarrow {\cal F}$, 
which is still associative but non-commutative and is defined by
\begin{gather}
  (F\star G)(\varphi):=
  \sum_{n=0}^{\infty}\frac{\hbar^n}{n!}
  \int\! dx_1\ldots dx_n dy_1\ldots dy_n 
  \frac{\delta^n F}{\delta\varphi(x_1)\cdots\delta\varphi(x_n)}\notag\\
  \cdot \prod_{i=1}^n H_m(x_i-y_i) 
  \frac{\delta^n G}{\delta\varphi(y_1)\cdots\delta\varphi(y_n)}\ .
  \label{*-product}
\end{gather}
There is a freedom in the choice of the 2-point function $H_m(x)$: 
it is required to differ from the Wightman 2-point function $\Delta^+_m(x)$
by a smooth and even function of $x$, to be Lorentz invariant and  
to satisfy the Klein Gordon equation.
The 'vacuum state' is the map (see \eqref{F(phi)})
\begin{equation}
  \omega_0\>:\>\mathcal{F} \longrightarrow \CC[[\hbar]]\ ,\,\,F\mapsto F(0)=f_0\ .
\label{vacuum}
\end{equation}

For the interacting quantum field $F_G$, $(F,G\in {\cal F}_{\rm loc})$ one makes the 
ansatz\footnote{With respect to 
factors of $\hbar$, our conventions ($R$) differ from \cite{Duetsch:2004dd} ($R^{DF}$),
namely $R(e_\otimes^G,F)=R^{DF}(e_\otimes^{G/\hbar},F)$. However, for the 
$T$-products we use the same conventions.}
of a \emph{formal power series} in the interaction $G$:
\begin{equation}
F_{G}=\sum_{n=0}^\infty\frac{1}{n!}R_{n,1}
\bigl(G^{\otimes n},F \bigr)\equiv  
R(e_\otimes^{G},F)\ .
\label{ansatz:intfield}
\end{equation}
The 'retarded product' $R_{n,1}$ is a {\bf linear} map, 
from ${\cal F}_{\rm loc}^{\otimes n}\otimes {\cal F}_{\rm loc}$ into ${\cal F}$ which is 
{\bf symmetric in the first $n$ variables}. The last expression in (\ref{ansatz:intfield})
is understood analogously to (\ref{R:map}). We interpret $R(A_1(x_1),...;A_n(x_n))\,$,
$A_1,...,A_n\in {\cal P}$, as $\mathcal{F}$-valued distributions on ${\cal D}(\MM^n)$,
which are defined by: $\int dx\, h(x)\,R(...,A(x),...):=R(...\otimes A(h)\otimes...)\>\>
\forall h\in {\cal D}(\MM)$.

Interacting fields are defined by the following axioms \cite{Duetsch:2004dd},
which are motivated by the principle that we want to maintain as much as possible 
of the classical structure in the process of quantization:\\
{\bf Basic axioms:}
\begin{description}
\item[Initial Condition.] 
$R(F)\equiv R_{0,1}(1,F)=F$.
\item[Causality.] $F_{G+H} = F_{G}\>\>$ if $\>\>\supp(\tfrac{\delta F}{\delta\varphi})\cap 
(\supp(\tfrac{\delta H}{\delta\varphi})+\bar V_+)=\emptyset$.
\item[GLZ Relation.] In the classical GLZ Relation we replace the Poisson 
bracket $\{\,\cdot\,,\,\cdot\}_{\rm cl}$ by 
\beq
\{\,\cdot\,,\,\cdot\}\=d \frac{1}{i\hbar}[\,\cdot\,,\,\cdot]_{\star}=
\{\,\cdot\,,\,\cdot\}_{\rm cl}+{\cal O}(\hbar)\label{Pb:QFT}
\eeq
(where $[H,\,F]_{\star}\equiv H\star F-F\star H$). This gives
\beq
\{F_{G},H_{G}\} = \frac{d}{d\lambda}\Big|_{\lambda=0}\, 
\left(F_{G+\lambda H}- H_{G+\lambda F}\right) \ .
\eeq
\end{description}
Based on these requirements, the retarded products $R_{n,1}$ can be constructed by 
induction on $n$. However, in each inductive step one is free to add a 
{\it local} functional, which corresponds to the usual renormalization ambiguity.  
This ambiguity is reduced by imposing the following further axioms:\\ 
{\bf Renormalization conditions:}
\begin{description}
\item[Unitarity.] $(F_{G})^{*} = F^{*}_{G^{*}}$
\item[Poincare Covariance.] For $L\in {\cal P}_+^\uparrow$ we set $\varphi_L(x):=\varphi(L^{-1}x)$ and
$h_L(x):=h(L^{-1}x),$ $h\in {\cal C},$ and define an automorphism
\beq
\beta_L\,:\,{\cal F}\rightarrow {\cal F}\,;\,\beta_L(\langle f_n,\varphi^{\otimes n}\rangle)=
\langle f_n,(\varphi_{L^{-1}})^{\otimes n}\rangle\ ,
\eeq
that is $(\beta_L\, F)(h)=F(h_{L^{-1}})$. ${\cal P}_+^\uparrow$-Covariance of the interacting fields means:

$\beta_L(F_G)=(\beta_L\,F)_{\beta_L\,G}\,$, $\,\forall L\in {\cal P}^\uparrow_+$.
\item[Field Independence.] 
The interacting field $F_G$ depends on $\varphi$ only through $F$ and $G$:
$\frac{\delta R}{\delta\varphi(x)}=0\,$. This condition is equivalent to the requirement that $R$
fulfills the causal Wick expansion \cite{Epstein:1973gw}, that is
\begin{gather}
R_{n-1,1}(A_1(x_1)\otimes \cdots \otimes A_{n-1}(x_{n-1}), A_n(x_n))=
\sum_{l_1, \ldots, l_n}\frac{1}{l_1!\cdots l_n!} \notag\\
 \times \omega_0\Bigl(R_{n-1, 1}\Big(\cdots \sum_{a_{i1}\ldots a_{il_i}}\frac{\del^{l_i}A_i}{\del(\del^{a_{i1}} \varphi)
\cdots \del(\del^{a_{il_i}}\varphi)}(x_i)\cdots\Big)\Bigr)\prod_{i=1}^{n}\prod_{j_i=1}^{l_i}\del^{a_{ij_i}}\varphi(x_i)\label{wickentP}
\end{gather}
with multi-indices $a_{ij_i}\in \mathbb{N}^d_0$.
\item[Field Equation.] 
\beq
\varphi_{G}(x) = \varphi(x) - \int\! \Delta^{\text{ret}}_{m}(x - y) 
\Bigl(\frac{\delta G}{\delta\varphi(y)}\Bigr)_G\, dy\quad ,\quad\forall G\in {\cal F}_{\rm loc}\ .
\label{FE}
\eeq
\item[Smoothness in $m$.]  Through the GLZ condition the interacting fields 
depend on the 2-point function $H_m$ and with that they depend on
the mass $m$ of the free field: $F_G\equiv (F_G)^{H_m}$. We require that the maps
  \begin{equation}
0\leq m\mapsto (F_G)^{H_m}\ ,\quad\quad F,G\in {\cal F}_{\rm loc}\ ,\label{smoothness:intfield}
\end{equation}
are smooth. In even dimensional spacetime this excludes the 2-point function 
$\Delta^+_m$ due to logarithmic singularities at $m=0$; more generally, 
homogeneous scaling of $H_m$ is not compatible 
with smoothness in $m\geq 0$. As in \cite{Duetsch:2004dd} we work with the 2-point function
$H^\mu_m$ (and the corresponding Feynman propagator) which is distinguished 
by almost homogeneous scaling \cite{BDF}. $H^\mu_m$ depends on an 
additional mass parameter $\mu>0$ and is explicitly given in Appendix A of 
\cite{Duetsch:2004dd}. For the corresponding star product,
retarded product and interacting fields we write 
$\star_{m,\mu}$, $R^{(m,\mu)}$ and $(F_G)^{(m,\mu)}
\equiv (F_G)^{H^\mu_m}$, respectively.
\item[$\mu$-Covariance.] The $\star$-products $(\star_{m,\mu})_{\mu>0}$ (and the
$\star$-product with respect to $\Delta^+_m$) are {\it equivalent}, that is there 
exists an invertible operator $(\mu_2/\mu_1)^\Gamma$ which intertwines these 
products (see e.g.~\cite{Duetsch:2000de}):
\beq
F\star_{m,\mu_2} G=({\mu_2}/{\mu_1})^{\Gamma^{(m)}}\Bigl(\Bigl(
({\mu_2}/{\mu_1})^{-\Gamma^{(m)}}F\Bigr)\star_{m,\mu_1}
\Bigl(({\mu_2}/{\mu_1})^{-\Gamma^{(m)}}G\Bigr)\Bigr)\ ,
\eeq
where $r^{\Gamma}\=d 1+\sum_{k=1}^\infty \frac{1}{k!}
(\mathrm{log}(r)\cdot\Gamma)^k$ (for $r>0$) and
\beq
\Gamma\equiv\Gamma^{(m)}\=d\int dx\>dy\> m^{d-2}\,\,f(m^2(x-y)^2)\>\frac{\delta^2}
{\delta\varphi(x)\delta\varphi(y)}\ .
\eeq
(The smooth function $f$ is explicitly given by formula (A.9) in \cite{Duetsch:2004dd}.)
The axiom $\mu$-Covariance requires that $(\mu_2/\mu_1)^\Gamma$ intertwines also the retarded 
products:\footnote{Given a linear map $f:\mathcal{V}\rightarrow \mathcal{V}$ (where $\mathcal{V}$
is a vector space) we define
\beq
\TT f\,:\,\TT\mathcal{V}\rightarrow\TT\mathcal{V}\,;\,(\TT f)(c\oplus\bigoplus_{j=1}^\infty
(v_{j1}\otimes...\otimes v_{jj}))=c\oplus\bigoplus_{j=1}^\infty
(f(v_{j1})\otimes...\otimes f(v_{jj}))\ .
\eeq
  }
\beq
R^{(m,\mu_2)}=({\mu_2}/{\mu_1})^{\Gamma^{(m)}}\circ
R^{(m,\mu_1)}\circ \TT({\mu_2}/{\mu_1})^{-\Gamma^{(m)}}\ .\label{mu-dep}
\eeq
\item[Scaling.] The \textit{mass dimension of a monomial} in $\mathcal{P}$ is 
defined by the conditions
\begin{equation}
  \mathrm{dim}(\d^a\varphi)=\frac{d-2}{2}+|a|\quad
\mathrm{and}\quad \mathrm{dim}(A_1A_2)=\mathrm{dim}(A_1)+
\mathrm{dim}(A_2)\label{UV-dim}
\end{equation}
for all monomials $A_1,A_2\in {\cal P}$. The mass dimension of a
\textit{polynomial} in $\mathcal{P}$ is the maximum of the mass dimensions
of the contributing monomials. We denote by ${\cal P}_{\text{hom}}$ the set
of all field polynomials which are homogeneous in the mass dimension.
A scaling transformation $\sigma_{\rho}$ is introduced as an 
automorphism of $\mathcal{F}$ (considered as an algebra with 
the classical product) by 
\begin{equation}
  \sigma_{\rho}(\langle f_n,\varphi^{\otimes n}\rangle)\=d
\rho^{\frac{n(2-d)}{2}}\int dx_1...dx_n\,f_n(x_1,...,x_n)\,\varphi(x_1/\rho)
... \varphi(x_n/\rho)\ . \label{sigma}
\end{equation}
For $A\in{\cal P}_\mathrm{hom}$ we obtain
$\rho^{\mathrm{dim}(A)}\sigma_\rho (A(\rho x))=A(x)\,$.
Our condition of {\bf almost homogeneous scaling} states that
\begin{equation}
\sigma_\rho\circ R^{(\rho^{-1}m,\mu)}_{n,1}\circ (\sigma_\rho^{-1})^{\otimes (n+1)}\ ,
\quad\quad n\in\NN_0\ ,\quad m\geq 0\ ,\quad\mu>0\ ,\label{scaling}
\end{equation}
is a polynomial in $({\rm log}\>\rho)$.
\end{description}
The construction of the retarded products 
proceeds in terms of the distributions $R(A_1(x_1),...;A_n(x_n))\,$,
$A_1,...,A_n\in {\cal P}$.
Since the retarded products depend only on the functionals (and not on how the 
latter are written as smeared fields (\ref{F_loc})),
they must satisfy the {\bf Action Ward Identity (AWI)} \cite{Duetsch:2004dd}:
\begin{equation}
  \d^x_{\mu}R_{n-1,1}(\ldots A_k(x)\ldots)= 
   R_{n-1,1}(\ldots,\d_{\mu}A_k(x),\ldots)\ .
\end{equation}
The AWI can simply be fulfilled by constructing $R(A_1(x_1),...;A_n(x_n))$
first only for balanced fields $A_k\in\mathcal{P}_{\text{bal}}\,\,\forall k$, and 
by using the AWI and linearity for the extension to general fields 
$A_k\in\mathcal{P}$.

The axioms Smoothness in $m$, $\mu$-Covariance and Scaling can be replaced by the
weaker axiom {\bf Scaling Degree}, which
requires  that 'renormalization may not make the interacting fields 
more singular' (in the UV-region). Usually this is formulated in terms of 
Steinmann's \textit{scaling degree} \cite{Steinmann:2000nr}:
 \begin{equation}
{\rm sd}(f)\=d {\rm inf}\{\delta\in \RR\>|\>\lim_{\rho\downarrow 0}
\rho^\delta f(\rho x)=0\},\quad f\in {\cal D}'(\RR^k)
\>\>\mathrm{or}\>\> f\in {\cal D}'(\RR^k\setminus \{0\}).\label{sd}
\end{equation}
Namely, one requires
\beq
{\rm sd}\Bigl(\omega_0(R(A_1,...;A_n))(x_1-x_n,...)\Bigr)\leq\sum_{j=1}^n
{\rm dim}(A_j)\ ,\quad\forall A_j\in {\cal P}_{\rm hom}\ ,\label{axiom-sd}
\eeq
where Translation Invariance is assumed.

In the inductive construction of the sequence $(R_{n,1})_{n\in \NN}$ (given in
\cite{Duetsch:2004dd}), the problem of renormalization appears as the 
extension of $\CC[[\hbar]]$-valued 
distributions from ${\cal D}(\RR^{dn}\setminus\{ 0\})$ to ${\cal D}(\RR^{dn})$.
This extension has to be done in the sense of formal power series 
in $\hbar$, that is individually in each order in $\hbar$. With that it holds
\beq
\lim_{\hbar\to 0} R=\Rcl\ .\label{R->R_cl}
\eeq
Namely, the GLZ Relation is the only axiom which depends explicitly on $\hbar$
and in the classical limit it goes over into the classical GLZ Relation, 
due to (\ref{Pb:QFT}).

The retarded product, having two different kinds of arguments, can be derived from 
the {\bf time ordered product} ('$T$-product') $T\,:\,
\TT {\cal F}_{\rm loc}\rightarrow {\cal F}$, which is totally symmetric i.e.~it has only 
one kind of arguments. The corresponding relation is Bogoliubov's formula: 
\beq
 R\bigl(e_\otimes^S\,,\,F\bigr)=-i\hbar\,\mathbf{S}(S)^{-1}
\star  \frac{d}{d\tau}\vert_{\tau=0}\, \mathbf{S}(S+\tau F)\ ,\quad 
 \mathbf{S}(S)\equiv T\bigl(e_\otimes^{iS/\hbar}\bigr)\ ,\label{R1-T}
\eeq
The axioms for retarded products translate directly into 
corresponding axioms for $T$-products, see Appendix E of \cite{Duetsch:2004dd}.
There is no axiom corresponding to the GLZ Relation. The latter can be interpreted 
as 'integrability condition' for the 'vector potential' $R\bigl(
e_\otimes^S\,,\,F\bigr)$, that is it ensures the existence of the
'potential' $\mathbf{S}(S)$  fulfilling (\ref{R1-T}); for details see \cite{BDF}.
(A derivation of the GLZ Relation from (\ref{R1-T}) is given in \cite{Duetsch:1998hf}.)

In \cite{Duetsch:2004dd} it is shown that there {\bf exist} retarded products which fulfill all 
axioms. The {\bf non-uniqueness} of solutions is characterized by the 'Main Theorem'; 
we use the version given in \cite{Duetsch:2004dd}:
\begin{thm} \label{maintheorem}
(a) Let $R$ and $\hat{R}$ be retarded products which fulfill the basic axioms
and the renormalization conditions Unitarity,
${\cal P}_+^\uparrow$-Covariance, Field Independence and Field Equation.
Then there exists a unique, symmetric and linear map
\begin{equation}\label{map-D}
D\>:\:\TT\mathcal{F}_{\mathrm{loc}}\longrightarrow \mathcal{F}_{\mathrm{loc}}
\end{equation}
with $D(1)=0$, $D(F)=F$ ($\forall F\in\mathcal{F}_{\text{loc}}$), 
such that for all $F,S\in {\cal F}_{\mathrm{loc}}$ the 
following intertwining relation holds (in the sense of 
formal power series in $\lambda$)
   \begin{equation}
        \hat{R}(e_{\otimes}^{\lambda S}, F) = 
        R(e_{\otimes}^{D(e_{\otimes}^{\lambda S})},
        D(e_{\otimes}^{\lambda S}\otimes F)) \ .
        \label{ren:intf}
   \end{equation}
In addition, $D$ satisfies the conditions:
\begin{enumerate}
\item $\mathrm{supp}\Big(\dfrac{\delta D(F_1\otimes \cdots \otimes F_n)}{\delta \varphi}\Big)
\subset \bigcap_{i\in \underline{n}}
\mathrm{supp}\Big(\frac{\delta F_i}{\delta \varphi}\Big), \quad 
F_i\in \mathcal{F}_{\mathrm{loc}}$
\item $\frac{\delta D(e_\otimes^F)}{\delta \varphi}=
D\Big(e_\otimes^F\otimes \frac{\delta F}{\delta \varphi}\Big)$
\item $D(e_\otimes^F\otimes \varphi(h))=\varphi(h), \quad 
\forall F\in \mathcal{F}_{\mathrm{loc}}$
\item $D(F^{\otimes n})^*=D((F^*)^{\otimes n})$
\item $\beta_L\circ D = D\circ\TT\beta_L\ , \quad \forall L\in \mathcal{P}_{+}^\uparrow$\label{endren}
\item (A) If $R^{(m,\mu)},\hat{R}^{(m,\mu)}$ are smooth in $m\ge 0$ and satisfy the 
axioms $\mu$-Covariance and Scaling, then the corresponding $D^{(m,\mu)}$ is also 
smooth in $m$, invariant under scaling
    \begin{equation}
    \sigma_\rho \circ D^{(\rho^{-1}m,\mu)}\circ\TT \sigma_\rho^{-1}=
    D^{(m,\mu)}\ .\label{D:scaling}
    \end{equation}
and $\mu$-covariant \footnote{The claim in \cite{Duetsch:2004dd} that $D^{(m,\mu)}$
is independent of $\mu$, is wrong.}
\beq
D^{(m,\mu_2)}=({\mu_2}/{\mu_1})^{\Gamma^{(m)}}\circ
D^{(m,\mu_1)}\circ \TT({\mu_2}/{\mu_1})^{-\Gamma^{(m)}}\ .
\eeq
(B) Alternatively, if $R$ and $\hat{R}$ satisfy the axiom Scaling degree, then 
\beq 
{\rm sd}\Bigl(\omega_0(D(A_1,...,A_n))(x_1-x_n,...)\Bigr)\leq\sum_{j=1}^n 
{\rm dim}(A_j)\ ,\quad\forall A_j\in {\cal P}_{\rm hom}\ .\label{scalingdeg:D} 
\eeq 
\noindent (b)Conversely, given $R$ and $D$ as above, equation (\ref{ren:intf}) gives a new 
retarded product $\hat{R}$ which satisfies the axioms.
\end{enumerate}
\end{thm}
Since the classical limit of the axioms has a unique solution (which is $\Rcl$), 
the map $D$ is  trivial to lowest order in $\hbar$, i.e.
\begin{equation}
D(e_\otimes^S)=S+\mathcal{O}(\hbar) \quad \mbox{and} 
\quad D(e_\otimes^S\otimes F)=F+\mathcal{O}(\hbar)\ .
\end{equation}

The relation (\ref{ren:intf}) can equivalently be expressed in terms 
of time ordered products,
\beq
\hat T(e_\otimes^{iS/\hbar})=T\Bigl(e_\otimes^{i\,D(\ets)/\hbar}\Bigr)\ ,\label{mainthm:T}
\eeq
where $T$ and $\hat T$ are the time ordered products belonging to $R$ and $\hat R$, resp.,
according to (\ref{R1-T}).

\section{Proper vertices for $T$-products}\setcounter{equation}{0}\label{propV}
Proper vertices\footnote{In the literature proper vertices (or the 'proper
interaction') are sometimes called 'effective vertices' 
(or 'effective interaction' resp.). However, differently to what we 
are doing here, the notion 'effective field theory' usually means an 
{\it approximation} to the perturbation series. For this reason we omit the word 
'effective' and use the terminology of Sect.~6-2-2 in 
\cite{IZ:1980}.}
are an old and standard tool in perturbative QFT.
In terms of $R$-products the basic idea is the following: 
a perturbative QFT can be rewritten as {\it classical}
field theory with a {\it non-local interaction} 
('proper interaction') which agrees to lowest order in $\hbar$ with the 
original local interaction. Since $\Rcl$ is the sum of all 
(connected) tree diagrams (as explained in Appendix A), this rewriting
means that we interpret each diagram
as tree diagram with non-local vertices ('proper vertices') given by the   
1-particle-irreducible (1PI) subdiagrams. This structural decomposition of Feynman 
diagrams can just as well be done for $T$-products and it is this latter form 
of proper vertices which is well known in the literature. 
Since $T$-products are totally symmetric, it is simpler to introduce proper 
vertices in terms of $T$-products than in terms of $R$-products and, 
hence, we work with the former (for the introduction of proper vertices 
for $R$-products see Appendix A).

A main motivation to introduce proper vertices is that the renormalization 
of an arbitrary diagram reduces to the renormalization of its 
1PI-subdiagrams. Indeed, due to the validity of the MWI for tree diagrams 
(i.e.~in classical FT), the MWI can equivalently be formulated 
in terms of proper vertices (i.e.~in terms of 1PI-diagrams), see Sect.~\ref{effMWI-QAP}.
This 'proper MWI' formally coincides with the usual formulation of Ward identities 
in the functional approach to QFT (for an overview see e.g.~\cite{Piguet:1995er}).
\subsection{Diagrammatics and definition of the 1-particle-irreducible
part $T^{\rm 1PI}$ of the time ordered product}\label{1part-irr}
To introduce proper vertices we need the tree part $T_{\rm tree}$
for {\it non-local entries} and, for later purpose, the 1PI part $T^{\rm 1PI}$
of the time ordered product $T$. The definition of $T_{\rm tree}$ can obviously be 
given in terms of Feynman diagrams; but in case of $T^{\rm 1PI}$ we are faced with the 
problem that for loop diagrams the decomposition of $T(A_1(x_1),...)$ into contributions 
of Feynman diagrams is {\it non-unique}, due to the local terms coming from renormalization.

To motivate the definition of $T^{\rm 1PI}$ we first study a 'smooth and symmetric 
$\star$-product': let $f\in {\cal C}^\infty (\RR^4,\CC)$ with $f(x)=f(-x)\, ,\,
\forall x$. We define $\star_f\,:\,{\cal F}\otimes {\cal F}\rightarrow {\cal F}$ 
by replacing in the definition (\ref{*-product}) of the $\star$-product the 2-point function
$H_m$ by $f$. This product, $\star_f$, is associative and commutative. 
By definition $\star_f$ satisfies 'Wick's theorem'. 
Due to that, $\star_{f,j=1}^n F_j\equiv F_1\star_f...\star_f F_n$ can uniquely 
be viewed as a sum of diagrams. In spite of the possible non-locality 
of the $F_j$'s, we symbolize each $F_j$ by {\it one} vertex. The contractions are 
symbolized by inner lines connecting the vertices, which are not oriented
due to $f(x)=f(-x)$. (We do not draw any external lines.)

The contribution of the connected diagrams (denoted by $(\star_{f,j=1}^n F_j)^c$) 
is obtained from $\star_{f,j=1}^n F_j$ by subtraction of the contributions 
of all disconnected diagrams; this gives the recursion relation (see e.g. \cite{Duetsch:2000nh})
\beq
(\star_{f,j=1}^n F_j)^c=\star_{f,j=1}^n F_j-
\sum_{|P|\geq 2}\prod_{J\in P}(\star_{f,j\in J} F_j)^c\ ,\label{star^c_f}
\eeq
where the sum runs over all partitions $P$ of $\{1,...,n\}$ in at least two 
subsets and $\prod$ means the classical product. One easily sees that the linked cluster 
theorem applies to $\star_f$: 
\beq
e_{\star_f}^F={\rm exp}_\bullet (e_{\star_f^c}^F)\ ,\quad\mathrm{where}\quad
e_{\star_f}^F\=d 1+\sum_{n=1}^\infty\frac{F^{\star_f\,n}}{n!}\ ,\quad
e_{\star_f^c}^F\=d \sum_{n=1}^\infty\frac{(F^{\star_f\,n})^c}{n!}
\label{lct:star}
\eeq
(with $F^{\star_f\,n}$ being the $n$-fold product $F\star_f...\star_f F$)
and ${\rm exp}_\bullet$ denotes the exponential function with respect to the 
classical product.

Analogously to (\ref{star^c_f}) the contribution of 
all 1PI-diagrams to $\star_{f,j=1}^n F_j$ (denoted by $(\star_{f,j=1}^n F_j)^{\rm 1PI}$),
is obtained from the connected diagrams $(\star_{f,j=1}^n F_j)^c$ by subtracting
the contributions of all connected one-particle-reducible diagrams. To formulate this
we need the contribution of all connected tree diagrams 
to $(\star_{f,j=1}^n F_j)^c$, which we denote by $(\star_{f,j=1}^n F_j)^c_{\rm tree}$.
This diagrammatic definition of $(\star_{f,j=1}^n F_j)^c_{\rm tree}$ fulfills 
the following unique and independent characterizations:
\begin{itemize}
\item {\it By recursion:} one easily finds that the connected tree diagrams
satisfy the recursion relation
\begin{gather}
(\star_{f,j=1}^{n+1} F_j)^c_{\rm tree}=\sum_{k=1}^n\int\! dx_1...dx_k\,
dy_1...dy_k\,\frac{\delta^k F_{n+1}}{\delta\varphi(x_1)...\delta\varphi(x_k)}\,\notag\\
\prod_{j=1}^k f(x_j-y_j)\,\frac{1}{k!}\,\sum_{I_1\sqcup...\sqcup I_k=\{1,...,n\}}
\frac{\delta}{\delta\varphi(y_1)}(\star_{f,j\in I_1} F_j)^c_{\rm tree}\cdot...\notag\\
\hspace{4.5cm}\cdot\frac{\delta}{\delta\varphi(y_k)}(\star_{f,j\in I_k} F_j)^c_{\rm tree}\ ,
\label{conntree:recursion}
\end{gather}
where $I_j\not=\emptyset\,\,\forall j\,$,  
$\sqcup$ means the disjoint union. 
(Note that in the sum over $I_1,...,I_k$ the succession of $I_1,...,I_k$
is distinguished and, hence, there is a factor of $\frac{1}{k!}$.)
\item {\it By the power in $\hbar$:} as explained in Sect.~5.2 
of \cite{Duetsch:2000nh} it holds
\beq
(\star_{f,j=1}^n F_j)^c={\cal O}(\hbar^{n-1})\quad\mathrm{for}\quad
F_1,...,F_n\sim\hbar^0\ ,\label{star^c_f:hbar}
\eeq
and the contribution of all tree diagrams is given by the 
terms of lowest order in $\hbar$
\beq
(\star_{f,j=1}^n F_j)^c_{\rm tree}=\hbar^{n-1}\,\,\lim_{\hbar\to 0}\,
\hbar^{-(n-1)}\,(\star_{f,j=1}^n F_j)^c\ .\label{star^c_f,tree}
\eeq
\end{itemize}
The contribution of all tree diagrams to $\star_{f,j=1}^n F_j$, which we denote by
$(\star_{f,j=1}^n F_j)_{\rm tree}$, is related to the connected tree diagrams 
by the linked cluster theorem: $e_{\star_{f,{\rm tree}}}^F=
{\rm exp}_\bullet (e_{\star_{f,{\rm tree}}^c}^F)$. (This follows immediately from 
(\ref{lct:star}) by selecting all tree diagrams.)

We now give the (above announced) unique recursive characterization of the
1PI-diagrams:
\beq
(\star_{f,j=1}^n F_j)^{\rm 1PI}=(\star_{f,j=1}^n F_j)^c-
\sum_{|P|\geq 2}\Bigl(\star_{f,J\in P}
(\star_{f,j\in J} F_j)^{\rm 1PI}\Bigr)^c_{\rm tree}\ .\label{star^1PI_f}
\eeq
The formulas (\ref{star^c_f}), (\ref{star^c_f:hbar}) and (\ref{star^c_f,tree})
hold also for the usual $\star$-product (i.e.~with $H_m$ instead of $f$)
\cite{Duetsch:2000nh}; but (\ref{star^1PI_f}) needs to be refined, because $H_m$
is not symmetrical and, hence, $\star_{H_m}$ is not commutative
(in particular $(F_1\star_{H_m}...\star_{H_m} F_n)^c_{\rm tree}$ is not symmetrical).

Turning to the {\it time ordered product} $T$, we will use (\ref{star^c_f}) 
and (\ref{star^1PI_f}) as motivation for the 
(recursive) {\it definition} of the connected part $T^c$ and the 
1PI-part $T^{\rm 1PI}$ of $T$, respectively. So we define \cite{Duetsch:2000nh}
\beq
T^c(\otimes_{j=1}^n F_j)\=d T(\otimes_{j=1}^n F_j)-
\sum_{|P|\geq 2}\prod_{J\in P}T^c(\otimes_{j\in J} F_j)\ .\label{T^c}
\eeq
It follows that $T$ and $T^c$ are related by the linked cluster theorem (\ref{lct:star}):
$T(e_\otimes^{iF})={\rm exp}_\bullet (T^c(e_\otimes^{iF}))$.

The following definition of the connected tree part $T^c_{\rm tree}$ of $T$ applies also 
to {\it non-local entries:}\footnote{In QCD the interaction $S=\kappa S_1+\kappa^2 S_2$
is a sum of a term of first order in the coupling 
constant $\kappa$, $S_1\sim\int\! gAA\d A$ $(g\in {\cal D}(\MM,\RR))$, 
and a term of second order in $\kappa$,
$S_2\sim\int\! g^2AAAA$. One can achieve that the order in $\kappa$ agrees with the 
order of the $T$- (or $R$-)product \cite{DHKS}. Namely, one starts with $T_1(S_1)$, 
the term $S_2$ is generated by a non-trivial renormalization of a certain tree diagram: 
in $T_2(S_1^{\otimes 2})\sim\int\! dx\,dy\,gAA(x)\,gAA(y)\,\d\d
\Delta^F(x-y)+...$ the propagator $\d^\mu\d^\nu\Delta^F(x-y)$ is replaced by
$\d^\mu\d^\nu\Delta^F(x-y)-1/2\,g^{\mu\nu}\,\delta(x-y)$. Due to the inductive 
procedure of causal perturbation theory this additional term propagates to higher 
orders such that this modified $T$-product, $T^N$, yields the same $S$-matrix:
$T^N(e_\otimes^{i\kappa S_1})=T(e_\otimes^{i(\kappa S_1+\kappa^2 S_2)})$ in the sense 
of formal power series in $\kappa$. (The corresponding renormalization map $D$
(\ref{map-D}) is given in \cite{Duetsch:2005}.)
Our definitions of $T^c_{\rm tree}$ (\ref{def:T^c_tree})
and $T_{\rm tree}$ (\ref{def:T_tree}) do not contain this
$1/2\,g^{\mu\nu}\,\delta$-term, in agreement with the definition of 
$R_{\rm cl}$ (\ref{retprod}). Generally, in this paper {\it all} terms $S_n$
of the interaction $S=\sum_{n\geq 1}\kappa^n S_n$ enter the 
perturbative construction of the $S$-matrix (or interacting field) already to 
first order of the $T$- (or $R$-)product.}
\beq
T^c_{{\rm tree},n}\>:\>{\cal F}^{\otimes n}\rightarrow {\cal F}\>;\>
T^c_{{\rm tree},n}(\otimes_{j=1}^n F_j)=
(F_1\star_{\Delta^F}...\star_{\Delta^F} F_n)^c_{\rm tree}\ ,\label{def:T^c_tree}
\eeq
i.e.~we replace in the definition of $(\star_f...)^c_{\rm tree}$ the smooth 
function $f$ by the Feynman propagator 
\begin{gather}
\Delta^F(z)\equiv\Delta^{F\,\mu}_m(z)=\Theta(z^0)\,H^\mu_m(z)+\Theta(-z^0)\,H^\mu_m(-z)\notag\\
=-i\,\Delta^{\rm ret}_m(z)+H^\mu_m(-z)=\Delta^F(-z)\ .\label{Feynprop}
\end{gather}
For tree diagrams the resulting expressions are well defined, since pointwise products of Feynman 
propagators do not appear. Obviously $T^c_{\rm tree}$ fulfills the recursion relation
(\ref{conntree:recursion}). In case of local entries another unique 
characterization of $T^c_{\rm tree}$ is possible: 
doing renormalization individually in each order in $\hbar$, $T^c$ satisfies
(\ref{star^c_f:hbar}) and $\hbar^{-(n-1)}\,T^c_{{\rm tree},n}$ is the classical limit
of $\hbar^{-(n-1)}\,T^c_n$ similarly to (\ref{star^c_f,tree}). (This is the translation of
(\ref{R->R_cl}) into $T$-products, see Sect.~5.2 of \cite{Duetsch:2000nh}.)
Analogously one defines the tree part $T_{\rm tree}$ of $T$ by
\beq
T_{{\rm tree},n}(\otimes_{j=1}^n F_j)\=d 
(F_1\star_{\Delta^F}...\star_{\Delta^F} F_n)_{\rm tree}\ ,
\quad F_j\in {\cal F}\,\,\forall j\ .\label{def:T_tree}
\eeq
Obviously the linked cluster theorem for $(\star_{f,{\rm tree}},\star_{f,{\rm tree}}^c)$
is valid also for $f=\Delta^F$:
$T_{\rm tree}(e_\otimes^{iF})={\rm exp}_\bullet (T^c_{\rm tree}(e_\otimes^{iF}))\,$.

Since $T^c_{\rm tree}$ is totally symmetric we may define $T^{1PI}$ in
analogy to (\ref{star^1PI_f}) by the recursive formula
\begin{gather}
T^{1PI}(F^{\otimes n})\=d T^c(F^{\otimes n})-
\sum_{k=2}^n\sum_{l_1+...+l_k=n\,,\,l_j\geq 1\,\,\forall j}\frac{n!}
{k!\,l_1!...l_k!}\cdot\notag\\
\cdot T^c_{\rm tree\,,\,k}\Bigl(T^{1PI}(F^{\otimes l_1})\otimes...
\otimes T^{1PI}(F^{\otimes l_k})\Bigr)\ .\label{T^1PI}
\end{gather}
The renormalization conditions listed in Sect.~3 are satisfied by $T^c_{\rm tree}\equiv T^{c\,(m,\mu)}_{\rm tree}\,$,
$\,T_{\rm tree}\equiv T^{(m,\mu)}_{\rm tree}$ and, provided that $T$ fulfills these conditions, also by 
$T^{1PI}\equiv T^{1PI\,(m,\mu)}$ (apart from the Field Equation). This can be verified by using the definitions 
(\ref{def:T^c_tree}), (\ref{def:T_tree}), (\ref{T^1PI}) and corresponding properties of the Feynman propagator.
Or, in case of local interactions, these properties can be derived for $T^c_{\rm tree}$ by the classical limit 
of $\hbar^{-(n-1)}\,T^c_n$; the linked cluster theorem implies then their validity for $T_{\rm tree}$.

\subsection{Definition and basic properties of the vertex function $\Gamma_T$}\label{vertexfunc}

Note that $T$ and $T_{\rm tree}$ satisfy the relations
\beq
T_{(\rm tree)}(1)=1\ ,\quad T_{(\rm tree)}(F)=F\ ,\quad
T_{(\rm tree)\, n+1}(1\otimes F_1\otimes ...)=
T_{(\rm tree)\, n}(F_1\otimes ...)\ ,\label{T(1)}
\eeq
which imply the following conclusions for $T_{\rm tree}$ and $T$
\begin{gather}
T_{(\rm tree)}\Bigl(e_\otimes^{\sum_{n=1}^\infty F_n\lambda^n}\otimes
\sum_{n=0}^\infty G_n\lambda^n\Bigr)=0\,\,\Longrightarrow\,\, G_n=0\,\,\forall n\label{Tps1}\\
T_{(\rm tree)}\Bigl(e_\otimes^{\sum_{n=1}^\infty F_n\lambda^n}\Bigr)=
T_{(\rm tree)}\Bigl(e_\otimes^{\sum_{n=1}^\infty G_n\lambda^n}\Bigr)
\,\,\Longrightarrow\,\, F_n=G_n\,\,\forall n\label{Tps2}
\end{gather}
(where $F_n,G_n\in {\cal F}$ are independent of $\lambda$),
as one obtains by proceeding by induction on the order in $\lambda$.

$R$- and $T$-products can be obtained from each other by Bogoliubov's
formula (\ref{R1-T}), which we will use in the more explicit form
\beq
R(e_\otimes^S,F)=\bar T(e_\otimes^{-iS/\hbar})\star 
T(e_\otimes^{iS/\hbar}\otimes F)\ ,\label{R-T}
\eeq
where the anti-chronological product $\bar T$ is defined by
\beq
\bar T(e_\otimes^{-iG})\equiv T(e_\otimes^{iG})^{-1}=
\sum_{n=0}^\infty \Bigl(1- T(e_\otimes^{iG})\Bigr)^{\star n}\ .
\label{bar-T}
\eeq
($T(e_\otimes^{iG})^{-1}$ is the inverse with respect to the $\star$-product.)
Although $R$ contains solely connected diagrams (see Appendix A), 
disconnected diagrams of $T$ and
$\bar T$ contribute to (\ref{R-T}). Unitarity reads
\beq
\bar T_{(\rm tree)}(e_\otimes^{-iG})=
T_{(\rm tree)}(e_\otimes^{iG^\ast})^\ast\ ;
\eeq
in this form it holds for the tree diagrams separately, 
where $\bar T_{\rm tree}$ is defined by
(\ref{def:T_tree}) with $\Delta^F$ replaced by the Anti-Feynman propagator 
$\Delta^{AF}(x)=\Delta^{F}(x)^*$. However, note that $T_{\rm tree}(e_\otimes^{iG})
\star T_{\rm tree}(e_\otimes^{iG^\ast})^\ast$ is not equal to $1$.
\medskip

We define the 'vertex function' $\Gamma_T$  
implicitly by the following Proposition:
\begin{prop}\label{prop:gamma}
There exists a {\bf totally symmetric} and {\bf linear} map
\beq
\Gamma_T\> :\> \TT {\cal F}_{\rm loc}\rightarrow {\cal F}
\label{gamma}
\eeq
which is uniquely determined by
\beq
T(e_\otimes^{iS/\hbar})=
T_{\rm tree}\Bigl(e_\otimes^{i\Gamma_T(e_\otimes^S)/\hbar}\Bigr)\ .
\label{Gamma}
\eeq
\end{prop}
To zeroth and first order in $S$ we obtain
\beq
\Gamma_T(1)=0\ ,\quad \Gamma_T(S)=S\ .\label{Gamma:0,1}
\eeq
The defining relation (\ref{Gamma}) also implies
\beq
T(e_\otimes^{iS/\hbar}\otimes F)=
T_{\rm tree}\Bigl(e_\otimes^{i\Gamma_T(e_\otimes^S)/\hbar}
\otimes \Gamma_T(e_\otimes^S\otimes F)\Bigr)\ .
\eeq
For $S=0$ this gives $F=\Gamma_T(1\otimes F)$.
The Proposition remains true if, in (\ref{Gamma}), we replace the
time ordered product $(T,T_{\rm tree})$ by the anti-chronological product
$(\bar T, \bar T_{\rm tree})$ and $i$ by $(-i)$; we denote the corresponding 
vertex function by $\bar\Gamma_{T}$.

\begin{proof}
We construct $\Gamma_T(\otimes_{j=1}^n F_j)$ by induction on $n$, starting with 
(\ref{Gamma:0,1}). Let $\Gamma_T$ of less than $n$ factors be constructed.
Then, (\ref{Gamma}) and the requirements total symmetry and linearity
determine $\Gamma_T(\otimes_{j=1}^n F_j)$ uniquely:
\beq
\Gamma_T(\otimes_{j=1}^n F_j)=(i/\hbar)^{n-1}\,T(\otimes_{j=1}^n F_j)-\sum_{|P|\geq 2}
(i/\hbar)^{|P|-1}\,T_{\rm tree}\Bigl(\bigotimes_{J\in P}
\Gamma_T(\otimes_{j\in J} F_j)\Bigr)\ ,\label{Gamma:recursion}
\eeq
where $P$ is a partition of $\{1,...,n\}$ in $|P|$ subsets $J$.
Obviously the so constructed $\Gamma_T$ is totally symmetric and linear.
\end{proof}

The proof applies also to the connected parts $(T^c,T^c_{\rm tree})$.
Hence, a totally symmetric and linear map $\Gamma_c$ can be defined analogously to
(\ref{gamma})-(\ref{Gamma}), that is by
\beq
T^c(e_\otimes^{iS/\hbar})=
T^c_{\rm tree}\Bigl(e_\otimes^{i\Gamma_c(e_\otimes^S)/\hbar}\Bigr)\ .
\label{Gamma_c}
\eeq
The linked cluster theorem for $T$ and 
$T_{\rm tree}$ and the definitions of $\Gamma_T$
and $\Gamma_c$ give
\beq
T_{\rm tree}\Bigl(e_\otimes^{i\Gamma_c(e_\otimes^S)/\hbar}\Bigr)
=T_{\rm tree}\Bigl(e_\otimes^{i\Gamma_T(e_\otimes^S)/\hbar}\Bigr)
\eeq
and with (\ref{Tps2}) we conclude
\beq
\Gamma_T=\Gamma_c\ .\label{Gamma:T-c}
\eeq
Therefore, on the right-hand side of (\ref{Gamma:recursion}) we may
replace the time ordered products $T$ and $T_{\rm tree\,,\,k}$
by their connected parts:

\begin{gather}
\Gamma_T(S^{\otimes n})=(i/\hbar)^{n-1}\,T^c(S^{\otimes n})-
\sum_{k=2}^n\sum_{\small \begin{array}c l_1+ \cdots +l_k=n\\l_j\geq 1\,\,\forall j\end{array}}\frac{(i/\hbar)^{k-1}\,n!}
{k!\,l_1!...l_k!}\cdot\notag\\
\cdot T^c_{\rm tree\,,\,k}\Bigl(\Gamma_T(S^{\otimes l_1})\otimes...
\otimes \Gamma_T(S^{\otimes l_k})\Bigr)\ .\label{Gamma:recursion^c}
\end{gather}
Now let $S\sim\hbar^0$ and $F\sim\hbar^0$. From (\ref{Gamma:recursion^c}) and $T^c(S^{\otimes n})
-T^c_{\rm tree}(S^{\otimes n})={\cal O}(\hbar^{n})$ we inductively conclude
\beq\label{properinter}
\Gamma_T(e_\otimes^S)=S+{\cal O}(\hbar)\ ,\quad \Gamma_T(e_\otimes^S
\otimes F)=F+{\cal O}(\hbar) .
\eeq
Motivated by this relation and (\ref{Gamma})
we call $\Gamma_T(e_\otimes^S)$ the 'proper interaction'.

By comparing the recursion relation (\ref{Gamma:recursion^c}) for $\Gamma_T$
with the recursive definition of $T^{\rm 1PI}$ (\ref{T^1PI}) we conclude:
\begin{cor}\label{G=T}
\beq
\Gamma_T(e_\otimes^{S})=(\hbar/i)\,T^{\rm 1PI}(e_\otimes^{iS/\hbar})\ .\label{Gamma:1PI}
\eeq
\end{cor}

Analogously to the Main Theorem it holds: 
\begin{lemma}\label{Gamma:rencond}
The validity of the renormalization conditions
for $T(\equiv T^{(m,\mu)})$ implies corresponding 
properties of $\Gamma_T(\equiv\Gamma_T^{(m,\mu)})$:
\begin{itemize}
\item {\bf ${\cal P}_+^\uparrow$-Covariance:}
$\beta_L\circ \Gamma_T=\Gamma_T\circ \TT\beta_L$ for all $L\in {\cal P}_+^\uparrow\,$;
\item {\bf Unitarity:}
$\Gamma_T(e_\otimes^S)^\star=\bar \Gamma_T(e_\otimes^{S^\star})\,$;
\item {\bf Field Independence:}
$\frac{\delta\,\Gamma_T(e_{\otimes}^S)}{\delta\varphi}=
\Gamma_T\Bigl(\frac{\delta S}{\delta\varphi}
\otimes e_{\otimes}^S\Bigr)\,$;
\item {\bf  Field Equation:}
$\Gamma_T(e_{\otimes}^{S}\otimes \varphi(h))=\varphi(h)\,$;
\item {\bf Smoothness in $m\geq 0$:}
$\Gamma_T^{(m,\mu)}$ is smooth in $m\geq 0\,$;
\item {\bf $\mu$-Covariance:} $\quad
\Gamma_{T}^{(m,\mu_2)}=({\mu_2}/{\mu_1})^{\Gamma}\circ
\Gamma_{T}^{(m,\mu_1)}\circ \TT({\mu_2}/{\mu_1})^{-\Gamma}\ $;
\item {\bf Almost homogeneous Scaling:} In contrast to the map $D$ of the Main Theorem, $\Gamma_T$
scales only {\it almost} homogeneously;
$\sigma_\rho \circ\Gamma_{T}^{(\rho^{-1}m,\mu)}
\circ \TT\,\sigma_\rho^{-1}=\Gamma_{T}^{(m,\mu)}+
{\cal O}({\rm log}\>\rho)$ is a polynomial in ${\rm log}\>\rho\,$.
\item If, instead of Smoothness in $m$, $\mu$-Covariance and Almost homogeneous Scaling, $T$
satisfies the axiom {\bf Scaling Degree}, then
\beq  
{\rm sd}\Bigl(\omega_0(\Gamma_T(A_1,...,A_n))(x_1-x_n,...)\Bigr)\leq\sum_{j=1}^n  
{\rm dim}(A_j)\ ,\quad\forall A_j\in {\cal P}_{\rm hom}\ .\label{scalingdeg:Gamma}  
\eeq  
\end{itemize}
\end{lemma}
\begin{proof} 
Each property can be proved for $\Gamma_T(S^{\otimes n})$ 
(or $\Gamma_T(S^{\otimes n}\otimes\varphi(h))$ resp.) by
induction on $n$: we work with the recursion relation (\ref{Gamma:recursion})
and use that $T$ and $T_{\rm tree}$ satisfy the corresponding axiom. 
In case of the property Scaling Degree we take into account that $\omega_0(T_{\rm tree}(...))$
is a tensor product of distributions $t_j$ and apply ${\rm sd}(\otimes_jt_j)=\sum_j {\rm sd}(t_j)$.

Only the Field Equation is somewhat more involved. We use that $T$ and $T_{\rm tree}$
fulfil the Field Equation and the Field Independence. This implies
\beq
T^c(\varphi(h)\otimes F_1\otimes...\otimes F_n)=\hbar\int\! dx\,dy\,h(x)\,
\Delta^F(x-y)\,\frac{\delta}{\delta\varphi(y)}T^c(F_1\otimes...\otimes F_n)
\label{feq:T}
\eeq
and the same equation for $T^c_{\rm tree}$. In the latter case $F_1,...,F_n$ 
may be non-local. (It is not necessary to work with 
the connected parts, but this simplifies the formulas.) With 
the recursion relation (\ref{Gamma:recursion^c}) and the inductive 
assumption we obtain
\begin{gather}
\Gamma_T(\varphi(h)\otimes F_1\otimes...\otimes F_n)=
T^c(\varphi(h)\otimes F_1\otimes...\otimes F_n)-
\sum_P T^c_{\rm tree}\Bigl(\varphi(h)\otimes
\bigotimes_{J\in P}\Gamma_T(F_J)\Bigr)\notag\\
=\hbar\int\! dx\,dy\,h(x)\,\Delta^F(x-y)\,\frac{\delta}{\delta\varphi(y)}
\Bigl(T^c(F_1\otimes...\otimes F_n)-\sum_P T^c_{\rm tree}\Bigl(
\bigotimes_{J\in P}\Gamma_T(F_J)\Bigr)\Bigr)=0\ ,\label{feq:G_T}
\end{gather}
where $F_J\equiv\otimes_{j\in J}F_j$ and $P$ runs through all partitions
of $\{1,...,n\}$.
\end{proof}

Analogously to the conventions for $R$- and $T$-products we sometimes write\\
$\int\! dx\, g(x)\, \Gamma_T(A(x)\otimes F_2...)$ for $\Gamma_T(\int\! dx\, g(x)\,
A(x)\otimes F_2...)$. Since $\Gamma_T$ depends only on the {\it functionals},
it fulfills the AWI: $\d^\mu_x\Gamma_T(A(x)\otimes F_2...)=
\Gamma_T(\d^\mu A(x)\otimes F_2...)$.

In the proper vertex formalism a finite renormalization $T\rightarrow \hat T$ 
of the $T$-products  is reflected in a finite renormalization $\Gamma_T
\rightarrow \hat\Gamma_T$ of the corresponding vertex functions.
To derive this we insert $\hat T(e_\otimes^{iS/\hbar})=
T_{\rm tree}\Bigl(e_\otimes^{i\hat\Gamma_T(e_\otimes^S)/\hbar}\Bigr)$ and
(\ref{Gamma}) into (\ref{mainthm:T}) and obtain
\beq
T_{\rm tree}\Bigl(e_\otimes^{i\hat\Gamma_T(e_\otimes^S)/\hbar}\Bigr)
=T_{\rm tree}\Bigl(e_\otimes^{i\Gamma_T
\Bigl(e_\otimes^{D(e_\otimes^S)}\Bigr)/\hbar}\Bigr)\ .
\eeq
(Note that the tree 
part $T_{\rm tree}$ is independent of the normalization of the $T$-product.)
By using (\ref{Tps2}) we conclude
\beq\label{renorm-gamma-T}
\hat\Gamma_T(e_\otimes^S)=\Gamma_T(e_\otimes^{D(e_\otimes^S)})\ .
\eeq
\subsection{Comparison with the literature}\label{comp-lit}
{\bf Definition of the 'vertex functional' (or 'proper function') $\Gamma$ in the literature}, 
see e.g.~\cite{Piguet:1995er}. Usually $\Gamma(h)\,,\,h\in {\cal S}(\MM,\RR)$, is defined
as the Legendre transformed $j\rightarrow h$ of the generating functional $Z(j)$
of the connected Green's functions (where $j$ is the 'classical source' of $\varphi$). 
With that $\Gamma(h)$ is the generating functional of the 
1PI-diagrams of $T(e_\otimes^{iS})$ (see \cite{Jona-Lasinio:1964}). 
We are going to express the latter fact in our formalism. 
To simplify the notations we study a scalar field $\varphi$
with free action $S_0(\varphi)=1/2\,\int\! dx\,((\d \varphi(x))^2-m^2\,(\varphi(x))^2)$.
Green's functions are obtained by the Gell-Mann Low formula 
\cite{Gell-MannLow}, which contains the adiabatic limit $g\to 1$:
\beq
G(x_1,...,x_n)=\lim_{g\to 1}\,\frac{\omega_0\Bigl(T(\varphi(x_1)...
\varphi(x_n)\,e^{iS(g)/\hbar})\Bigr)}{\omega_0\Bigl(T(e^{iS(g)/\hbar})\Bigr)}\ ,
\eeq
where $S(g)=\sum_{n=1}^\infty\kappa^n\int\! dx\,(g(x))^n\,{\cal L}_n(x)$
and $\kappa$  is the coupling constant. All diagrams with vacuum-subdiagrams
are divided out. These diagrams are disconnected and, hence, not of interest 
for our purposes. Namely, to obtain the vertex functional $\Gamma$ one selects  
all diagrams of $G(x_1,...,x_n)$ which are 1PI after amputation 
of the external legs. The contribution of these diagrams is given by
\begin{gather}
G^{\rm 1PI}(x_1,...,x_n)=\lim_{g\to 1}\,\int\! dy_1...dy_n\,\Delta_F(x_1-y_1)...
\Delta_F(x_n-y_n)\cdot\notag\\
\cdot\omega_0\Bigl(\frac{\delta^n}{\delta\varphi(y_1)...
\delta\varphi(y_n)}\,T^{\rm 1PI}(e^{iS(g)/\hbar})\Bigr)+\delta_{n,2}\,\Delta^F(x_1-x_2)\ .
\label{G1PI}
\end{gather}
>From this expression $\Gamma(h)$ is obtained by replacing each external leg
$\Delta_F(x_l-y_l)$ by the classical field $h(y_l)$. 
In addition one multiplies with $(-i)/n!$ and sums over $n\geq 1$:\footnote{Usually it is assumed that
$\lim_{g\to 1}\omega_0(\varphi_{g{\cal L}}(x))=0$,
which implies $\frac{\delta\Gamma(h)}{\delta h(x)}\vert_{h=0}=0$,
i.e.~the sum runs only over $n\geq 2$.}
\begin{gather}
\Gamma(h)=S_0(h)+\frac{\hbar}{i}\,
\lim_{g\to 1}\,\sum_{n\geq 1}\frac{1}{n!}\int\! dy_1...dy_n\,
h(y_1)...h(y_n)\cdot\notag\\
\cdot\omega_0\Bigl(\frac{\delta^n}{\delta\varphi(y_1)...
\delta\varphi(y_n)}\,T^{\rm 1PI}(e^{iS(g)/\hbar})\Bigr)\ .\label{Gamma:literature}
\end{gather}
Note that the term $S_0(h)$ comes from the
$\delta_{n,2}\,\Delta^F$-term in (\ref{G1PI}).

{\bf Relation to our proper interaction $\Gamma_T(e_\otimes^S(g))$:}
We compare (\ref{Gamma:literature}) with the Taylor expansion in $\varphi$ 
of $\Gamma_T(e_\otimes^{S(g)})$:
\begin{gather}
\Gamma_T(e_\otimes^{S(g)})=\sum_{n\geq 0}\frac{1}{n!}\int\! dy_1...dy_n\,
\varphi(y_1)...\varphi(y_n)\,
\omega_0\Bigl(\frac{\delta^n}{\delta\varphi(y_1)...
\delta\varphi(y_n)}\,\Gamma_T(e_\otimes^{S(g)})\Bigr)
\end{gather}
and use Corollary \ref{G=T}. This yields
\beq\label{G=lGt}
\Gamma(h)=S_0(h)+\lim_{g\to 1}\,\Bigl(\Gamma_T(e_\otimes^{S(g)})(h)-
\omega_0\bigl(\Gamma_T(e_\otimes^{S(g)})\bigr)\Bigr)\ .
\eeq
(On the right-hand side the functionals $S_0\,,\,\Gamma_T(e_\otimes^{S(g)})\in {\cal F}$ 
are evaluated on the classical field configuration $h\in {\cal S}(\MM,\RR)$.)

\section{The Master Ward Identity}\label{MWI}
\subsection{The classical MWI in the off-shell formalism}\setcounter{equation}{0}\label{classMWI}
In \cite{Duetsch:2002yp} the MWI for on-shell fields (i.e.~the retarded products are restricted
to the solutions of the free field equation(s)) was derived in the framework of classical field theory. Since here, we work throughout in a general off-shell formalism \cite{Duetsch:2004dd},
we shall derive an off-shell version of the classical MWI. In addition, we give 
an equivalent formulation of the classical MWI in terms of $T_\mathrm{tree}$-products that will be useful for the proper field formulation of the MWI in Sect.~\ref{effMWI-QAP}.


The classical off-shell MWI follows from the factorization \eqref{interfactor} and the definition
of the retarded wave operators. Let $\mathcal{J}$ be the ideal generated by the free field equation(s),
\beq
\mathcal{J}\defi \Big\{\sum_{n=1}^{N} \int\! dx_1\ldots dx_n \,\varphi(x_1) \cdots \varphi(x_{n-1})
\frac{\delta S_0}{\delta \varphi(x_n)} f_{n}(x_1, \ldots, x_n)\Big\}\subset \mathcal{F}\ ,\nonumber
\eeq
with $N<\infty$ and the $f_{n}$'s being defined as in \eqref{F(phi)}. Every 
$A\in \mathcal{J}$ can be written as
\begin{equation}
A\defi \int\! dx\, Q(x) \ffgx\ ,
\end{equation}
where $Q$ is of the form
\begin{equation}\label{Q(x)}
Q(x)=\sum_{n=0}^N\int\! dx_1\ldots dx_n\, \varphi(x_1) \cdots
\varphi(x_n) f_{n+1}(x_1, \ldots, x_n, x)\ .
\end{equation}
Note that in the present framework of classical field 
theory $Q$ does not need to be a local functional. Given 
$A\in \mathcal{J}$ we introduce a corresponding derivation \cite{Duetsch:2002yp}
\begin{equation}\label{defdeltaA}
\dA \defi \int\! dx\, Q(x)\frac{\delta}{\delta \varphi(x)}.
\end{equation}
>From the defining property of the retarded wave operators Def.~\ref{def:waveop} \emph{(ii)} we obtain
\begin{eqnarray}
(A+\delta_A S)\circ r_{S_0+S, S_0}&=&\int\! dx\, Q(x)\circ r_{S_0+S, S_0}\frac{\delta(S_0+S)}{\delta \varphi(x)}\circ r_{S_0+S, S_0}\nonumber\\
&=& \int\! dx\, Q(x)\circ r_{S_0+S, S_0}\frac{\delta S_0}{\delta \varphi(x)},\label{klMWI}
\end{eqnarray}
which reads perturbatively
\begin{equation}\label{klMWIIdeal}
R_{\mathrm{cl}}(e_\otimes^S, A+\delta_A S)=\int\! dx\,
R_{\mathrm{cl}}\big(e_\otimes^S, Q(x)\big)\frac{\delta S_0}{\delta
\varphi(x)}\in \mathcal{J}.
\end{equation}
This is the MWI written in the general off-shell formalism. Indeed, by restricting \eqref{klMWIIdeal} on solutions of the free field equation, the right-hand side vanishes and we obtain the on-shell version of the MWI, as it was derived in \cite{Duetsch:2002yp}. Note that for the simplest case $Q=1$ the MWI reduces to the off-shell version of the (interacting) field equation
\begin{equation}\label{klfeldgloff}
R_{\mathrm{cl}}\Big(e_\otimes^S, \frac{\delta(S_0+S)}{\delta \varphi(x)}\Big)=\frac{\delta S_0}{\delta \varphi(x)},
\end{equation} 
which is an alternative formulation of the axiom Field Equation in Sect.~\ref{pQFT}.

The classical field equation \eqref{klfeldgloff} can be expressed in the 
time ordered formalism:
\beq
T_\mathrm{tree}\Bigl(e_\otimes^{iS}\otimes\frac{\delta (S_0+S)}{\delta\varphi(x)}\Bigr)
=\frac{\delta S_0}{\delta\varphi(x)}
\cdot T_\mathrm{tree}\Bigl(e_\otimes^{iS}\Bigr)\ .
\eeq
This identity holds even for non-local entries and can be obtained easily by 
using the definition of $T_\mathrm{tree}$ given in \eqref{def:T_tree} and the 
fact that $\Delta^F$ is a Green's function of the Klein Gordon operator. 
Similarly to $R_\mathrm{cl}$, the tree diagrams of the time ordered product
factorize (cf.~\cite{BD-qap}), that is
\beq
T_{\rm tree}\Bigl(e_\otimes^{iS}\otimes F\Bigr)\cdot
T_{\rm tree}\Bigl(e_\otimes^{iS}\otimes G\Bigr)=
T_{\rm tree}\Bigl(e_\otimes^{iS}\otimes F G\Bigr)
\cdot T_\mathrm{tree}\Bigl(e_\otimes^{iS}\Bigr)\ .\label{factorization}
\eeq

We now multiply the field equation for $T_{\rm tree}$ with $T_{\rm tree}(e_\otimes^{iS}\otimes Q(x))$. 
This yields the MWI in the time ordered formalism:
\beq
T_{\rm tree}\Bigl(e_\otimes^{iS}\otimes (A+\delta_A S)\Bigr)=
\int \!dx\,
T_{\rm tree}\Bigl(e_\otimes^{iS}\otimes Q(x)\Bigr)\cdot
\frac{\delta S_0}{\delta\varphi(x)}\ .\label{MWI-tree}
\eeq

We point out that the MWI for $T_{\rm tree}$ (\ref{MWI-tree}) 
holds also for non-local entries $S$, $Q(x)$ and $A$.

\subsection{Structure of possible anomalies of the MWI in QFT}\label{anomalMWI}

The classical MWI was derived for arbitrary interaction $S\in \mathcal{F}$ and 
arbitrary $A\in \mathcal{J}$. For {\it local} functionals 
$A\in \mathcal{J}_{\mathrm{loc}} \equiv \mathcal{J}\cap\mathcal{F}_{\mathrm{loc}}$ 
and $S\in \mathcal{F}_{\mathrm{loc}}$ it
can be transferred formally into pQFT (by the replacement $R_\mathrm{cl}\rightarrow R$),
where it serves as an additional, highly non-trivial renormalization condition.
It is impossible to fulfill this condition for all $A\in \mathcal{J}_{\mathrm{loc}}$. We aim 
to find a general expression for the possible violations ('anomalies') of the MWI. 
Later we will use this result as starting point for a proof of relevant cases of the MWI.
This procedure is motivated by algebraic renormalization, where
the QAP serves as the crucial input to study the possibility to fulfill some Ward identities
(see \cite{Piguet:1995er} and references cited therein). 

The main insight into the structure of possible anomalies of the MWI is the fact that they can be 
expressed in terms of a \emph{local} interacting field:
\begin{thm}\label{satzQAP}
Given a retarded product $R$ fulfilling the basic axioms 
Initial Condition, Causality and GLZ Relation and given a local functional
\beq
A=\int\! dx\, h(x)\, Q(x)\frac{\delta S_0}{\delta \varphi(x)}\in \mathcal{J}_{\mathrm{loc}}\ , 
\quad h\in \mathcal{D}(\mathbbm{M})\ ,\quad Q\in \mathcal{P}\ ,\label{A:loc}
\eeq
there exists a unique, linear and symmetric map 
\begin{eqnarray}\label{AbbDelta}
\Delta_A:\mathbb{T}\mathcal{F}_{\mathrm{loc}}&\longrightarrow &\mathcal{F}_{\mathrm{loc}}\\
F_1\otimes \cdots \otimes F_n &\longmapsto & \Delta_A(F_1\otimes \cdots \otimes F_n) \nonumber
\end{eqnarray}
which is implicitly defined by the 'anomalous MWI'
\begin{equation}\label{retQAP}
R\big(\ets, A+\delta_A S+\Delta_A(\ets)\big)=\int\! dy\, h(y) R(\ets, Q(y))\frac{\delta S_0}{\delta \varphi(y)}.
\end{equation}
As a consequence of \eqref{retQAP} the map $\Delta_A$ has the following properties:
\begin{eqnarray}
&(i)&\Delta_A \,\,\mbox{depends linearly on} \,\,A\ ;\nonumber\\
&(ii)&\mbox{locality expressed by the two relations:}\nonumber\\
&(iia)& \omega_0( \Delta_A(\otimes_{j=1}^n F_j))=0\,\,\mbox{if}\,\,
\cap_{i=1}^n\mathrm{supp}\Big(\frac{\delta F_i}{\delta \varphi}\Big)\cap
\mathrm{supp}\Big(\frac{\delta A}{\delta \varphi}\Big)=\emptyset\ ,\nonumber\\
&(iib)&\mathrm{supp}\Big(\dfrac{\delta \Delta_A(\otimes_{j=1}^n F_j)}{\delta \varphi}\Big)
\subset \cap_{i=1}^n
\mathrm{supp}\Big(\frac{\delta F_i}{\delta \varphi}\Big)\cap
\mathrm{supp}\Big(\frac{\delta A}{\delta \varphi}\Big)
;\nonumber\\
&(iii)&\Delta_A(1)=0\ ;\nonumber\\
&(iv)&\Delta_A\equiv 0 \, \Leftrightarrow \, R(\ets, A+\delta_A S)=
\int\! dx \, h(x)\, R(e_\otimes^S, Q(x))\,\frac{\delta S_0}{\delta \varphi(x)}\ ,
\forall S\in \mathcal{F}_{\mathrm{loc}};\nonumber\\
&(v)&\Delta_A(F_1\otimes\cdots\otimes F_n)=\mathcal{O}(\hbar) \qquad \forall n>0, \, F_i\sim \hbar^0\ ,
\label{propdelta}
\end{eqnarray}
and\\
\hspace*{0.6cm} (vi) We set $\Delta_A^n\equiv\Delta_A\vert_{\mathcal{F}_{\mathrm{loc}}^{\otimes n}}\ $.
For $g_j\in{\cal D}(\MM)\,,\,\,L_j\in {\cal P}$ it holds
\begin{gather}
\Delta_A^n(L_1(g_1)\otimes...\otimes L_n(g_n))=\int\! dx_1...dx_n\,dy\,g_1(x_1)...g_n(x_n)\,h(y)\notag\\
\cdot\Delta^n(L_1(x_1)\otimes...\otimes L_n(x_n);Q(y))\ ,\label{kernel}
\end{gather}
\hspace{1.3cm} where the distributional kernel $\Delta^n(L_1(x_1)\otimes...;Q(y))$ is inductively given by
\begin{gather}
\Delta^n(L_1(x_1)\otimes...\otimes
L_n(x_n);Q(y))=-R\Bigl(\otimes_{j=1}^n L_j(x_j);Q(y)
\cdot\frac{\delta S_0}{\delta\varphi(y)}\Bigr)\notag\\
-\sum_{l=1}^nR\Bigl(\otimes_{j(\not= l)}
L_j(x_j);Q(y)\sum_a(\d^a\delta)(x_l-y)\,
\frac{\d L_l}{\d(\d^a\varphi)}(x_l)\Bigr)\notag\\
-\sum_{I\subset\{1,...,n\}\,,\,I\not=\emptyset}
R\Bigl(\otimes_{i\in I} L_i(x_i);\Delta^{|I^c|}(\otimes_{j\in I^c} L_j(x_j);Q(y))\Bigr)\notag\\
+R\Bigl(\otimes_{j=1}^n L_j(x_j);Q(y)\Bigr)\cdot \frac{\delta
S_0}{\delta\varphi(y)}\ . \label{kernel1}
\end{gather}
\end{thm}
Note that \eqref{retQAP} differs from the MWI \eqref{klMWIIdeal} only by the local term 
$\Delta_A(e_\otimes^S)$, which clearly depends on the chosen normalization of the retarded 
products. Therefore, property $(iv)$ means that the MWI for $A$ is fulfilled if and only if 
the corresponding map $\Delta_A$ vanishes identically.
\begin{proof}
To show the existence and uniqueness of $\Delta_A$
we construct its components $\Delta_A^n$ 
by induction on $n$ using \eqref{retQAP}. In this inductive procedure we also prove
the properties $(i)-(iii)$ and $(vi)$.
To lowest order in $S$ the condition \eqref{retQAP} gives 
$\Delta_A(1)\defi 0$. Given $n>0$, we assume the existence and uniqueness
of linear and symmetrical maps $\Delta_A^k:\mathcal{F}_{\mathrm{loc}}^{\otimes k} 
\rightarrow \mathcal{F}_{\mathrm{loc}}$, $0<k<n$, which depend linearly on $A$, are local and satisfy $(vi)$, 
such that \eqref{retQAP} is fulfilled to all lower orders in $S$:
\begin{gather}
R(S^{\otimes k}, A)+k R(S^{\otimes k-1}, \delta_A S)+\sum_{l=0}^k \binom{k}{l}R(S^{\otimes k-l}, \Delta^l_A(S^{\otimes l}))=\notag \\
\int\! dx \,h(x) R(S^{\otimes k}, Q(x))\frac{\delta S_0}{\delta \varphi(x)}\label{indhypo}
\end{gather}
for all $k<n$. We define $\Delta^n_A$ in terms of the inductively known $\Delta_A^k\ ,\ k<n$:
\begin{eqnarray}
\Delta^n_A(F_1\otimes \cdots \otimes F_n)&\defi& \int\! dx\,h(x) 
R(F_1\otimes \cdots \otimes F_n, Q(x))\frac{\delta S_0}{\delta \varphi(x)}\nonumber\\
&-&\Big(R(F_1\otimes \cdots \otimes F_n, A)+\sum_{k=1}^n R(\otimes_{i\in \underline{n}\backslash\{k\}}F_i, \delta_A F_k)\nonumber\\
&&+\sum_{I\sqcup J=\underline{n},\, |J|<n} R(\otimes_{i\in I}F_i, \Delta^{|J|}_A(\otimes_{j\in J}F_j))\Big)\ ,
\label{defdelta}
\end{eqnarray}
where we used the notation $\underline{n}\defi \{1, \ldots, n\}$ and $I\sqcup J$ for the disjoint union of 
$I$ and $J$. For $F_1=\ldots =F_n=S$ the formula (\ref{defdelta}) agrees with \eqref{retQAP} to order $n$ in $S$.
Obviously, $\Delta^n_A$ is linear and symmetrical and it is uniquely determined by these properties and \eqref{retQAP}.
We also see that $\Delta^n_A$ is linear in $Q$ resp.~$A$.

The main task is to show the locality $(ii)$ of the right-hand side of \eqref{defdelta} 
(which also implies $\Delta^n_A(F_1\otimes \cdots\otimes F_n)\in \mathcal{F_\mathrm{loc}}$). To this end we denote by
\begin{eqnarray*}
M_{n,1}(L_1(x_1), \ldots, L_n(x_n);Q(y))&\defi& R\left(\otimes_{i\in \underline{n}}L_i(x_i), Q(y)\tfrac{\delta S_0}{\delta \varphi(y)}\right)\\
&+&\sum_{k=1}^n R\left(\otimes_{i\in \underline{n}\backslash\{k\}}L_i(x_i),Q(y)\tfrac{\delta L_k(x_k)}{\delta \varphi(y)}\right)\\
&+&\hspace{-1cm}\sum_{I\sqcup J=\underline{n}\,,\,|J|<n}R\left(\otimes_{i\in I}L_i(x_i), \Delta^{|J|}(\otimes_{j\in J}L_j(x_j);Q(y))\right)
\end{eqnarray*}
the distributional kernel of the three terms in brackets in equation \eqref{defdelta},
where $F_i=\int\! dx \,g_i(x) L_i(x)$ ($g_i\in \mathcal{D}(\mathbbm{M})$, $L_i\in \mathcal{P}$) and
we use (\ref{kernel}) for $\Delta^{|J|}_A$ to lower orders $|J|<n$.
We will show that $M_{n,1}(L_1(x_1), \ldots ;Q(y))$ coincides outside the total 
diagonal $\mathbbm{D}_{n+1}=\{(x_1, \ldots, x_{n+1})\in \mathbb{M}^{n+1}, x_1=\cdots =x_{n+1}\}$ 
with the distributional kernel $R(L_1(x_1)\otimes \cdots \otimes L_n(x_n), Q(y))
\frac{\delta S_0}{\delta \varphi(y)}$ of the first term in \eqref{defdelta}. 
This shows 
\beq
\mathrm{supp}\>\Delta^n(\otimes_{j=1}^n L_j;Q)\subset\mathbbm{D}_{n+1}\ ,\label{suppDelta}
\eeq
where $\Delta^n(\otimes_{j=1}^n L_j;Q)$ is defined by (\ref{kernel1}). (By construction 
$\Delta^n(\otimes_{j=1}^n L_j;Q)$ is the distributional kernel of
$\Delta^n_A(L_1(g_1)\otimes...)$ (\ref{defdelta}), which proves (vi).)
The support property \eqref{suppDelta} implies locality $(iia)$. To derive the the second locality statement $(iib)$
we additionally use that
\beq
\frac{\delta R(\otimes_{j=1}^{k-1} A_j(x_j);A_k(x_k))}{\delta\varphi(z)}=0\quad\mbox{if}
\quad z\not= x_j\,\,\forall j=1,...,k\ ,\nonumber
\eeq
as explained in \cite{BD-qap}. By means of \eqref{kernel1} we conclude 
\beq
\mathrm{supp}\>\frac{\delta\Delta^n(\otimes_{j=1}^n L_j;Q)}{\delta\varphi}\subset\mathbbm{D}_{n+2}\ ,\nonumber
\eeq
which is equivalent to locality $(iib)$. It follows also $\Delta^n_A(\otimes_{j=1}^n L_j(g_j))\in \mathcal{F_\mathrm{loc}}$.

We turn to the proof of \eqref{suppDelta}.
Since $M_{n,1}$ has retarded support
\begin{equation}
\mathrm{supp}(\delta M_{n,1}/\delta\varphi)\subset\{ (x_1, \ldots, x_n, y)\in \mathbb{M}^{n+1}; 
x_1, \ldots, x_n\in y+\overline{V}_{-}\}
\end{equation}
and is symmetrical under permutation of the first $n$ entries, it is thereby sufficient 
to consider the particular case where $x_n\in y+\bar{V}_{-}\backslash \{y\}$. In this case it holds
\begin{eqnarray*}
M_{n,1}(L_1(x_1), \ldots ;Q(y))&=&R\left(\otimes_{i\in \underline{n-1}}L_i(x_i)\otimes L_n(x_n), Q(y)\tfrac{\delta S_0}{\delta \varphi(y)}\right)\nonumber\\
&+&\sum_{k=1}^{n-1} R\left(\otimes_{i\in \underline{n}\backslash\{k\}}L_i(x_i)\otimes L_n(x_n),Q(y)\tfrac{\delta L_k(x_k)}{\delta \varphi(y)}\right)\nonumber\\
&+&\hspace{-0.4cm}\sum_{I\sqcup J=\underline{n-1}}R\big(\otimes_{i\in I}L_i(x_i)\otimes L_n(x_n), \Delta^{|J|}(\otimes_{j\in J}L_j(x_j);Q(y))\big)
\end{eqnarray*}
where the locality of the distributions $\Delta^{|J|}(\otimes_{j\in J}L_j(x_j);Q(y))$ for $|J|<n$ is used.
Now we apply the GLZ Relation to each term and take the support property of the retarded products
into account. This yields
\begin{gather}
M_{n,1}(L(x_1), \ldots ;Q(y))=\sum_{I\sqcup J=\underline{n-1}}\Big\{R(\otimes_{i\in I}L_i(x_i), L_n(x_n)), R(\otimes_{j\in J}L_j(x_j), Q(y)\tfrac{\delta S_0}{\delta \varphi(y)})\Big\}\notag\\
+\sum_{k=1}^{n-1}\sum_{I\sqcup J\sqcup\{k\}=\underline{n-1}}\Big\{R(\otimes_{i\in I}L_i(x_i), L_n(x_n)), R(\otimes_{j\in J}L_j(x_j),
Q(y)\tfrac{\delta L_k(x_k)}{\delta \varphi(y)})\Big\}\notag\\
+\sum_{I\sqcup J=\underline{n-1}}\sum_{K\sqcup L=I}\Big\{R\left(\otimes_{i\in K} L_i(x_i), L_n(x_n)\right),
R\big(\otimes_{l\in L} L_l(x_l),\Delta^{|J|}(\otimes_{j\in J}L_j(x_j);Q(y))\big)\Big\}\label{distM}.
\end{gather}
To transform the last term we use the induction hypothesis \eqref{indhypo} written in distributional form,
\begin{gather}
\sum_{L\sqcup J=H}R(\otimes_{l\in L}L_l(x_l), \Delta^{|J|}(\otimes_{j\in J}L_j(x_j);Q(y)))
=R(\otimes_{i\in H}L_i(x_i), Q(y))\tfrac{\delta S_0}{\delta \varphi(y)}\notag\\
\hspace{-2.5cm}-R(\otimes_{i\in H}L_i(x_i), Q(y)\tfrac{\delta S_0}{\delta \varphi(y)})-
\sum_{j\in H}R(\otimes_{i\in H\backslash \{j\}}L_i(x_i), Q(y)\frac{\delta L_j(x_j)}{\delta \varphi(y)})
\end{gather}
for all $H \subset\underline{n-1}$. After rearranging the sums the last term in \eqref{distM} is equal to
\begin{gather}
\hspace{-0.5cm}-\sum_{I\sqcup J=\underline{n-1}}\Big\{R(\otimes_{i\in I}L_i(x_i), L_n(x_n)), R(\otimes_{j\in J}L_j(x_j), Q(y)\tfrac{\delta S_0}{\delta \varphi(y)})\Big\}\notag\\
-\sum_{k=1}^{n-1}\sum_{I\sqcup J\sqcup\{k\}=\underline{n-1}}\Big\{R(\otimes_{i\in I}L_i(x_i), L_n(x_n)), R(\otimes_{j\in J}L_j(x_j),
Q(y)\tfrac{\delta L_k(x_k)}{\delta\varphi(y)})\Big\}\notag\\
+\sum_{I\sqcup J=\underline{n-1}}\Big\{R(\otimes_{i\in I}L_i(x_i), L_n(x_n)), R(\otimes_{j\in J}L_j(x_j), Q(y))\tfrac{\delta S_0}{\delta
\varphi(y)}\Big\}\label{distM3}.
\end{gather}
The last term in \eqref{distM3} can be transformed by using the identity
\begin{equation}\label{stkom}
\{F, P(x)\tfrac{\delta S_0}{\delta \varphi(x)}\}=\{F, P(x)\}\tfrac{\delta S_0}{\delta \varphi(x)}\,,\quad F, P(x)\in \mathcal{F}
\end{equation}
(which follows from the fact that the 2-point function $H_m$ is a 
solution of the free field equation) and by applying the GLZ Relation:
\begin{eqnarray*}
\lefteqn{\sum_{I\sqcup J=\underline{n-1}}\Big\{R(\otimes_{i\in I}L_i(x_i), L_n(x_n)), R(\otimes_{j\in J}L_j(x_j), Q(y))\tfrac{\delta S_0}{\delta
\varphi(y)}\Big\}}\\
&&\hspace{1cm}=\sum_{I\sqcup J=\underline{n-1}}\big\{R(\otimes_{i\in I}L_i(x_i), L_n(x_n)), R(\otimes_{j\in J}L_j(x_j), Q(y))\big\}\tfrac{\delta S_0}{\delta
\varphi(y)}\\
&&\hspace{1cm}=R(L_1(x_1)\otimes\cdots\otimes L_n(x_n), Q(y))\tfrac{\delta S_0}{\delta \varphi(y)}.
\end{eqnarray*}
Summing up the first two terms in \eqref{distM} and the first two terms in \eqref{distM3} 
cancel and we get the desired result
\begin{equation}
M_{n,1}(L_1(x_1), \ldots Q(y))=R(L_1(x_1)\otimes\cdots\otimes L_n(x_n), Q(y))\frac{\delta S_0}{\delta \varphi(y)}
\end{equation}
$\forall(x_1, \ldots, x_n,y)\notin \mathbb{D}_{n+1}$ which proves \eqref{suppDelta}.


The conclusion '$\Rightarrow$' of property $(iv)$ is obvious from \eqref{retQAP} and '$\Leftarrow$'
follows inductively for $\Delta_A^n(S^{\otimes n})$, because the right-hand side of (\ref{defdelta})
(for $F_1=\ldots =F_n=S$) vanishes in that case if $\Delta_A^k\equiv 0\quad\forall k<n\ .$
And $\Delta_A(\ets)=0\,\,\forall S$ implies $\Delta_A\equiv 0$ by the polarization identity.

The crucial property that $\Delta_A(F_1\otimes\cdots\otimes F_n)=\mathcal{O}(\hbar)$ for all 
$n>0$ and $F_i\sim \hbar^0$, follows immediately from the validity of the classical MWI 
and $\lim_{\hbar\rightarrow 0}R=R_\mathrm{cl}$ by using property $(iv)$.
\end{proof}

Mostly we will omit the index $n$ of $\Delta_A^n$ and its kernel $\Delta^n$.

Up to here we only assumed that the $R$-product satisfies the basic axioms. If it
fulfills renormalization conditions, then corresponding properties of $\Delta_A$ are implied.
\begin{lemma}\label{Delta:RC}
\begin{itemize}
\item[(i)] The axioms ${\cal P}_+^\uparrow$-Covariance, Unitarity, Field Independence and Field Equation, respectively, imply 
corresponding properties of $\Delta_A$:
\begin{itemize}
\item {\bf ${\cal P}_+^\uparrow$-Covariance} $\beta_L\,\Delta_A(\ets)=\Delta_{\beta_L A}(e_\otimes^{\beta_L S})\,$,
 $\,\forall L\in {\cal P}^\uparrow_+\,$;
\item {\bf Unitarity} $\Delta_A(\ets)^\star=\Delta_{A^\star}(e_\otimes^{S^\star})\,$;
\item {\bf Field Independence} 
\beq 
\frac{\delta}{\delta\varphi (x)}\Delta_A(S^{\otimes
n})=\Delta_{A_x}(S^{\otimes n}) +n\,\Delta_A(S^{\otimes
(n-1)}\otimes \frac{\delta S}{\delta\varphi (x)})\ , \label{FI} 
\eeq
where $A_x$ is obtained from $A$ (\ref{A:loc}) by replacing $Q(y)$ by
$\frac{\delta Q(y)}{\delta\varphi (x)}$: 
\beq A_x\=d \int\! dy\,
h(y)\, \frac{\delta Q(y)}{\delta\varphi (x)}\, \frac{\delta
S_0}{\delta\varphi(y)}\in{\cal J}_{S_0}\ ; \eeq
\item {\bf Field Equation}
\beq
\Delta_A\equiv 0 \qquad \mathrm{if}\qquad A=\int\! dx\, h(x)\frac{\delta S_0}{\delta \varphi(x)}, 
\quad \forall\,h\in \mathcal{D}(\mathbb{M})\ ;
\eeq
\end{itemize}
\item[(ii)] Let $g_j\in{\cal D}(\MM)\,,\,\,L_j\in {\cal P}$. 
Assuming that the $R$-products satisfy the axioms Translation Invariance and Field Independence, 
there exist linear maps $P^n_a\,:\,\mathcal{P}^{\otimes (n+1)}\rightarrow\mathcal{P}$ (where
$a$ runs through a finite subset of $(\NN_0^d)^n$), which are symmetric in the first $n$ factors,
such that $\Delta_A^n$ can be written as
\begin{gather}
\Delta_A^n(L_1(g_1)\otimes\ldots\otimes L_n(g_n))=\notag\\
\sum_{a\in (\NN_0^d)^n}\int\! dx\,h(x)\,\d^{a_1}g_1(x)\ldots\d^{a_n}g_n(x)\,
P^n_a(L_1\otimes\ldots\otimes L_n;Q)(x)\ .\label{Delta_A:localization}
\end{gather}
\end{itemize}
\end{lemma}
Part (ii) can be proved without assuming Field Independence of the $R$-products, see \cite{BD-qap}.

\begin{proof} $(i)$ We prove all properties for the components $\Delta_A^n$ of $\Delta_A$ by induction 
on $n$. We verify that the right-hand side of (\ref{defdelta}) fulfills the assertion by using the 
pertinent property of the $R$-product and the inductive assumption.  

In case of the Field Independence we additionally take 
\begin{gather}
\frac{\delta A}{\delta\varphi (x)}=A_x+\int\! dy\, h(y)\,  Q(y)\,
\frac{\delta^2 S_0}{\delta\varphi(y)\,\delta\varphi (x)}\ ,\label{dA}\\
\frac{\delta (\delta_AS)}{\delta\varphi (x)}= \delta_{A_x}S+
\delta_A\Bigl(\frac{\delta S}{\delta\varphi (x)}\Bigr)\label{ddeltaS}
\end{gather}
into account.

Field Equation: for $A=\int\! dx\, h(x)\frac{\delta S_0}{\delta \varphi(x)}$ 
the MWI reduces to the off-shell field
equation in differential form, which is equivalent to 
the integrated version (\ref{FE}). The assertion follows by 
means of property $(iv)$ of $\Delta_A$ in Theorem \ref{satzQAP}.

$(ii)$  The statement is a simplified version of Proposition 4.3 in \cite{Duetsch:2004dd}.
The proof, which is given there in words, is carried out here explicitly, since these formulas will be 
used below in the proof of part $(ii)$ of Prop.~\ref{prop:massdimbound}.

The Field Independence (\ref{FI}) translates into a
corresponding relation for the distributional kernel (\ref{kernel}):
\begin{gather}
\frac{\delta}{\delta\varphi(x)}\Delta(\otimes_{j=1}^n L_j(x_j);Q(y))=\notag\\
\sum_{l=1}^n\Delta\Bigl(\otimes_{j(\not=l)} L_j(x_j)
\otimes\frac{\delta L_l(x_l)}{\delta\varphi(x)};Q(y)\Bigr)+
\Delta\Bigl(\otimes_{j=1}^n L_j(x_j);\frac{\delta
Q(y)}{\delta\varphi(x)}\Bigr)\ .
\end{gather}
We insert this identity into the Taylor expansion of
$\Delta(\otimes_{j=1}^n L_j(x_j); Q(y))$ with respect to $\varphi$.
This yields the causal Wick expansion
\begin{gather}
\Delta(\otimes_{j=1}^n
L_j(x_j);Q(y))=\sum_{l_1,...;l}\sum_{a_{ij_i},a_j}
C^{l_1...;l}_{a_{11}...a_{1l_1},...;a_1...a_l}\notag\\
\cdot\omega_0\Bigl(\Delta\Bigl(\otimes_{i=1}^n\frac{\d^{l_i}L_i}
{\d(\d^{a_{i1}}\varphi)...\d(\d^{a_{il_i}}\varphi)}(x_i);\frac{\d^{l}Q}
{\d(\d^{a_1}\varphi)...\d(\d^{a_l}\varphi)}(y)\Bigr)\Bigr)\notag\\
\cdot\prod_{i=1}^n\Bigl(\prod_{j_i=1}^{l_i}\d^{a_{ij_i}}\varphi(x_i)\Bigr)
\cdot\prod_{j=1}^l\d^{a_j}\varphi(y)\ , \label{CW}
\end{gather}
where $C^{l_1...;l}_{a_{11}...a_{1l_1},...;a_1...a_l}$ is a
combinatorial factor. Now we crucially use the locality $(iia)$ of
$\omega_0\Bigl(\Delta\Bigl(\frac{\d^{l_1}L_1}
{\d...}(x_1)\otimes...;\frac{\d^{l}Q}{\d...}(y)\Bigr)\Bigr)$.
Hence this distribution is of the form
\beq \omega_0\Bigl(\Delta\Bigl(\frac{\d^{l_1}L_1}
{\d...}(x_1)\otimes...;\frac{\d^{l}Q}{\d...}(y)\Bigr)\Bigr)= \sum_b
C_b\,(\d^b\delta)(x_1-y,...,x_n-y)\ ,\label{Delta:local} \eeq
with constant numbers $C_b$ (due to Translation Invariance) which depend on the field polynomials in
the argument of $\Delta$. 
The term with index $b$ gives the following contribution to
$\Delta_A(L_1(g_1)\otimes...\otimes L_n(g_n))$ (\ref{kernel}):
\begin{gather}
(-1)^{|b|}\int\! dx_1...dx_n\,dy\,\delta(x_1-y,...,x_n-y)\notag\\
\cdot h(y)\cdot\d^b_{x_1...x_n}\Bigl(g_1(x_1)...g_n(x_n)\cdot
\prod_{i=1}^n\Bigl(\prod_{j_i=1}^{l_i}\d^{a_{ij_i}}\varphi(x_i)\Bigr)
\cdot\prod_{j=1}^l\d^{a_j}\varphi(y)\Bigr)\ , \label{b-term}
\end{gather}
where we have omitted constant factors. By reordering the sums we write 
$\Delta_A(L_1(g_1)\otimes...\otimes L_n(g_n))$ in the form (\ref{Delta_A:localization}).
Since $\Delta_A(L_1(g_1)\otimes...\otimes L_n(g_n))$ is multilinear in the fields $L_1,\ldots ,L_n,Q$ and 
symmetric in $L_1,\ldots ,L_n$, the maps $P_a^n$ must satisfy corresponding properties.
\end{proof}

We turn to the scaling behaviour of $\Delta_A$. Assuming that the $R$-products satisfy the axioms
Smoothness in $m\geq 0$, $\mu$-Covariance and almost homogeneous Scaling (\ref{scaling}), the map $(S^{\otimes n}
\otimes A)\mapsto \Delta_A^{(m,\mu)}(S^{\otimes n})$ does not 
scale almost homogeneously in general.
Namely, for non-vanishing mass we are faced with $m$-dependent 
inhomogeneous polynomials: $S_0^{(m)}\equiv S_0\not\in {\cal P}_\mathrm{hom}\,,\,
\frac{\delta S_0^{(m)}}{\delta\varphi}\not\in {\cal P}_\mathrm{hom}\,$ and, hence, 
$A^{(m)}\equiv A$ is in general not in ${\cal P}_\mathrm{hom}$. In typical applications of the MWI
(see e.g.~the $O(N)$-model in Sect.~(\ref{O(N)model}) or current conservation in QED) the simplification appears 
that $Q_j$ is independent of $m$ and the $m$-dependent terms of $\frac{\delta S_0}{\delta\varphi_j}$ cancel in
$A=\int dx\,\sum_jQ_j(x)\,\frac{\delta S_0}{\delta\varphi_j(x)}$. But even with that, 
$\sigma_\rho\,\Delta^{(\rho^{-1}m,\mu)}_{\sigma_\rho^{-1}\,A}((\sigma_\rho^{-1}\,S)^{\otimes n})$
is in general not a polynomial in $(\mathrm{log}\,\rho)$: proceeding inductively
the first term on the right-hand side of 
(\ref{defdelta}) does not scale almost homogeneously due to the mass term in $\frac{\delta S_0}{\delta\varphi}$.\footnote{In
detail: from the derivation of the classical MWI (\ref{klMWI})-(\ref{klMWIIdeal}) it follows that the scaling 
of the right-hand side of the anomalous MWI (\ref{retQAP}) must be such that the mass $m$ in $R^{(m,\mu)}$ and
in $\frac{\delta S^{(m)}_0}{\delta \varphi(y)}$ has the same value that is the scale transformation reads
$\int\! dy\, h(y) \sigma_\rho\circ R^{(\rho^{-1}m,\mu)}(e_\otimes^{\sigma_\rho^{-1}\,S},\sigma_\rho^{-1}\,Q(y))
\frac{\delta S^{(\rho^{-1}m)}_0}{\delta \varphi(y)}\ $.}

For this reason, we assume here the axiom {\bf Scaling Degree}
(\ref{axiom-sd}) instead of Smoothness in $m$, $\mu$-Covariance and almost homogeneous Scaling. 

With this weaker assumption we are going to derive a corresponding 
scaling degree property of $\Delta_A$ and
an upper bound for the mass dimension of
$\Delta_A(S^{\otimes n})$ which does not depend on $n$
if $S$ is a renormalizable interaction. For the latter
purpose we have to define the mass dimension of a local
{\it functional}. 
We use that every $F\in {\cal F}_{\rm loc}$ can uniquely be written in the form
\beq F=\sum_i\int\! dx\,g_i(x)\,P_i(x)\ ,\quad\quad g_i\in {\cal
D}(\MM)\,,\,\,P_i\in {\cal P}_{\rm bal}\ .\label{F-balanced} \eeq
With that we define
\beq {\rm dim}\, F\=d {\rm max}_i\,{\rm dim}\,P_i\label{dim(F)} \eeq
This definition is minimal in the following sense:
\begin{lemma}\label{minimal} Let
\beq F=\sum_i\int\! dx\,g_i(x)\,B_i(x)\ ,\quad\quad g_i\in {\cal
D}(\MM)\,,\,\,B_i\in {\cal P}\ .\label{F-local} \eeq
Then it holds
\beq {\rm dim}\, F\leq {\rm max}_i\,{\rm dim}\,B_i \eeq
\end{lemma}
\begin{proof}
Every $B_i$ can uniquely be written as $B_i=\sum_j
p_{ij}(\d)\,P_{ij}\,,\,\,P_{ij}\in {\cal P}_{\rm bal}$,
where $p_{ij}(\d)$ is a polynomial in the partial derivatives \cite{Duetsch:2004dd}. Inserting
this into (\ref{F-local}) and shifting the derivatives to the test
function we obtain the unique representation (\ref{F-balanced}) of
$F$. The assertion follows from ${\rm dim}\,P_{ij}\leq {\rm
dim}\,B_i\,\,\,\forall i,j$.
\end{proof}

>From $A=\int\!dy\, h(y)\, Q(y)\,\frac{\delta S_0}{\delta \varphi(y)}$ we conclude
\beq {\rm dim}(A)\leq {\rm dim}(Q)+{\rm dim}(\frac{\delta
S_0}{\delta\varphi})\quad {\rm where}\quad {\rm dim}(\frac{\delta
S_0}{\delta\varphi})= {\rm dim}((\square+m^2)\varphi)=\frac{d+2}{2}\
.\label{dim(A)} \eeq
Analogously the relation
$\delta_A S=\int\! dy\, h(y)\, Q(y)\,\frac{\delta S}{\delta\varphi(y)}$ implies
\beq {\rm dim}(\delta_A S)\leq {\rm dim}(Q)+{\rm dim}(S)-{\rm
dim}(\varphi)= {\rm dim}(Q)+{\rm dim}(S)-\frac{d-2}{2}\
.\label{dim(d_AS)} \eeq
With these tools we are ready to formulate and prove the following Proposition. 
\begin{prop}\label{prop:massdimbound}
\begin{itemize}
\item[(i)]{\bf Scaling Degree.} If the $R$-products fulfill the axiom Scaling Degree (\ref{axiom-sd}),
then the scaling degree of the 'vacuum expectation value'
of $\Delta(L_1(x_1)\otimes...;Q(y))$ is bounded by
\beq {\rm sd}\,\omega_0\Bigl(\Delta(\otimes_{j=1}^n
L_j(x_j);Q(y))\Bigr)\leq \sum_{i=1}^n {\rm dim}\,L_i+{\rm
dim}\,Q+\frac{d+2}{2}\ . \label{scaldeg} \eeq
\item[(ii)] {\bf Mass Dimension.} If the $R$-products fulfill the axioms Field Independence and
Scaling Degree (\ref{axiom-sd}), then 
the mass dimension of $\Delta_A(F_1\otimes...\otimes F_n)$ is bounded by
\beq {\rm dim}\,\Delta_A(F_1\otimes...\otimes F_n)\leq \sum_{i=1}^n
{\rm dim}(F_i)+{\rm dim}\,Q+\frac{d+2}{2}-d\, n\ .\label{massdim}
\eeq
For a renormalizable interaction, that is ${\rm dim}(S)\leq d$, this
implies
\beq {\rm dim}\,\,\Delta_A(e_{\otimes}^S)\leq {\rm
dim}\,Q+\frac{d+2}{2}\ .\label{massdimbound} \eeq
\end{itemize}
\end{prop}
Note that for a renormalizable interaction the upper bounds on
the mass dimension of $A$ (\ref{dim(A)}), $\delta_AS$
(\ref{dim(d_AS)}) and $\Delta_A(e_{\otimes}^S)$ agree, as one
expects because the sum $(A+\delta_AS+\Delta_A(e_{\otimes}^S))$
appears in Theorem \ref{satzQAP}.
\begin{proof}
(i) Proceeding by induction on $n$ we apply $\omega_0$ to (\ref{kernel1}) and
estimate the scaling degree
of the resulting terms on the right-hand side by using
\beq {\rm
sd}\,\omega_0\Bigl(R(\otimes_{j=1}^{n-1}L_j(x_j);L_n(x_n))\Bigr)\leq
\sum_{j=1}^n {\rm dim}\,L_j\ , \eeq
which follows from (\ref{axiom-sd}). (In contrast to (\ref{axiom-sd}) we do not assume $L_j\in 
{\cal P}_\mathrm{hom}$.) For example we obtain
\begin{gather}
{\rm sd}\,\omega_0\Bigl(R(\otimes_{j(\not= l)}
L_j(x_j);Q(y)\,(\d^a\delta)(x_l-y)\,
\frac{\d L_l}{\d(\d^a\varphi)}(x_l))\Bigr)\notag\\
\leq\sum_{j(\not= l)} {\rm dim}\,L_j+{\rm dim}\,Q+{\rm dim}
\Bigl(\frac{\d L_l}{\d(\d^a\varphi)}\Bigr)+{\rm sd}(\d^a\delta)\notag\\
=\sum_{j=1}^n {\rm dim}\,L_j+{\rm dim}\,Q+\frac{d+2}{2}\ .
\end{gather}
The same bound results for all other terms. (The vacuum expectation
value of the last term vanishes.)

(iv) We may assume $F_j=L_j(g_j)$ with  $L_j$ a balanced field, $L_j\in {\cal P}_\mathrm{bal}$.
With that it holds ${\rm dim}(L_j)={\rm dim}(F_j)$. We use the formulas derived in the proof of part $(ii)$
of Lemma \ref{Delta:RC}, in particular the causal Wick expansion of $\Delta(\otimes_{j}
L_j(x_j);Q(y))$ (\ref{CW}). (In the formulas given below the indices have precisely the same meaning as in that proof.)
Looking at (\ref{Delta:local}), we get an upper bound for the scaling degree of the left-hand side by means of 
(\ref{scaldeg}). This implies that the range of $b$ on the right-hand side of (\ref{Delta:local}) is
restricted by
\begin{gather}
|b|+d\,n\leq \sum_{i=1}^n \Bigl({\rm dim}\,L_i-
\sum_{j_i=1}^{l_i}\Bigl(|a_{ij_i}|+\frac{d-2}{2}\Bigr)\Bigr)\notag\\
+{\rm
dim}\,Q-\sum_{j=1}^{l}\Bigl(|a_j|+\frac{d-2}{2}\Bigr)+\frac{d+2}{2}\ .\label{bound:b}
\end{gather}
All terms of $\Delta_A(L_1(g_1)\otimes...\otimes L_n(g_n))$ are of the form (\ref{b-term})
where $|b|$ is bounded by (\ref{bound:b}). Due to Lemma \ref{minimal}
the mass dimension of the functional (\ref{b-term}) is bounded by
\begin{gather}
\sum_{i=1}^n\sum_{j_i=1}^{l_i}\Bigl(|a_{ij_i}|+\frac{d-2}{2}\Bigr)
+\sum_{j=1}^{l}\Bigl(|a_j|+\frac{d-2}{2}\Bigr)+|b|\notag\\
\leq \sum_{i=1}^n {\rm dim}\,L_i+{\rm dim}\,Q+\frac{d+2}{2}-d\,n\ .
\end{gather}
This implies the assertion (\ref{massdim}).
\end{proof}
\subsection{The MWI in the proper field formalism, and the Quantum Action Principle}\label{effMWI-QAP}

In the literature the Quantum Action Principle (QAP) is formulated in terms of time ordered products. Hence, to be able to compare 
our results with the QAP, we first rewrite Theorem~\ref{satzQAP} using $T$-products:
\begin{lemma}\label{satzQAP-T}
Using the assumptions and notations of Theorem \ref{satzQAP} the anomalous MWI \eqref{retQAP} can equivalently 
be written in terms of the time-ordered product: 
\begin{equation}\label{retQAP-T} 
T\big(e_\otimes^{iS/\hbar}\otimes (A+\delta_A S+\Delta_A(\ets))\big)=\int\! dy\, h(y) T(e_\otimes^{iS/\hbar}\otimes Q(y))\frac{\delta S_0}{\delta \varphi(y)}.
\end{equation}
\end{lemma}
\begin{proof}
Using the formula of Bogoliubov \eqref{R-T}
and the identity
\beq
(F\star G)\cdot\frac{\delta S_0}{\delta \varphi}=F\star \big(G \cdot \frac{\delta S_0}{\delta\varphi}\big)\qquad \forall F, G\in \mathcal{F}
\eeq
(cf.~(\ref{stkom})), where $\cdot$ denotes the classical product, we obtain
\begin{gather}
\bar T(e_\otimes^{-iS/\hbar})\star T\big(e_\otimes^{iS/\hbar}\otimes (A+\delta_A S+\Delta_A(e_\otimes^S))\big)=\notag\\
\bar T(e_\otimes^{-iS/\hbar})\star \bigg(\int\! dy \,h(y)\,
T(e_\otimes^{iS/\hbar}\otimes Q(y))\cdot\frac{\delta S_0}{\delta\varphi(y)}\bigg)\ .
\end{gather}
Multiplication by $T(e_\otimes^{iS/\hbar})$ from the left yields \eqref{retQAP-T}.
\end{proof}

As shown in (\ref{MWI-tree}) the MWI is satisfied by $T_{\rm tree}$, i.e.~it can be violated only by the contribution of the loop diagrams. 
To concentrate the study of the solvability of the MWI to that terms for which the MWI is a non-trivial condition,
we reformulate the anomalous MWI \eqref{retQAP} within the proper field formalism introduced in Sect.~\ref{propV}.
\begin{cor}\label{satzeffQAP-T}
The anomalous MWI in Theorem \ref{satzQAP} can equivalently be expresses in terms of the vertex function
$\Gamma_T$ (defined in Prop.~\ref{prop:gamma}): 
\begin{equation}\label{effQAP-T}
\int\! dx\, h(x)\Gamma_T(e_\otimes^S\otimes Q(x))\frac{\delta(S_0+\Gamma_T(e_\otimes^S))}{\delta \varphi(x)}=\Gamma_T\big(e_\otimes^S\otimes (A+\delta_A S + \Delta_A(\ets))\big).
\end{equation}
\end{cor}
\begin{proof}
Applying \eqref{Gamma} on both sides of \eqref{retQAP-T} we obtain
\begin{eqnarray*}
\lefteqn{T_{\mathrm{tree}}\Big(e_\otimes^{i\Gamma_T(e_\otimes^S)/\hbar}\otimes \Gamma_T(e_\otimes^S\otimes (A+\delta_A
S+\Delta_A(e_\otimes^S))\big)\Big)}\\
&\hspace{2cm}&=\int\! dy\, h(y)T_{\mathrm{tree}}\Big(e_\otimes^{i\Gamma_T(e_\otimes^S)/\hbar}\otimes \Gamma_T\big(e_\otimes^S\otimes
Q(y)\big)\Big)\frac{\delta S_0}{\delta \varphi(y)}.
\end{eqnarray*}
On the right-hand side we use the classical MWI in terms of 
$T$-products \eqref{MWI-tree} and obtain
\begin{equation*}
\cdots=T_{\mathrm{tree}}\Big(e_\otimes^{i\Gamma_T(e_\otimes^S)/\hbar}\otimes \int\! dy\, h(y)\Gamma_T(e_\otimes^S\otimes Q(y))\frac{\delta
(S_0+\Gamma_T(e_\otimes^S))}{\delta \varphi(y)}\Big),
\end{equation*}           
which leads, by means of \eqref{Tps1}, to the assertion \eqref{effQAP-T}.
\end{proof}

To compare this result with the literature, we introduce $\Gamma(S_0,S)\defi S_0+\Gamma_T(e_\otimes^S)$ and, 
motivated by (\ref{properinter}) and (\ref{G=lGt}), interpret $\Gamma(S_0,S)$ as
the \emph{proper total action} associated with the total classical action $S_{\mathrm{tot}}=S_0+S$. 
In addition, we introduce 'insertions' (see \cite{Piguet:1995er}) in our algebraic formalism:
let $P\in \mathcal{C}^\infty(\MM,\mathcal{P})$ and $\rho\in
\mathcal{D}(\mathbbm{M})$ some testfunction (in the literature often called 'external field'). We define the
'insertion of $P(x)$', usually denoted\footnote{The dot does not mean the classical product here!} by $P(x)\cdot
\Gamma(S_0, S)$, as follows:\footnote{Note, 
that in the setting of causal perturbation theory the introduction of external fields in order 
to express 'insertions' or nonlinear symmetry transformations is -- in contrast to conventional perturbation theory -- not necessary.}   
\beq 
P(x)\cdot \Gamma(S_0, S)\defi
\frac{\delta}{\delta\rho(x)}\Big|_{\rho\equiv 0}\Gamma\Big(S_0, S+\int\!
dx \rho(x) P(x)\Big) = \Gamma_T\big(e_\otimes^S\otimes P(x)\big). 
\eeq 
We denote by  $S'\defi S+\int\! dx \rho(x) Q(x)$, $\rho \in \mathcal{D}(\mathbbm{M})$ the modified classical interaction 
containing a coupling to the external field $\rho$, and write $\Delta_A(e_\otimes^S)=
\int\! dx\,h(x) \tilde\Delta(e_\otimes^S;Q(x))$ with $\tilde\Delta(e_\otimes^S;Q(x))
\in {\cal D}(\MM,{\cal P})$ (Lemma \ref{Delta:RC}\emph{(ii)}).
Using this notation and the fact that the anomalous MWI
\eqref{effQAP-T} is valid for all $h\in \mathcal{D}(\mathbbm{M})$, we rewrite (\ref{effQAP-T}) in terms of the proper total action:
\beq\label{QAP}
\frac{\delta \Gamma(S_0, S')}{\delta \rho(x)}\frac{\delta \Gamma(S_0, S')}{\delta \varphi(x)}\Big|_{\rho\equiv 0}=\Delta(x)\cdot \Gamma(S_0, S),
\eeq
where the local field $\Delta(x)$ is given by $\Delta(x)\defi Q(x)
\frac{\delta(S_0+S)}{\delta \varphi(x)}+\tilde\Delta(e_\otimes^S;Q(x))$. 
This formulation of the anomalous MWI is formally equivalent to the general formulation of the QAP in \cite{Piguet:1995er}. Note that due to 
property \emph{(v)} in Theorem \ref{satzQAP} and \eqref{properinter}, the $\hbar$-expansion of the right-hand side of \eqref{QAP} 
starts with
\beq
\Delta(x)\cdot \Gamma(S_0, S)=Q(x)\frac{\delta(S_0+S)}{\delta \varphi(x)}+\mathcal{O}(\hbar)\equiv \frac{\delta (S_0+S')}{\delta \rho(x)}\frac{\delta (S_0+S')}{\delta \varphi(x)}\Big|_{\rho= 0}+\mathcal{O}(\hbar),
\eeq
which is completely analogous to the expansion of the right-hand side of the QAP in \cite{Piguet:1995er}. 
Moreover, for a renormalizable interaction $S$, the mass dimension of the local insertion $\Delta(x)$ is 
bounded by $\mathrm{dim}Q+\frac{d+2}{2}=\mathrm{dim}Q-\mathrm{dim}\varphi+d$ (Prop.~\ref{prop:massdimbound}),
in agreement with \cite{Piguet:1995er}.

\subsection{Removal of violations of the MWI}\label{fulfillmentMWI}
\subsubsection{Fulfillment of the MWI to first order in $S$}
Before we are going to investigate the question whether the MWI can be 
fulfilled in a concrete model (i.e.~for a given 
$A\in \mathcal{J}_{\mathrm{loc}}$ and a given interaction 
$S\in {\cal F}_{\rm loc}$), we will show that the 
MWI can always be fulfilled to first order in the 
interaction $S$.\footnote{E.g.~the axial anomaly of QED
is of second order in the interaction.} Apart from the interest in its own, 
this result will be needed in our proof
of the Ward identities of the scalar $O(N)$-model (Sect.~\ref{O(N)model}).
\begin{prop}\label{MWI:order1} 
Given $A=\int\! dy\, h(y) Q(y)\frac{\delta S_0}{\delta \varphi(y)}$
(with $Q\in \mathcal{P}$) and $S\in {\cal F}_{\rm loc}$ 
there exists a retarded product $R_{1,1}$ which fulfills all 
renormalization conditions (specified in Sect.~\ref{pQFT}) and
\begin{equation}\label{claim}
R_{1,1}(S, A)+\delta_A S=\int\! dy\,h(y) R_{1,1}(S, Q(y))
\frac{\delta S_0}{\delta \varphi(y)}\ .
\end{equation}
\end{prop}
\begin{proof}
We start with $R$-products satisfying all renormalization conditions. 
For simplicity we work in the proper field formalism.
>From the anomalous MWI (Cor.~\ref{satzeffQAP-T}) we obtain to first order in $S$
\beq\label{aMWI first order}
\Delta_A(S)=-\int\! dy\, h(y)\Big(\Gamma_T\big(S\otimes Q(y)\frac{\delta S_0}{\delta \varphi(y)}\big)-\Gamma_T\big(S\otimes Q(y)\big)\frac{\delta S_0}{\delta \varphi(y)}\Big)\ .
\eeq  
By inserting the Wick expansion (\ref{wickentP}) of $\Gamma_T$ (which holds since $\Gamma_T$ fulfills Field Independence), 
we find that the terms containing no contraction of $\frac{\delta S_0}{\delta \varphi}$ cancel. It remains:
\beq\label{aMWI first order again}
\Delta_A(S)=-\int\! dx\,dy\, g(x)h(y) \sum_{\overline{\mathcal{L}}\subset \mathcal{L}, \overline{Q}\subset Q} 
\omega_0\Big(\Gamma_T\big(\overline{\mathcal{L}}(x)\otimes \overline{Q}(y)
\frac{\delta S_0}{\delta \varphi(y)}\big)\Big) \underline{\mathcal{L}}(x)\underline{Q}(y)
\eeq
where we write $S=\int\! dx\, g(x) \mathcal{L}(x)$, $g\in \mathcal{D}(\mathbbm{M})$ 
and sum over all 
subpolynomials $\overline{P}=\frac{\del^{l}P}{\del(\del^{a_{1}} \varphi)
\cdots \del(\del^{a_{l}}\varphi)}$ of $P\in \mathcal{P}$ (for $P={\cal L},\,Q$), 
denoting by $\underline{P}=\frac{1}{l!}\prod_{j=1}^{l}\del^{a_{j}}\varphi$ 
the corresponding  'counterpart' (see (\ref{wickentP})). Due to locality of $\Delta_A(S)$ the distributions 
$\omega_0\Big(\Gamma_T(\overline{\mathcal{L}}(x)\otimes \overline{Q}(y)\frac{\delta S_0}{\delta \varphi(y)})\Big)$ 
are supported on the diagonal $\{(x,y)\vert x=y\}$, cancellations of non-local terms on the right-hand side of 
(\ref{aMWI first order again}) are impossible.
We are searching a finite renormalization $\Gamma_T\rightarrow\hat\Gamma_T$ which removes the violating term
$\Delta_A(S)$. Due to (\ref{renorm-gamma-T}) 
a renormalization of $\Gamma_T(S\otimes F)$ must be of the form
\beq
\hat{\Gamma}_T(S\otimes F)=\Gamma_T(S\otimes F)+D(S\otimes F)\, , \quad \forall F\in \mathcal{F}_{\mathrm{loc}}\ ,
\eeq
with $D$ satisfying the properties listed in the Main Theorem. To obtain a field independent $D$ we construct it from its vacuum
expectation values by the causal Wick expansion (\ref{wickentP}). Obviously, setting
\begin{equation}\label{D-MWI:order1}
\omega_0\Big(D\big(\overline{\mathcal{L}}(x)\otimes \overline{Q}(y)\frac{\delta S_0}{\delta \varphi(y)}\big)\Big)
\defi -\omega_0\Big(\Gamma_T\big(\overline{\mathcal{L}}(x)\otimes \overline{Q}(y)\frac{\delta S_0}{\delta \varphi(y)}\big)\Big)\ ,
\end{equation}
the corresponding $\hat\Gamma_T$ satisfies the MWI to first order in $S$.
Here we essentially use that the distribution on the right-hand side has local support. One
verifies easily that $D$ (and hence the resulting $\hat\Gamma_T$) satisfies all renormalization conditions.
\end{proof}
 
\subsubsection{Removal of possible anomalies by induction on 
the order in $\hbar$}\label{fulfillMWI}
The formal equivalence of the anomalous MWI (written in the form 
of Cor.~\ref{satzeffQAP-T}) and the QAP makes it possible to apply basic 
techniques of algebraic renormalization within the framework of 
causal perturbation theory. The main idea underlying algebraic renormalization 
is to start with an arbitrary renormalization prescription; 
with that the considered Ward identities are in general broken.
Proceeding by induction on the power of $\hbar$, 
the question whether one can find a finite 
renormalization which removes the possible anomaly 
(i.e.~the local terms violating the Ward identity) 
then amounts -- by means of the QAP -- to purely algebraic (often cohomological) problems.


We will apply this strategy to the MWI in the proper field formalism (which is given
by (\ref{effQAP-T}) with $\Delta_A(e_\otimes^S)=0$): given $A\in \mathcal{J}_\mathrm{loc}$ 
and $S\in {\cal F}_{\rm loc}$, we are going to investigate whether possible anomalies of 
the corresponding MWI can be removed by finite renormalizations of $\Gamma_T$
(\ref{renorm-gamma-T}).
Proceeding by induction on the power of $\hbar$, we assume
that for a given map $\Gamma^{(k)}_T$ (which satisfies the renormalization conditions
Unitarity, ${\cal P}_+^\uparrow$-Covariance, Field 
Independence, Field Equation and Scaling Degree,
see Lemma \ref{Gamma:rencond}) the MWI is violated only by terms of order $\hbar^{k}$; 
that is in the anomalous MWI (Cor.~\ref{satzeffQAP-T}) for $\Gamma^{(k)}_T$, 
\begin{gather}
\Gamma^{(k)}_T\Bigl(e_\otimes^S\otimes (A+\delta_AS)\Bigr)+\Gamma^{(k)}_T(\ets\otimes\Delta_A^{(k)}(e_\otimes^S))
= \notag\\
\int\! dy\,h(y)\,\Gamma^{(k)}_T\Bigl(e_\otimes^S\otimes Q(y)\Bigr)\, \frac{\delta
(S_0+\Gamma^{(k)}_T(e_\otimes^S))}{\delta\varphi(y)}\ ,\label{MWI kth order} 
\end{gather}
the violating term can be written as
\beq
\Gamma^{(k)}_T(\ets\otimes\Delta_A^{(k)}(e_\otimes^S))=\Delta_A^{(k)}(e_\otimes^S)+{\cal O}(\hbar^{k+1})\ ,\quad
\Delta_A^{(k)}(e_\otimes^S)={\cal O}(\hbar^{k})\ ,\label{MWI-Delta-Gamma:k-1}
\eeq
where \eqref{properinter} is used.
Note that, due to Theorem \ref{satzQAP} \emph{(v)} and \eqref{properinter}, any vertex function $\Gamma_T$ fulfills this 
assumption (\ref{MWI kth order})-(\ref{MWI-Delta-Gamma:k-1})
for $k=1$ and, hence, can be used as $\Gamma^{(1)}_T$. 

To fulfill the MWI to $k^{\mathrm{th}}$ order in $\hbar$ (and to maintain the other renormalization conditions), we have to find a 
renormalization map $D^{(k)}$ (with the properties $(i)$-$(v)$ and 
$(vi)(B)$ of the Main Theorem) such that
\begin{eqnarray}
\lefteqn{\Gamma^{(k+1)}_T\Bigl(e_\otimes^S\otimes (A+\delta_AS)\Bigr)+{\cal
O}(\hbar^{k+1})=}\nonumber\\
&&\hspace{1cm} \int\! dy\,
h(y)\,\Gamma^{(k+1)}_T\Bigl(e_\otimes^S\otimes Q(y)\Bigr)\, \frac{\delta
(S_0+\Gamma^{(k+1)}_T(e_\otimes^S))}{\delta\varphi(y)}\ ,
\label{MWI-Delta-Gamma:k}
\end{eqnarray}
with $\Gamma_T^{(k+1)}$ given by
\beq
\Gamma_T^{(k+1)}(e_\otimes^F)=\Gamma_T^{(k)}\Big(e_\otimes^{D^{(k)}(e_\otimes^F)}\Big) \qquad \forall F\in \mathcal{F_\mathrm{loc}}\ .
\eeq
To maintain the validity of the MWI to lower orders in $\hbar$ we choose $D^{(k)}$ to be of the form
\beq\label{renormmap}
D^{(k)}(e_\otimes^F)=F+D^{(k)}_{>1}(e_\otimes^F),\,\quad D^{(k)}_{>1}(e_\otimes^F)={\cal O}(\hbar^k)\ ,\quad\forall F\in
{\cal F}_{\rm loc}\ ,\ F\sim\hbar^0\ .
\eeq
This implies
\beq e_\otimes^{D^{(k)}(e_\otimes^F)}=e_\otimes^F+e_\otimes^F\otimes_\mathrm{sym}
D^{(k)}_{>1}(e_\otimes^F)+{\cal O}(\hbar^{k+1})\ \eeq
(where $\otimes_\mathrm{sym}$ denotes the symmetrized tensor product) and
\begin{gather}
\Gamma^{(k+1)}_T(e_\otimes^S)=\Gamma^{(k)}_T(e_\otimes^S)+D^{(k)}_{>1}(e_\otimes^S)
+{\cal O}(\hbar^{k+1})\ ,\\
\Gamma^{(k+1)}_T(e_\otimes^S\otimes F)=\Gamma^{(k)}_T(e_\otimes^S\otimes F)+
D^{(k)}_{>1}(e_\otimes^S\otimes F) +{\cal O}(\hbar^{k+1}).
\end{gather}
We insert the latter two equations into our requirement
(\ref{MWI-Delta-Gamma:k}) and use the inductive assumption (\ref{MWI kth order})-(\ref{MWI-Delta-Gamma:k-1}).
It results
\begin{eqnarray}\label{renormcon}
\Delta_A^{(k)}(e_\otimes^S)&=&D^{(k)}_{>1}\bigl(e_\otimes^S\otimes (A+\delta_A S)\bigr)-\delta_A D^{(k)}_{>1}(e_\otimes^S)\nonumber\\
&&-\int\! dy\,h(y)\,D^{(k)}_{>1}(e_\otimes^S\otimes Q(y))\, \tfrac{\delta
(S_0+S)}{\delta\varphi(y)}+{\cal O}(\hbar^{k+1})\,.
\end{eqnarray}
The violating term $\Delta_A^{(k)}(e_\otimes^S)$ is 
inductively given by $\Gamma_T^{(k)}$. 
If we succeed to find a corresponding map $D_{>1}^{(k)}$ fulfilling \eqref{renormcon}
and the properties of a renormalization map, then the pertinent finite renormalization removes 
the 'anomaly' $\Delta_A^{(k)}(e_\otimes^S)$.  
However, since $D_{>1}^{(k)}$ appears in \eqref{renormcon} several times with different arguments it 
seems almost impossible to discuss the existence of solutions in general.
\subsubsection{Assumption: localized off-shell version of  
Noether's Theorem}\label{sect:symmcurrent}
In various important applications of the MWI the search for solutions $D_{>1}^{(k)}$
of (\ref{renormcon}) is simplified due to the validity of the following assumption. 
In the given model the total action\footnote{$\kappa$ denotes the coupling constant.}
\begin{equation*}
S_0+S(g)\,\,\, \mathrm{(with}\,\,\, S(g)=\sum_{n\geq 1}\kappa^n\,S_n(g)\,,\,\,
S_n(g)=\int dx\,(g(x))^n\,{\cal L}_n(x)\,,\,\,g\in {\cal D}(\MM)\,,
\,{\cal L}_n\in {\cal P}\mathrm{)}
\end{equation*}
is invariant with respect to the symmetry transformation
\begin{equation*}
\delta_A=\int dy\,h(y)\,Q(y)\,
\frac{\delta}{\delta\varphi (y)}=\sum_{n\geq 0}\kappa^n\,\delta_{A_n}\,\,
\mathrm{(corresponding\>to}\,\, A=\sum_{n\geq 0}\kappa^n\,A_n\,,\,\,
A_n\in\mathcal{J}_{\rm loc}\mathrm{)}
\end{equation*}
in the following way: there exist
\begin{itemize}
\item a current $j^\mu(g)=\sum_{n\geq 0}\kappa^n\,j^\mu_n(g)\,,\,\,\,
j^\mu_n(g)(x)=(g(x))^n\,j_n^{\mu}(x)$
(with $j_n^{\mu}\in {\cal P}$) and 
\item a 'Q-vertex'\footnote{The name 'Q-vertex' is due to 'perturbative gauge
invariance' \cite{DHKSII}, which is related to BRST-symmetry.}
${\cal L}^{(1)\,\mu}(g)=\sum_{n\geq 1}\kappa^n\,{\cal L}^{(1)\,\mu}_n(g)\,,\,\,\,
{\cal L}^{(1)\,\mu}_n(g)(x)=(g(x))^{n-1}\,{\cal L}^{(1)\,\mu}_n(x)$
(with ${\cal L}^{(1)\,\mu}_n\in {\cal P}$)
\end{itemize}
such that 
\begin{equation}\label{symmetrycond} 
\delta_A(S_0+S(g))\equiv A+\delta_A S(g)=\int dx\,h(x)\Bigl(\del_\mu j^\mu(g)(x)
-{\cal L}^{(1)\,\mu}(g)(x)\,\del_\mu g(x)\Bigr) 
\end{equation} 
in the sense of formal power series in $\kappa$. To zeroth order in $\kappa$ the 
assumption (\ref{symmetrycond}) reads
\begin{equation}\label{symmetrycond:0}  
\delta_{A_0}S_0\equiv A_0=\int dx\,h(x)\del_\mu j_0^\mu(x)\ ,  
\end{equation} 
where $j_0$ is the symmetry current of the underlying free theory.
This simplifying assumption can be interpreted as the 
validity of an off-shell version of Noether's Theorem for the case that the
interaction and the symmetry transformation are localized.
It is satisfied e.g.~for the scalar $O(N)$-model treated in Sect.~\ref{O(N)model} and
for the BRST-symmetry of (massless) Yang-Mills theories, massive spin-1 fields
and massless spin-2 fields (gravity). 
For the interacting scalar $O(N)$-model (see Sect.~\ref{O(N)model}) 
the simplifications $A=A_0$, $\delta_A S=0$, ${\cal L}^{(1)} =0$ and $j=j_0$ 
appear. However, for the BRST-symmetry $\delta_A$ and $j$ are
generically non-trivial deformations of 
$\delta_{A_0}$ and $j_0$, and $\delta_A S$ and ${\cal L}^{(1)}$ do not vanish.
\medskip

\noindent {\it Example: BRST-symmetry.} We are going to verify that the above
mentioned models satisfy the assumption (\ref{symmetrycond}). Our argumentation is 
based on conservation of the classical BRST-current. For constant couplings
(i.e.~$g(x)=1\>\>\forall x$) there is a conserved Noether current $j=
\sum_{n\geq 0}\kappa^n\,j_n$, due to the BRST-invariance of the total action.
We use these $j_n$'s to construct the BRST-current of the corresponding model
with localized coupling $\kappa g(x)$ ($g\in {\cal D}(\MM)$): we set
$j(g)(x):=\sum_{n\geq 0}(\kappa g(x))^n\,j_n(x)$. The violation of the conservation 
of $j(g)$ is expressed in terms of the $Q$-vertex \cite{Duetsch:1998hf, Duetsch:2001sw,
Duetsch:2002yp}: in Sects.~3 and 4 of
\cite{Duetsch:2005} it is shown that for the considered models there 
exists a $Q$-vertex ${\cal L}^{(1)}(g)$ such that
\beq
R_{\rm cl}\Bigl(e_\otimes^{S(g)},\int dx\,h(x)(\del j(g)(x)-{\cal L}^{(1)\,\mu}(g)(x)
\,\del_\mu g(x))\Bigr)\in {\cal J}\ .\label{currconserv}
\eeq
Proceeding by induction on the order in $\kappa$ and using the MWI (as it is 
worked out in formulas (190)-(191) and (152)-(157) of \cite{Duetsch:2002yp})
one finds that (\ref{currconserv}) is equivalent to the sequence of relations
\beq
A_0:=\int dx\,h(x)\,\del j_0(x)\in {\cal J}_{\rm loc}\label{cc:0}
\eeq
and
\beq
A_n:=-\sum_{l=0}^{n-1}\delta_{A_l}S_{n-l}(g)+\int dx\,h(x)\Bigl(\del j_n(g)(x)
-{\cal L}^{(1)}_n(g)(x)\,\,\del g(x)\Bigr)\in {\cal J}_{\rm loc}\label{cc:n}
\eeq
for $n\geq 1$. This yields our assumption: with (\ref{cc:0}) the condition (\ref{symmetrycond:0})
holds true and (\ref{cc:n}) implies
$A_n=\delta_{A_n}S_0$ and with that (\ref{cc:n}) gives (\ref{symmetrycond})
to $n$th order in $\kappa$. $\delta_A=\sum_{n\geq 0}\kappa^n\,\delta_{A_n}$ is
a localized version of the usual BRST-transformation.

We point out that the nilpotency of the BRST-transformation is not directly 
used.  Every model fulfilling the local current conservation (\ref{currconserv})
(with $S(g)$, $j(g)$ and ${\cal L}^{(1)}(g)$ of the above form) satisfies our 
assumption \eqref{symmetrycond} if the symmetry transformation $\delta_A$ 
is defined by (\ref{cc:0})-(\ref{cc:n}).
\medskip

Assuming now the validity of \eqref{symmetrycond}, we are going 
to derive a simplified version of \eqref{renormcon}. For shortness
and coincidence with the notations of the preceding Sects.~we write
$S$, $j$ and ${\cal L}^{(1)}$ for $S(g)$, $j(g)$ and ${\cal L}^{(1)}(g)$, respectively.
For the time being, we additionally assume that the testfunction $h$ satisfies 
$h(x)=1$ for all $x\in \mathrm{supp}(\delta\,S/\delta\varphi)$. 
With that $\Delta_A^{(k)}(e_\otimes^S)$ is independent of the 
choice of $h$ within this class and $h$
can be replaced by the number $1$ (see Lemma \ref{Delta:RC}\emph{(ii)}). 
$h$ does not appear also on the right-hand side of  \eqref{renormcon}: namely,
due to our assumption and the locality of the map $D^{(k)}_{>1}$ 
(see Theorem \ref{maintheorem}\emph{(i)}) we obtain
\beq
D^{(k)}_{>1}\Bigl(e_\otimes^S\otimes(A+\delta_A S)\Bigr)=\int\! dy\,\Bigl(\del_\mu^y 
\underbrace{D^{(k)}_{>1}\big(e_\otimes^S\otimes j^\mu(y)\big)}_{=0\,
\mathrm{for}\,y\notin \mathrm{supp}(\delta S/\delta\varphi)}
- D^{(k)}_{>1}\big(e_\otimes^S\otimes {\cal L}^{(1)\,\mu}(y)\big)\,\d_\mu g(y)\Bigr)
\eeq
and, using again the locality of $D^{(k)}_{>1}$, \eqref{renormcon} simplifies to the condition
\begin{gather}
\Delta^{(k)}(e_\otimes^S)=-\int\! dy\, \Big(D^{(k)}_{>1}\big(e_\otimes^S\otimes Q(y)\big)
\tfrac{\delta(S_0+S)}{\delta \varphi(y)}+ D^{(k)}_{>1}
\big(e_\otimes^S\otimes {\cal L}^{(1)}(y)\big)\,\d g(y)\Big)\notag\\
-\delta D^{(k)}_{>1}(e_\otimes^S)+{\cal O}(\hbar^{k+1})\ ,\label{renormcongrenz}
\end{gather}
where $\Delta^{(k)}(e_\otimes^S)\=d \Delta_A^{(k)}(e_\otimes^S)\vert_{h\equiv 1}$ and
$\delta$ is the non-localized version of $\delta_A$:
\beq
\delta\=d\int dx\, Q(x)\frac{\delta}{\delta \varphi(x)}.
\eeq
It is much easier to find a solution $D^{(k)}_{>1}$ for (\ref{renormcongrenz}) than 
for \eqref{renormcon}, since in (\ref{renormcongrenz}) the localization of the 
derivation $\delta_A$ is removed (i.e.~$h$ is replaced by $1$) and since the 
$D^{(k)}_{>1}(e_\otimes^S\otimes \d j)$-term vanishes.

However, we want to solve the MWI for general $h\in {\cal D}(\MM)$.\footnote{For example this is used in the 
derivation of our version of the QAP (\ref{QAP}) from the anomalous MWI.} 
To investigate whether \eqref{renormcongrenz} is also sufficient for the more involved condition 
\eqref{renormcon} for arbitrary $h$, let a solution $D^{(k)}_{>1}$ of \eqref{renormcongrenz} be given. We point out that the map $D^{(k)}_{>1}$ is not completely determined by that, only the 
combination of $D^{(k)}_{>1}(e_\otimes^S)$, $D^{(k)}_{>1}(e_\otimes^S\otimes Q)$ and 
$D^{(k)}_{>1}(e_\otimes^S\otimes {\cal L}^{(1)})$ 
appearing on the right-hand side of (\ref{renormcongrenz})
is fixed. We claim that, given such a $D^{(k)}_{>1}$, there exists a linear map 
$K^\mu:\TT\mathcal{F}_\mathrm{loc}\rightarrow \mathcal{D}(\mathbbm{M},\mathcal{P})$
with the property
\begin{gather}
\Delta_A^{(k)}(e_\otimes^S)=\int\! dy\,
h(y)\biggl(\del_\mu K^\mu(e_\otimes^S)(y)
-D^{(k)}_{>1}(e_\otimes^S\otimes Q(y))\, \tfrac{\delta
(S_0+S)}{\delta\varphi(y)}\notag\\
-D^{(k)}_{>1}\big(e_\otimes^S\otimes {\cal L}^{(1)}(y)\,\d g(y)\big)\biggr)
-\delta_A D^{(k)}_{>1}(e_\otimes^S)+{\cal O}(\hbar^{k+1})\label{renormconre}
\end{gather}
for all $h\in\mathcal{D}(\mathbbm{M})$ and with $K^\mu(1)=0$. The latter condition
is compatible with (\ref{renormconre}), because to zeroth order in $\kappa$ the condition
(\ref{renormconre}) reduces to $\int h\,\d K(1)={\cal O}(\hbar^{k+1})$, due to
$D^{(k)}_{>1}(1\otimes F)=D^{(k)}_{>1}(F)=0$ ($\forall F$, see (\ref{renormmap})).

To show the existence of $K^\mu$, we first prove 
the following Lemma, which describes the difference between $\delta_A$ and 
$\delta$ with respect to their action on local functionals: 
\begin{lemma}\label{lemma:derivloc}
Let be given $F\in {\cal F}_\mathrm{loc}\,$, $l\in \mathcal{D}(\mathbbm{M}, \mathcal{P})$ and
a localized derivation $\delta_h=\int\! dx\, h(x) Q(x)\frac{\delta}{\delta \varphi(x)}$, $h\in \mathcal{D}(\mathbbm{M})$, $Q\in \mathcal{P}$, such that the corresponding non-localized derivation $\delta=\int\! dx\, Q(x)\frac{\delta}{\delta \varphi(x)}$ satisfies
\begin{equation}\label{varglobal}
\delta F=\int\! dx\, l(x).
\end{equation}
Then there exists a $k^\mu \in \mathcal{D}(\mathbbm{M},\mathcal{P})$ such that the following localized version of \eqref{varglobal} holds true:
\beq\label{varlocal}
\delta_h F=\int\! dx\, h(x)\,(l(x)+ \del_\mu k^\mu(x))\ .
\eeq
\end{lemma}
\begin{proof}
Let $f\in \mathcal{D}(\mathbbm{M}, \mathcal{P})$ with $F=\int\! dy\, f(y)$.
Carrying out the functional derivative in \eqref{varglobal} we conclude
\beq
l(x)=\sum_{a\in \mathbbm{N}_0^d} \del^a Q(x) \frac{\del f}{\del (\del^a \varphi)}(x)+\del_\mu k_1^\mu(x)
\eeq
for some $k_1^\mu \in \mathcal{D}(\mathbbm{M}, \mathcal{P})$. On the other hand we obtain
\begin{eqnarray}
\delta_h F&=&
\int\! dx\, \sum_{a\in \mathbbm{N}_0^d} \del^a \big(h(x)Q(x)\big) \frac{\del f}{\del (\del^a \varphi)}(x)\nonumber\\
&=&\int\! dx\, h(x) \sum_{a\in \mathbbm{N}_0^d} \del^a Q(x) \frac{\del f}{\del (\del^a \varphi)}(x)+\int\! dx\,h(x)\del_\mu k_2^\mu(x)
\end{eqnarray}
for some other $k_2^\mu \in \mathcal{D}(\mathbbm{M}, \mathcal{P})$. Hence, setting $k^\mu=-k_1^\mu+k^\mu_2$ we obtain the assertion \eqref{varlocal}.
\end{proof}

To prove (\ref{renormconre}) we use that $\Delta_A^{(k)}(e_\otimes^S)$ can be written as 
\beq
\Delta_A^{(k)}(e_\otimes^S)=\int\! dx\, h(x)\,\tilde\Delta^{(k)}(e_\otimes^S;Q(x))\quad\mathrm{with}\quad
\tilde\Delta^{(k)}(e_\otimes^S;Q(x))\in\mathcal{D}(\mathbbm{M}, \mathcal{P})\ ,
\eeq
due to Lemma \ref{Delta:RC}\emph{(ii)}. Hence, (\ref{renormcongrenz})
can be written in the form $\delta\,F=\int l+{\cal O}(\hbar^{k+1})$
(\ref{varglobal}) with $F=D^{(k)}_{>1}(e_\otimes^S)$ and
\beq
-l(x)=D^{(k)}_{>1}(e_\otimes^S\otimes Q(x))\, \tfrac{\delta (S_0+S)}{\delta\varphi(x)}
+D^{(k)}_{>1}\big(e_\otimes^S\otimes {\cal L}^{(1)}(x)\big)\,\d g(x)
+\tilde\Delta^{(k)}(e_\otimes^S;Q(x))\in\mathcal{D}(\mathbbm{M}, \mathcal{P})\ . 
\eeq
With that Lemma \ref{lemma:derivloc} yields our assertion (\ref{renormconre}).

We conclude that a solution of \eqref{renormcon} can be obtained from 
a solution of \eqref{renormcongrenz} by setting
\beq\label{D-extension}
D^{(k)}_{>1}\big(e_\otimes^S\otimes j^\mu(y)\big)\defi K^\mu(e_\otimes^S)(y)\ ,
\eeq
provided this does not lead to any contradictions with the partial fixing of 
$D$ in terms of $D^{(k)}_{>1}(e_\otimes^S)$,
$D^{(k)}_{>1}(e_\otimes^S\otimes Q)$ and $D^{(k)}_{>1}(e_\otimes^S\otimes {\cal L}^{(1)})$. 
Due to the causal Wick expansion (\ref{wickentP}) and the Field equation 
$D^{(k)}_{>1}(e_\otimes^S\otimes \partial^a\varphi)=0$ (see Theorem \ref{maintheorem},\emph{(iii)}),
this is the case whenever the intersection of the subpolynomials of $j$ with the 
subpolynomials of $Q$, ${\cal L}^{(1)}$ or $\mathcal{L}_n$ ($\forall n\in\NN$) contains only  
numbers and terms which are linear in the field $\varphi$ itself or partial derivatives 
thereof.\footnote{In QED this condition is not satisfied; one has to discuss
the individual cases to see that the definition \eqref{D-extension} 
does not lead to contradictions (see e.g.~\cite{Duetsch:1998hf}).}
Note that \eqref{D-extension} satisfies the condition $D^{(k)}_{>1}(j)=0$.
\medskip

\noindent {\it Remark:} If $Q$ is linear in $\varphi$, the maintenance 
of the Field Equation requires $D^{(k)}_{>1}(e_\otimes^S\otimes Q)=0$
(Theorem \ref{maintheorem}).
With that the right-handed side of \eqref{renormcongrenz} vanishes to first order
in $\kappa$ up to terms of order $\hbar^{k+1}$. That is the condition 
\eqref{renormcongrenz} can only be satisfied if $\Delta^{(k)}(S_1)=
{\cal O}(\hbar^{k+1})$. Hence, following the proof of Proposition \ref{MWI:order1},
we first perform a finite renormalization which maintains the considered
renormalization conditions and removes the term $\sim\hbar^k$ of 
$\Delta^{(k)}_A(S_1)$. This can be done such that 
$\Delta^{(k)}_A(e_\otimes^S)={\cal O}(\hbar^{k})$ is preserved. Namely,
since $\Delta^{(k)}_A(S_1)={\cal O}(\hbar^{k})$ the vacuum expectation
values on the right-hand side of (\ref{aMWI first order again}) are of order
$\hbar^k$ and, hence, the pertinent renormalization map $D$ (\ref{D-MWI:order1})
can be chosen of the form (\ref{renormmap}).

\subsubsection{Proof of the Ward identities in the $O(N)$ scalar field model}\label{O(N)model}
In the case of compact internal symmetry groups, covariance can be obtained 
by integration over the group. To illustrate the developed formalism we proceed alternatively.

We will prove that an off-shell generalization of the
Ward identities expressing current conservation in a scalar $O(N)$-model can be 
fulfilled to all orders of perturbation theory. Our strategy is based partially on 
techniques of algebraic renormalization described in detail in \cite{Piguet:1995er}.

We consider a multiplet of $N$ scalar fields
$\varphi_i(x)$, $i=1, \ldots, N$, transforming under the defining
representation of $O(N)$ -- the group of orthogonal $N\times N$ matrices --\footnote{By setting $\varphi=\frac{1}{\sqrt{2}}(\varphi_1+i \varphi_2)$ and
$\bar{\varphi}=\frac{1}{\sqrt{2}}(\varphi_1-i \varphi_2)$ the $O(2)$ model can be seen to be equivalent to the well
known $U(1)$ model of a complex scalar field $\varphi$.}
\begin{equation}\label{trafo}
\varphi_i \longrightarrow A_{ij} \varphi_j \quad , \quad A\in
O(N).\footnote{repeated indices are summed over;}
\end{equation}
Let $\{X^a\,\vert\, a=1,...,\frac{1}{2} N(N-1)\}$ be a basis of the Lie algebra $\mathfrak{o}(N)$ of $O(N)$,
and $f^{abc}$ the corresponding structure constants, 
\begin{equation}\label{lie}
[X^a, X^b]=f^{abc}X^c\ .
\end{equation}

The dynamics of our model is given by the free action
\begin{equation}
S_0=\frac{1}{2}\int\! dx \,\big(\del^\mu \varphi_i(x)\del_\mu
\varphi_i(x)-m^2 \varphi_i(x)\varphi_i(x)\big)
\end{equation}
and the localized, $O(N)$-invariant interaction
\begin{equation}
S=\int\! dx \,g(x) \big(\varphi_i(x) \varphi_i(x)\big)^2\quad
\mathrm{with}\quad g \in \mathcal{D}(\mathbb{M})\ .
\end{equation}
Since the free action
$S_0$ is invariant under the transformation \eqref{trafo}, there exist $\frac{1}{2} N(N-1)$
conserved Noether currents $j^a_\mu=X^a_{ij}\varphi_j \del_{\mu}
\varphi_i$, i.e.~the local functionals $A^a\defi \int\! dx\, h(x)\,
\del^\mu j^a_\mu(x)$ (with arbitrary $h\in \mathcal{D}(\mathbb{M})$) are elements of the ideal $\mathcal{J}$ generated by the free field equations:
\begin{equation}
A^a=
\delta_{A^a}\,S_0\in \mathcal{J}\ ,\quad\mathrm{with}\quad
\delta_{A^a}=\int\! dx \,h(x)\, X^a_{ij}\,\varphi_j(x)\,\frac{\delta}{\delta
\varphi_i(x)}\ .\label{delta}
\end{equation}
Essential simplifications of this model are the validity of (\ref{symmetrycond})  
in simplified form and additionally that $Q$ (\ref{A:loc}) is linear in $\varphi$.

The conservation of the interacting currents
$(j^a_\mu)_S=R(e_\otimes^S, j^a_\mu)$ follows from the MWI for the given $A^a$ and interaction $S$: 
\begin{equation}\label{WI}
\int\! dx \,h(x)\del^\mu R(e_\otimes^S, j^a_\mu(x))\equiv R(e_\otimes^S,
A^a)=\int\! dx\, h(x) R(e_\otimes^S, X^a_{ij} \varphi_j(x))\frac{\delta
S_0}{\delta \varphi_i(x)}\ .
\end{equation}
Regarding the question whether \eqref{WI} can be fulfilled to all orders, we start with an $R$-product satisfying the renormalization conditions Unitarity, ${\cal P}_+^\uparrow$-Covariance,
Field Independence, Field Equation and Scaling Degree. Then, the Ward
identities \eqref{WI} may be violated; however, Theorem \ref{satzQAP} guarantees the 
existence of local maps $\Delta_{A^a}: \TT\mathcal{F}_\mathrm{loc} \rightarrow \mathcal{F}_\mathrm{loc}$ such that
\begin{equation}
R\big(e_\otimes^S,A^a+\Delta_{A^a}(e_\otimes^S)\big)=\int\! dx \,h(x) R(e_\otimes^S, X^a_{ij} \varphi_j(x))\frac{\delta
S_0}{\delta \varphi_i(x)}\,.
\end{equation}

To find a finite renormalization of the $R$-product which removes $R\big(e_\otimes^S,\Delta_{A^a}(e_\otimes^S)\big)$, 
we follow the technique described in the preceding sections: 
we assume the existence of a $\Gamma_T^{(k)}$ such that \eqref{WI} is 
fulfilled up to terms of order $\hbar^{k}$, i.e.
\begin{equation}
\Gamma_T^{(k)}\Bigl(e_\otimes^S\otimes
A^a\Bigr)+\Delta_{A^a}^{(k)}(e_\otimes^S)
 +{\cal O}(\hbar^{k+1})=
\int\! dy\, h(y)\,X^a_{ij}\, \varphi_j(y)\,
\frac{\delta (S_0+\Gamma_T^{(k)}(e_\otimes^S))}{\delta\varphi_i(y)}\
\label{renWI}
\end{equation}
where $\Delta_{A^a}^{(k)}(e_\otimes^S)={\cal O}(\hbar^k)$. 
(The Field Equation for $\Gamma_T^{(k)}$ (Lemma \ref{Gamma:rencond}) is taken into account.) 
First we perform a finite renormalization which maintains 
$\Delta_{A^a}^{(k)}(e_\otimes^S)={\cal O}(\hbar^k)$ and the mentioned 
renormalization conditions and which removes the terms $\sim\hbar^k$ of
$\Delta_{A^a}^{(k)}(S)$.
Due to the requirement $D^{(k)}_{>1}(e_\otimes^S\otimes\varphi)=0$
(Theorem \ref{maintheorem},\emph{(iii)}), the condition (\ref{renormcongrenz}) simplifies to
\beq\label{consistency1}
\Delta_a^{(k)}(e_\otimes^S)=-\delta_a D^{(k)}_{>1}(e_\otimes^S)+{\cal O}(\hbar^{k+1})\ ,
\eeq
where we set $\Delta^{(k)}_{a}(e_\otimes^S)\defi \Delta^{(k)}_{A^a}(e_\otimes^S)\big|_{h\equiv 1}$ 
and $\delta_a\defi \int\! dx\, X^a_{ij}\,\varphi_j(x)\frac{\delta}{\delta
\varphi_i(x)}$. To fulfill \eqref{WI} up to terms of order $\hbar^{k+1}$, we have to solve \eqref{consistency1} and 
to extend the definition of this $D^{(k)}_{>1}$ in such a way that condition \eqref{renormcon} 
(for general $h$) holds true. 
The latter can be done by means of \eqref{D-extension}, because the intersection of the non-trivial subpolynomials
of $j_\mu^a$ with the subpolynomials of $Q$ or $(\varphi_i\varphi_i)^2$ 
is a subset of $\CC\,\varphi$.

It remains to show the solvability of \eqref{consistency1}. For this 
purpose we temporarily restrict the functionals (\ref{functionals})
in \eqref{renWI} to the space $\mathcal{D}(\mathbb{M})$ of
compactly supported test functions on Minkowski space. This permits us
to perform the limit\footnote{This limit is done as follows: let $h\in {\cal D}(\MM)$ such that there is a neighbourhood $U$
of $0(\in\MM)$ with $h\vert_U=1$. Then we replace $h(x)$ by $h_\epsilon(x)\equiv h(\epsilon x)$ ($\,\epsilon >0$) and perform 
the limit $\epsilon\to 0$.}
$h\rightarrow 1$ in \eqref{renWI}, ending up with the equation
\begin{equation}\label{grenzWI}
\delta_{a}(S_0+\Gamma_T^{(k)}(e_\otimes^S))=\Delta^{(k)}_{a}(e_\otimes^S)+\mathcal{O}(\hbar^{k+1}).
\end{equation}
Furthermore, using \eqref{lie} we obtain the identity
\begin{equation}
[\delta_a, \delta_b]=f_{abc}\delta_c\ ,
\end{equation}
which we insert into $[\delta_a, \delta_b](S_0+\Gamma_T^{(k)}(e_\otimes^S))$. This yields the \emph{consistency condition}
\begin{equation}\label{conscon}
\delta_a \Delta^{(k)}_{b}(e_\otimes^S)-\delta_b
\Delta^{(k)}_{a}(e_\otimes^S)=f_{abc}\,\Delta^{(k)}_{c}(e_\otimes^S)+\mathcal{O}(\hbar^{k+1})\ .
\end{equation}
Due to the compact support of $S$, and the locality of $\Delta^{(k)}_{a}$, each term in 
\eqref{conscon} has compact support as well. Therefore, this equation holds true
on the entire configuration space $\mathcal{C}(\mathbb{M}, \mathbb{R})$, i.e.~the
restriction of the functionals to ${\cal D}(\MM)$ can be omitted.

The consistency condition \eqref{conscon} is the cocycle condition in the Lie algebra cohomology corresponding 
to the Lie algebra generated by the derivations $\{\delta_a\}$ acting on ${\cal F}_\mathrm{loc}$. 
Trivial solutions are the coboundaries,
\beq\label{consistencysolution} 
\Delta_a^{(k)}(e_\otimes^S)=-\delta_a \hat{\Delta}^{(k)}(e_\otimes^S)+\mathcal{O}(\hbar^{k+1}) 
\eeq
for some linear, symmetric and local map $\hat{\Delta}^{(k)}:
\TT\mathcal{F}_\mathrm{loc}\rightarrow \mathcal{F}_\mathrm{loc}$.
If $\Delta^{(k)}_{a}(e_\otimes^S)$ is a coboundary, the condition
\eqref{consistency1} can be solved by setting 
\beq
D^{(k)}_{>1}(1)\defi 0\ ,\quad
D^{(k)}_{>1}(S)\defi 0\quad\mathrm{and}\quad
D^{(k)}_{>1}(S^{\otimes j})\defi \hat{\Delta}^{(k)}(S^{\otimes j})\quad\forall j\geq 2\ ,
\eeq
due to $\Delta_a^{(k)}(S)=\mathcal{O}(\hbar^{k+1})$.

Hence, we only have to show that the present cohomology is trivial. 
In the literature (see e.g.~\cite{Piguet:1995er} and references therein) 
it is shown that every Lie algebra cohomology 
corresponding to some semi-simple Lie group and some finite dimensional representation is trivial.
This result applies to our problem. Namely, $O(N)$ is semi-simple for $N>2$ and,  
since the mass dimension of $\Delta^{(k)}_{a}(e_\otimes^S)$ is bounded \eqref{massdimbound}, 
the anomaly terms indeed span a finite dimensional representation of $\mathfrak{o}(N)$. 
It does not matter that our functionals are {\it local}.
Note that $D^{(k)}_{>1}(e_\otimes^S)$ is not uniquely defined by this procedure.
\section{Conclusions and Outlook}\setcounter{equation}{0}
In algebraic renormalization the QAP is used to remove possible anomalies of Ward identities by induction on $\hbar$.
We have worked out an analogous procedure for the MWI in the different framework of causal perturbation theory. The 
main difference is that we work solely with compactly supported interactions 
$S$ and localized symmetry transformations $\delta_A$.\footnote{Algebraic 
renormalization applies to global and local symmetries; examples for local symmetries
which have been dealt are
current algebras of $\sigma$-models and current algebras of gauge theories in which one 
keeps external fields (e.g.~antifields).} 
Our main result gives a crucial insight into the structure of possible anomalies of the MWI, 
in particular with respect to the deformation parameter $\hbar$, 
and allows the transfer of techniques from algebraic renormalization into causal perturbation theory. 

This yields a general method to fulfill the MWI for a given model. A first non-trivial application is worked 
out (Sect.~\ref{O(N)model}). The developed method seems to be applicable to many
models (as suggested by \cite{Duetsch:2001sw}, \cite{Barnich:2000zw} and \cite{Piguet:1995er}).
Together with the powerful tool of BRST cohomology it should  
make possible a proof of that cases of the MWI which are needed for the
construction of the net of local observables of Yang-Mills type QFTs. (This would complete the construction given in
\cite{Duetsch:2001sw}.) A main advantage of this approach to quantum Yang-Mills theories is that there seems 
to be no serious obstacle for the generalization to curved spacetimes where the techniques developed for scalar fields
in \cite{Brunetti:1999jn} and \cite{Hollands:2004yh} can be used. 
For recent and far-reaching progress in the construction of renormalized quantum Yang-Mills fields 
in curved spacetime see \cite{Hollands:2007}; this paper uses a generalization of the off-shell Master BRST Identity
(i.e.~the MWI for the symmetry transformation $\delta_{A_0}$ with $A_0$ given by (\ref{cc:0}), see 
\cite{Duetsch:2002yp,Duetsch:2001sw} for the on-shell version) to models with antifields.

\appendix
\section{Proper vertices for $R$-products}\setcounter{equation}{0}\label{propVR}
\subsection{Definition and basic properties}
Before we introduce proper vertices in terms of $R$-products, we shortly 
consider the diagrammatics of retarded products $R(A_1(x_1),...;A_n(x_n))$.

For the 
{\bf unrenormalized} expressions (i.e.~for $x_i\not=x_j\,,\,\,\forall i\not= j$)
the diagrammatic interpretation is unique. One can show that there are two kinds 
of inner lines which symbolize $\Delta^{\rm ret}$ and $H^\mu_m$, respectively, 
and are oriented. For tree diagrams only $\Delta^{\rm ret}$ appears and all 
inner lines are pointing to the distinguished vertex $A_n(x_n)$. Solely 
connected diagrams contribute to $R$; and $\Rcl$ is 
precisely the contribution of all tree diagrams. Both statements follow from the 
inductive construction of the $(R_{n,1})_{n\in\NN}$ \cite{Duetsch:2004dd}. The decisive
step is the GLZ Relation: $\{F_G,H_G\}=...$ . In the quantum case 
there is at least one contraction between 
$F_G$ and $H_G$, and in classical FT there 
is precisely one contraction in $\{F_G,H_G\}_{\rm cl}$. 

For the {\bf renormalized} retarded product we use these results as definition of the 
connected and tree part: $R^c\equiv R$ and $R_{\rm tree}\equiv \Rcl$.
\medskip

Analogously to (\ref{T(1)})-(\ref{Tps2}) the property $R_{({\rm cl})}(1,F)=F$
implies the following conclusions 
\begin{gather}
R_{({\rm cl})}\Bigl(e_\otimes^{\sum_{n=1}^\infty F_n\lambda^n}\,,\,
\sum_{n=0}^\infty G_n\lambda^n\Bigr)=0\,\,\Longrightarrow\,\, 
G_n=0\,\,\,\,\forall n\ ,\label{Rps1}\\
R_{({\rm cl})}\Bigl(e_\otimes^{\sum_{n=1}^\infty F_n\lambda^n}\,,\,
\sum_{n=0}^\infty \frac{\delta F_n}{\delta\varphi}\lambda^n\Bigr)=
R_{({\rm cl})}\Bigl(e_\otimes^{\sum_{n=1}^\infty G_n\lambda^n}\,,\,
\sum_{n=0}^\infty \frac{\delta G_n}{\delta\varphi}\lambda^n\Bigr)\notag\\
\wedge\quad\omega_0(F_n)=\omega_0(G_n)\,\,\forall n\,\quad\Longrightarrow\,\quad
F_n=G_n\quad\forall n\ ,\label{Rps2}
\end{gather}
which hold for the $R$-products of classical FT ($\Rcl$) and
of QFT ($R$). In classical FT the statement (\ref{Rps1}) can be proved also 
non-perturbatively: $0=\Rcl(e_\otimes^{F(\lambda)},G(\lambda))=G(\lambda)\circ
r_{S_0+F(\lambda),S_0}$ implies $G(\lambda)=G(\lambda)\circ r_{S_0+F(\lambda),S_0}
\circ r_{S_0,S_0+F(\lambda)}=0$.
\medskip

The concept of proper vertices has a clear physical interpretation
when applied to $R$-products, since $R_{\rm tree}=\Rcl$. As explained in 
Sect.~2 the entries of $\Rcl$ may be {\it non-local}. 
We want to rewrite an interacting  QFT-field
$R(e_\otimes^S,F)$ as a classical field $\Rcl(e_\otimes^{\Gamma_R(e_\otimes^S)}, 
\gr(e_\otimes^S, F)\Big)$ where the 'proper interaction'
$\Gamma_R(e_\otimes^S)$ and the 'proper retarded field'
$\gr(e_\otimes^S, F)$ are non-local and agree to lowest order in $\hbar$
with the original local functionals $S$ and $F$ respectively. This is indeed possible:
\begin{prop}
\begin{itemize}
\item[(a)]
There exist
\begin{itemize}
\item a totally symmetric and linear map
\beq
\Gamma_R\> :\>\TT\mathcal{F}_\mathrm{loc}\longrightarrow\mathcal{F}\label{Gamma_R}
\eeq
\item and a linear map
\beq
\gr\>:\>\TT\mathcal{F}_\mathrm{loc}\otimes\mathcal{F}_\mathrm{loc}
\longrightarrow\mathcal{F}\label{retGamma}
\eeq
which is totally symmetric in the former entries (i.e.~
$\gr(\otimes_{j=1}^n F_{\pi j},F)=\gr(\otimes_{j=1}^n F_{j},F)$), 
\end{itemize}
which are uniquely determined by the conditions
\begin{gather}
R(e_\otimes^S, F)=\Rcl\Big(e_\otimes^{\Gamma_R(e_\otimes^S)}, \gr(e_\otimes^S, F)\Big)\ ,
\label{defeff}\\
\gr(e_\otimes^S, \varphi(h))=\varphi(h)\ ,\quad\quad
h\in\mathcal{D}(\mathbb{M})\ ,\label{triv}\\
\Gamma_R(\mathbf{1})=0\ ,\quad\quad\omega_0(\Gamma_R(e_\otimes^S))=0\ .\label{O(G)}
\end{gather}
\item[(b)] $\Gamma_R$ and $\gr$ are related by
\begin{equation}\label{efffeldgl}
\frac{\delta \Gamma_R(e_\otimes^S)}{\delta \varphi(x)}=
\gr\left(e_\otimes^S, \frac{\delta S}{\delta \varphi(x)}\right)\ ,
\end{equation}
that is, with (\ref{O(G)}), $\Gamma_R$ is uniquely determined by $\gr$.
\end{itemize}
\end{prop}
Compared with the defining condition (\ref{Gamma}) for $\Gamma_T$, there is more 
flexibility in (\ref{defeff}) since it contains two kinds of 'vertex functions', 
$\Gamma_R$ and $\gr$. To define the latter {\it uniquely}, we additionally require 
(\ref{triv}) and (\ref{O(G)}).

\begin{proof}
(b) First we show that the defining conditions for $\Gamma_R\,,\,\gr$ given in part (a)
imply the statement in part (b). The off-shell field equation 
\begin{equation}\label{wwfeldgl}
R_{(\mathrm{cl})}\left(\ets, \frac{\delta S_0}{\delta \varphi(x)}\right)=\frac{\delta S_0}{\delta \varphi(x)}-R_{(\mathrm{cl})}\left(\ets, \frac{\delta S}{\delta \varphi(x)}\right),
\end{equation}
holds for $R$ (QFT) and $R_{\mathrm{cl}}$ (classical FT), in the latter case even 
for non-local entries. With that and using the conditions (\ref{defeff}) and
(\ref{triv}) and finally (\ref{Rps1}) we obtain the assertion (\ref{efffeldgl}).

(a) By expanding (\ref{defeff}) in powers of $S$ and using (\ref{efffeldgl}) and
(\ref{O(G)}) we find an inductive construction of $\Gamma_R$ and $\gr$ in terms of 
$R$ and $\Rcl$:
\begin{subequations}\label{ent}
\begin{equation}\label{ent0}
\gr(1,F)=R(F)\equiv F\ ,
\end{equation}
\beq\label{g-gr:1}
\frac{\delta \Gamma_R(S)}{\delta \varphi(x)}=\gr\Bigl(1,
\frac{\delta S}{\delta \varphi(x)}\Bigr)=\frac{\delta S}{\delta \varphi(x)}
\quad\Longrightarrow \Gamma_R(S)=S\ ,
\eeq
\begin{equation}\label{ent1}
\gr(S, F)=R(S, F)-\Rcl(S, F)
\end{equation}
\begin{equation}\label{g-gr:2}
\frac{\delta \Gamma_R(S^{\otimes 2})}{\delta \varphi(x)}=
2\gr\left(S, \frac{\delta S}{\delta \varphi(x)}\right)=
2R\left(S,\frac{\delta S}{\delta \varphi(x)}\right)-
2\Rcl\left(S,\frac{\delta S}{\delta \varphi(x)}\right)\ ,
\end{equation}
\begin{equation}\label{ent2}
\gr(S^{\otimes 2}, F)=R(S^{\otimes 2},F)-\Rcl(S^{\otimes 2},F)-
\Rcl(\Gamma_R(S^{\otimes 2}), F)-2\Rcl(S, \gr(S, F))\ .
\end{equation}
\end{subequations}
We explicitly see that $\gr$ is not totally symmetric, it is retarded with respect 
to the last entry.

Now let $\Gamma_R(\ets)$ and $\gr(\ets, F)$ be constructed up to order $n$ and 
$(n-1)$ respectively in $S$. Then, the condition (\ref{defeff}) determines
$\gr(S^{\otimes n}, F)$ uniquely:
\begin{equation*}
\gr(S^{\otimes n}, F)=R(S^{\otimes n}, F)-\sum_{k=1}^n {n \choose k}
\Rcl\left(\frac{d^k}{d\lambda^k}\Big|_{\lambda=0}
e_\otimes^{\Gamma_R(e_\otimes^{\lambda S})},\gr(S^{\otimes n-k}, F)\right)\end{equation*}
\begin{equation}\label{allent}
=R(S^{\otimes n}, F)-\sum_{k=1}^n {n \choose k}
\sum_{j=1}^k \hspace{-0.2cm} 
\sum_{\begin{array}{c} \scriptstyle{l_1,\ldots ,l_j=1}\\ 
\scriptstyle{l_1+\cdots+l_j=k} \end{array}}^k \hspace{-0.5cm}
\frac{1}{j!l_1!\cdots l_j!} \, 
\Rcl\left(\bigotimes_{i=1}^j\Gamma_R(S^{\otimes l_i}), \gr(S^{\otimes n-k}, F)\right).
\end{equation}
>From that and with (\ref{efffeldgl}) and (\ref{O(G)}) we uniquely 
get $\Gamma_R(S^{\otimes n+1})$.
\end{proof}
\noindent{\it Remark:} The roles of the conditions (\ref{triv}) and (\ref{efffeldgl})
can be exchanged. In the list (\ref{defeff})-(\ref{O(G)}) of defining conditions, 
(\ref{triv}) can be replaced by (\ref{efffeldgl}). Then, (\ref{triv}) can be derived 
from (\ref{defeff}), (\ref{O(G)}) and (\ref{efffeldgl}) analogously to 
(\ref{feq:T}) and (\ref{feq:G_T}): proceeding inductively we use (\ref{allent}),
the integrated field equation for $R$ and $\Rcl$ and (\ref{efffeldgl}).
\medskip

Following the construction (\ref{ent})-(\ref{allent}) we inductively 
prove the following properties of  $\Gamma_R\,,\,\gr$:
\begin{itemize}
\item {\bf $\hbar$-dependence}:
\begin{equation}
\left.
\begin{array}{rl}
\gr(\ets, F)=F+\mathcal{O}(\hbar)&\\
\Gamma_R(\ets)=S+\mathcal{O}(\hbar)&
\end{array}\right\} \qquad \mbox{if} \quad F, S \sim \hbar^0.\label{Gamma_R:hbar}
\end{equation}
\item {\bf ${\cal P}_+^\uparrow$-Covariance, Unitarity
($\Gamma_R(e_\otimes^S)^\ast=\Gamma_R(e_\otimes^{S^\ast})$
and similarly for $\gr$), Field Independence, 
Smoothness in $m\geq 0$, $\mu$-Covariance and Almost homogeneous Scaling
of $\Gamma_R\equiv\Gamma_R^{(m,\mu)}$ and  $\gr\equiv\gr^{(m,\mu)}\,$
(or, alternatively, Scaling Degree).}
\end{itemize}
In the proof of (\ref{Gamma_R:hbar}) we use $R=\Rcl +{\cal O}(\hbar)\,,\,\Rcl\sim\hbar^0$. The other properties 
rely on the validity of the corresponding axioms for $R$ and $\Rcl$, analogously to Lemma \ref{Gamma:rencond}.

We point out that $\Gamma_R(\ets\otimes\varphi(h))$ differs in general from 
$\varphi(h)$, in contrast to the Field equation for $\Gamma_T$ and (\ref{triv}).
Namely, inserting (\ref{defeff})-(\ref{triv}) into the GLZ-relation for 
$[R(\ets,\varphi(h)),R(\ets,\varphi(g))]$ we obtain
\begin{gather}
\frac{i}{\hbar}\,\Bigl[\Rcl\Bigl(e_\otimes^{\Gamma_R(\ets)},\varphi(h)\Bigr)\,,\,
\Rcl\Bigl(e_\otimes^{\Gamma_R(\ets)},\varphi(g)\Bigr)\Bigr]=\notag\\
\Rcl\Bigl(e_\otimes^{\Gamma_R(\ets)}\otimes\Gamma_R(\ets\otimes\varphi(g)),
\varphi(h)\Bigr)-\Rcl\Bigl(e_\otimes^{\Gamma_R(\ets)}\otimes
\Gamma_R(\ets\otimes\varphi(h)),\varphi(g)\Bigr)\ .
\end{gather}
Due to the GLZ Relation for $\Rcl$, the left-hand side is equal to
\begin{gather}
\Rcl\Bigl(e_\otimes^{\Gamma_R(\ets)}\otimes\varphi(g),
\varphi(h)\Bigr)-\Rcl\Bigl(e_\otimes^{\Gamma_R(\ets)}\otimes\varphi(h),\varphi(g)\Bigr)\notag\\
+\frac{i}{\hbar}\,\Bigl[\Rcl\Bigl(e_\otimes^{\Gamma_R(\ets)},\varphi(h)\Bigr)\,,\,
\Rcl\Bigl(e_\otimes^{\Gamma_R(\ets)},\varphi(g)\Bigr)\Bigr]_{\star^{(\geq 2})}\ ,
\end{gather}
where $\frac{i}{\hbar} [\cdot,\cdot]_{\star^{(\geq 2})}\equiv 
\frac{i}{\hbar} [\cdot,\cdot]_{\star}-\{\cdot,\cdot\}_\mathrm{cl}\,$. The assertion follows from 
the non-vanishing of the $[\cdot,\cdot]_{\star^{(\geq 2})}$-term.

Finally we study a finite renormalization $R\rightarrow \hat R$.
To express the corresponding renormalizations of $\Gamma_R$ and $\gr$ in terms of the 
corresponding map $D$ of the Main Theorem we insert the defining relation 
(\ref{defeff}) into both sides of (\ref{ren:intf}).
In the resulting equation
\beq
\Rcl\Bigl(e_\otimes^{\hat{\Gamma}_R(e_\otimes^S)}\,,
\,\hat{\Gamma}_\mathrm{ret}(\ets, F)\Bigr)=
\Rcl\Bigl(e_\otimes^{\Gamma_R\Bigl(e_\otimes^{D(e_\otimes^S)}\Bigl)}\,,\,
\gr\Bigl(e_\otimes^{D(e_\otimes^S)}, D(e_\otimes^S\otimes F)\Bigr)\Bigr)\label{ReffD}
\eeq
we choose $F=\frac{\delta S}{\delta\varphi}$. With that we may replace
$\hat{\Gamma}_\mathrm{ret}(\ets, F)$ by 
$\frac{\delta \hat{\Gamma}_R(e_\otimes^S)}{\delta\varphi}$
and\\
$\gr\Bigl(e_\otimes^{D(e_\otimes^S)}, D(e_\otimes^S\otimes F)\Bigr)$
by $\frac{\delta \Gamma_R(e_\otimes^{D(e_\otimes^S)})}{\delta\varphi}$.
By means of (\ref{Rps2}) we conclude
\beq
\hat{\Gamma}_R(e_\otimes^S)=\Gamma_R(e_\otimes^{D(e_\otimes^S)})\ .\label{grenbed}
\eeq
We insert this into (\ref{ReffD}) and apply (\ref{Rps1}). This yields
\beq
\hat{\Gamma}_\mathrm{ret}(e_\otimes^S, F)=
\gr\Bigl(e_\otimes^{D(e_\otimes^S)}, D(e_\otimes^S \otimes F)\Bigr)\ .\label{grrenbed}
\eeq
\subsection{Comparison of the vertex functions in terms of $T$- and $R$-products}
\label{comp-vertfunc}
The vertex functions $\Gamma_T$ and $\Gamma_R$ 
defined in terms of $T$- and $R$-products, respectively, are both totally symmetric,
nevertheless they do not agree. This follows from  
the different forms of the unitarity property or, alternatively, from the non-validity of 
$\Gamma_T(\ets\otimes\varphi(h))=\varphi(h)$ for $\Gamma_R$. We are going to compare $\Gamma_T$ 
with $\Gamma_R$ to lowest orders in $S$.

By using the definitions of $\Gamma_R,\Gamma_{\rm ret}$ (\ref{defeff}) 
and $\Gamma_T$ (\ref{Gamma}), as well as (\ref{R-T})-(\ref{bar-T})
we obtain
\begin{gather}
R_{\rm cl}\Bigl(e_\otimes^{\Gamma_R(e_\otimes^S)},
\Gamma_{\rm ret}(e_\otimes^S,F)\Bigr)=\notag\\
\sum_{n=0}^\infty \Bigl(1- T_{\rm tree}
\Bigl(e_\otimes^{i\Gamma_T(e_\otimes^S)/\hbar}\Bigr)\Bigr)^{\star n}\star
T_{\rm tree}\Bigl(e_\otimes^{i\Gamma_T(e_\otimes^S)/\hbar}
\otimes \Gamma_T(e_\otimes^S\otimes F)\Bigr)\ .\label{G:R-T}
\end{gather}
If we interpret $\Gamma_R,\Gamma_{\rm ret}$ and $\Gamma_T$
as one vertex, then the l.h.s.~contains solely tree diagrams, 
but on the r.h.s.~there appear also loop diagrams! 
This indicates that the relation of
$\Gamma_R$ to $\Gamma_T$ is rather involved. 
To zeroth and first order in $S$ we obtain
\begin{gather}
\Gamma_{\rm ret}(1,F)=\Gamma_T(1\otimes F)=F\ ,\label{G:R-T:0}\\
\Gamma_{\rm ret}(S,F)+R_{\rm cl}(S,F)=i/\hbar\,\>T_{\rm tree}(S\otimes F)
+\Gamma_T(S\otimes F)-i/\hbar\,\>S\star F\label{G:R-T:1}
\end{gather}
The terms $\sim\hbar^{-1}$ and $\sim\hbar^0$ of (\ref{R-T}) read
\beq
R_{\rm cl}(S,F)=i/\hbar\,\>T_{\rm tree}(S\otimes F)
-i/\hbar\,\>S\star^{(\leq 1)} F\ ,\label{R-T:tree2}
\eeq
where
\begin{gather}
S\star^{(\leq 1)} F=
  \sum_{n\leq 1}\frac{\hbar^n}{n!}
  \int\! dx_1\ldots dx_n dy_1\ldots dy_n 
  \frac{\delta^n S}{\delta\varphi(x_1)\cdots\delta\varphi(x_n)}
\notag\\
  \cdot \prod_{i=1}^n H^\mu_m(x_i-y_i) 
  \frac{\delta^n F}{\delta\varphi(y_1)\cdots\delta\varphi(y_n)}\ .
\end{gather}
In the same way we define $S\star^{(\geq 2)} F$
(i.e.~$S\star F=S\star^{(\leq 1)} F+S\star^{(\geq 2)} F$) and e.g.
$S\star^{(2)} F$. With that (\ref{G:R-T:1}) reads
\begin{gather}
\Gamma_T(S\otimes F)=\Gamma_{\rm ret}(S,F)+i/\hbar\,\>
S\star^{(\geq 2)}F\notag\\
=\frac{1}{2}\Bigl(\Gamma_{\rm ret}(S,F)+\Gamma_{\rm ret}(F,S)+i/\hbar\,\,(
S\star^{(\geq 2)}F+F\star^{(\geq 2)}S)\Bigr)\ .\label{G:R-T:1a}
\end{gather}
Using additionally the Field Independence of $\Gamma_T$
and (\ref{g-gr:2}) we find
\beq
\frac{\delta}{\delta\varphi}\Bigl(\Gamma_T(S\otimes F)-
\Gamma_R(S\otimes F)\Bigr)=\frac{i}{\hbar}\,\Bigl(S\star^{(\geq 2)}
\frac{\delta F}{\delta\varphi}+
F\star^{(\geq 2)}\frac{\delta S}{\delta\varphi}\Bigr)\ .\label{G:R-T:1b}
\eeq

Selecting the terms of second order in $S$ from (\ref{G:R-T}) we find
\begin{gather}
1/2\,\Gamma_{\rm ret}(S^{\otimes 2},F)+\Rcl(S,\Gamma_{\rm ret}(S,F))
+1/2\,\Rcl(S^{\otimes 2},F)+1/2\,
\Rcl(\Gamma_R(S^{\otimes 2}),F)=\notag\\
1/2\,\Gamma_T(S^{\otimes 2}\otimes F)+i/\hbar\,\,T_{\rm tree}(S\otimes\Gamma_T(S\otimes F))
+i/2\hbar\,\,T_{\rm tree}(\Gamma_T(S^{\otimes 2})\otimes F)\notag\\
-1/2\hbar^2\,T_{\rm tree}(S^{\otimes 2}\otimes F)
-i/\hbar\,\,S\star\Gamma_T(S\otimes F)+1/\hbar^2\,\,S\star T_{\rm tree}(S\otimes F)\notag\\
-i/2\hbar\,\,\Gamma_T(S^{\otimes 2})\star F+1/2\hbar^2\,\,
T_{\rm tree}(S^{\otimes 2})\star F-1/\hbar^2\,\,S\star S\star F\ .\label{G:R-T:2}
\end{gather}
To  simplify this formula and to eliminate all vertex functions $\Gamma_{\rm ret}\,,\,
\Gamma_R\,,\,\Gamma_T$ with two arguments, we use
(\ref{G:R-T:1a}), (\ref{g-gr:2}), the Field Independence, as well as
\beq
\Rcl(G,H)=\int\! dx\,dy\,\frac{\delta H}{\delta\varphi(x)}\,
\Delta^{\rm ret}(x-y)\,\frac{\delta G}{\delta\varphi(y)}
\eeq 
and the corresponding expression for $T^c_{\rm tree}(G,H)$,
in addition (\ref{R-T:tree2}) and the corresponding identity
\begin{gather}
1/2\,\, R_{\rm cl}(S^{\otimes 2},F)=\frac{1}{\hbar^2}\,\Bigl(-1/2\,
T_{\rm tree}(S^{\otimes 2}\otimes F)\notag\\
+(S\star T(S\otimes F))_{\rm tree}+1/2\,(T(S^{\otimes 2})\star F)_{\rm tree}
+(S\star S\star F)_{\rm tree}\Bigr)\ ,
\end{gather}
and also
\beq
T_{\rm tree}(S\otimes F)-i\hbar\,\Gamma_T(S\otimes F)=T(S\otimes F)
\eeq
(which follows from Corollary \ref{G=T} and (\ref{T^1PI})). It results
\begin{gather}
1/2\,\Bigl(\Gamma_{\rm ret}(S^{\otimes 2},F)-
\Gamma_T(S^{\otimes 2}\otimes F)\Bigr)=\notag\\
\frac{i}{\hbar}\int\! dx\,dy\,\frac{\delta F}{\delta\varphi(x)}\,\Delta^{\rm ret}(x-y)\,
\Bigl(S\star^{(\geq 2)}\frac{\delta S}{\delta\varphi(y)}\Bigr)+\frac{i}{\hbar}\,
\Rcl(S\,,\,S\star^{(\geq 2)}F)\notag\\
-\frac{1}{\hbar^2}\sum_{a+b=3}(S\star^{(\geq a)}S\star^{(\geq b)}F)^c+
\frac{1}{\hbar^2}\Bigl((S\star^{(\geq 2)}T(S\otimes F))^c
-(S\star^{(2)}(S\cdot F))_{\rm tree}\Bigr)\notag\\
+\frac{1}{2\hbar^2}\Bigl((T(S^{\otimes 2})\star^{(\geq 2)}F)^c-
((S\cdot S)\star^{(2)}F)_{\rm tree}\Bigr)\ .
\label{G:R-T:2a}
\end{gather}
In comparison with (\ref{G:R-T:2}) a main simplification is that
on the right-hand side solely connected diagrams contribute and the cancellation of all 
tree diagrams is obvious (i.e.~the right-hand side is manifestly of order $\hbar$). 

We have not succeeded to generalize the results (\ref{G:R-T:1a}), (\ref{G:R-T:1b})
and (\ref{G:R-T:2a}) to a general formula relating $\Gamma_T$ to $\Gamma_{\rm ret}$
or $\Gamma_R$.
\medskip

\noindent{\bf Acknowledgment:}
This paper is to a large extent based on the diploma thesis of one of us (F.B.) 
\cite{Brennecke:2005}, which was supervised by Klaus Fredenhagen.
We profitted from discussions with him in many respects: 
he gave us important ideas, technical help and also suggestions for 
the presentation of the material. We are grateful also to 
Raymond Stora for valuable and detailed comments on the manuscript, which we 
used to improve some formulations.

\bibliography{biblio}

\end{document}